\documentclass[letterpaper,notitlepage,nofootinbib,superscriptaddress]{revtex4-1} 

\usepackage[utf8]{inputenc}
\usepackage{amsmath,amssymb,amsfonts}
\usepackage{graphicx,wrapfig}
\usepackage{dcolumn}
\usepackage{bm}
\usepackage{color}
\usepackage{bbold}
\usepackage{url}
\usepackage{slashed}
\usepackage[svgnames]{xcolor}
\usepackage[english]{babel}
\usepackage{microtype}

\usepackage[colorlinks=true,linkcolor=blue,urlcolor=blue,citecolor=blue]{hyperref}

\usepackage{color}

\definecolor{darkgreen}{rgb}{0,0.65,0}

\newcommand{\be}{\begin{equation}}
\newcommand{\ee}{\end{equation}}
\newcommand{\ba}{\begin{eqnarray}}
\newcommand{\ea}{\end{eqnarray}}
\newcommand{\sub}[1]{	\begin{subequations}
			#1
		     	\end{subequations} }

\newcommand{\la}{\langle}
\newcommand{\ra}{\rangle}
\begin{document}

\title{Forces inside hadrons: 
pressure, surface tension, mechanical radius, and all that}

\author{Maxim V.~Polyakov}
	\affiliation{Petersburg Nuclear Physics Institute, 
		Gatchina, 188300, St.~Petersburg, Russia}
	\affiliation{Institut f\"ur Theoretische Physik II, 
		Ruhr-Universit\"at Bochum, D-44780 Bochum, Germany}
\author{Peter~Schweitzer}
	\affiliation{Department of Physics, University of Connecticut, 
		Storrs, CT 06269, USA}

\begin{abstract}
The physics related to the form factors of the energy momentum 
tensor spans a wide spectrum of problems, and includes gravitational 
physics, hard exclusive reactions, hadronic decays of heavy quarkonia, 
and the physics of exotic hadrons described as hadroquarkonia. It also 
provides access to the ``last global unknown property:'' the $D$-term.
We review the physics associated with the form factors of the 
energy-momentum tensor and the $D$-term, their interpretations 
in terms of mechanical properties, their applications, and the 
current experimental status. 
\end{abstract}

\date{May 2018} 

\maketitle

\tableofcontents


\section{Introduction}

Such fundamental properties of a particle as mass and spin can be viewed 
as the response of a particle to external gravitational field (change of 
space-time metric). As the gravity couples to matter via the energy-momentum 
tensor (EMT), the mass and the spin of a particle can be obtained from 
the matrix elements of EMT  \cite{Kobzarev:1962wt,Pagels:1966zza}.
However, the EMT matrix elements contain one more fundamental information: 
the $D$-term \cite{Polyakov:1999gs}.\footnote{%
	The name ``$D$-term'' was coined in Ref.~\cite{Kivel:2000fg} and can 
	be traced back to the notation chosen in Ref.~\cite{Polyakov:1999gs}.
	The similarity to the ``$D$-term'' used in the terminology of 
	supersymmetric theories is mostly accidental \cite{Kivel:2000fg}.}  
The corresponding quantity is related to the 
variation of the spatial components of the space-time metric, i.e.\ 
to spatial deformations. Therefore, the $D$-term is expressed in 
terms of the stress-tensor which characterizes the distribution of forces 
inside the system (see, e.g. \cite{LLv7}).  
This allows one to define ``mechanical properties''
of hadrons \cite{Polyakov:2002yz}.

The $D$-term is a characteristics which is as fundamental as the mass and 
the spin of a particle.  For free particles (and also for Goldstone bosons)
its value is fixed by general principles --- in the same way as, e.g.\ the 
gyromagnetic ratio of the free electron is fixed.
In strongly interacting systems the $D$-term is sensitive to correlations 
in the system. For example, the baryon $D$-term behaves as $\sim N_c^2$ 
whereas all other global observables (mass, magnetic moments, axial charge, 
etc.) behave at most as $\sim N_c$ in the limit of a large number of
colors $N_c$.

The $D$-term is a contribution to unpolarized generalized parton 
distributions (GPDs) in the region $-\xi \le x \le \xi$ \cite{Polyakov:1999gs}
which can be accessed through studies of hard exclusive reactions
\cite{Mueller:1998fv,Ji:1996ek,Radyushkin:1996nd,Radyushkin:1996ru,
Ji:1996nm,Collins:1996fb,Radyushkin:1997ki,Vanderhaeghen:1998uc,
Belitsky:2001ns,Goloskokov:2005sd,Kumericki:2007sa,Kumericki:2009uq,
Mueller:2013caa,Kumericki:2015lhb,Muller:2014wxa},
see \cite{Ji:1998pc,Radyushkin:2000uy,Goeke:2001tz,Diehl:2003ny,
Belitsky:2005qn,Boffi:2007yc,Guidal:2013rya,Kroll:2014tma} 
for reviews.
The $D$-term determines the asymptotics of GPDs in the limit of 
renormalization scale $\mu\to\infty$ \cite{Goeke:2001tz}, 
appears in Radon transforms \cite{Teryaev:2001qm}, and emerges 
as a subtraction constant in fixed-$t$ dispersion relations for 
DVCS amplitudes\footnote{It seems for the first time this was 
	shown in Ref.~\cite{Polyakov:2002yz} for dispersion relations 
	at small Bjorken $x$ (see Eq.~(4) of that paper).}
\cite{Teryaev:2005uj,Anikin:2007yh,Diehl:2007jb,Radyushkin:2011dh}. 

The physics of EMT form factors is important not only for the description 
of hadrons in strong gravitational fields and hard exclusive processes. 
It is also {relevant} for the QCD description of hadronic decays of heavy 
quarkonia \cite{Novikov:1980fa,Voloshin:1980zf,Voloshin:1982eb}, 
played an important role for the physics of light Higgs bosons 
(at a time when a light Higgs was not excluded)
\cite{Leutwyler:1989tn,Donoghue:1990xh},
and for the quantification of the picture of pentaquarks and tetraquarks 
with hidden charm as hadroquarkonia \cite{Eides:2015dtr}.

The purpose of this article is to provide a concise review of the physics 
associated with the $D$-term and EMT form factors, their interpretation 
in terms of mechanical properties, their applications, and the current
experimental status.

\section{The EMT of QCD}
\label{Sec-2:EMT-in-QCD-general-definitions}

The EMT of QCD can be obtained by varying the action $S_{\rm grav}$ of QCD 
coupled to a weak classical torsionless gravitational background field with 
respect to the metric $g^{\mu\nu}(x)$ of this curved background field according 
to 
\be\label{Eq:EMT-from-gravity}
	\hat{T}_{\mu\nu}(x)=\frac{2}{\sqrt{-g}}\,
	\frac{\delta S_{\rm grav}}{\delta g^{\mu\nu}(x)}
\ee
where $g$ denotes the determinant of the metric  
{(the signature of the metric we use is $+---$)}. This procedure 
(see e.g.\ App.~E of \cite{Belitsky:2005qn} for a pedagogical description) 
yields a symmetric Belinfante-Rosenfeld EMT. The quark and gluon 
contributions to the total EMT operator are given by 
\sub{\label{eq:EMT-QCD}
\ba
	T^{\mu\nu}_q &=& \frac{1}{4}\overline{\psi}_q\biggl(
	-i\overset{ \leftarrow}{\cal D}{ }^\mu\gamma^\nu
	-i\overset{ \leftarrow}{\cal D}{ }^\nu\gamma^\mu
	+i\overset{\rightarrow}{\cal D}{ }^\mu\gamma^\nu
	+i\overset{\rightarrow}{\cal D}{ }^\nu\gamma^\mu\biggr)\psi_q
	-g^{\mu\nu}\overline{\psi}_q\biggl(
	-\frac{i}{2}\,\overset{ \leftarrow}{\slashed{\cal D}}{ }
	+\frac{i}{2}\,\overset{\rightarrow}{\slashed{\cal D}}{ }
	{\,-\,m_q}\biggr)\psi_q \\
	T^{\mu\nu}_g &=& F^{a,\mu\eta}\,{F^{a,}}_{\eta}{ }^\nu+\frac14\,g^{\mu\nu}
	F^{a,\kappa\eta}\,{F^{a,}}_{\kappa\eta}
\ea}
Here $\overset{\rightarrow}{\cal D}_\mu=\overset{\rightarrow}\partial_\mu+ig\,t^aA_\mu^a$ 
and $\overset{\leftarrow}{\cal D}_\mu = \overset{\leftarrow}\partial_\mu-ig\,t^aA_\mu^a$ 
with arrows indicating which fields are differentiated, 
$F^a_{\mu\nu}=\partial_\mu A^a_\nu-\partial_\nu A^a_\mu-g\,f^{abc}A^b_\mu A^c_\nu$
and the SU(3) color group generators satisfy the algebra 
$[t^a,t^b]=i\,f^{abc}t^c$ and are normalized as 
${\rm tr}\,(t^at^b)=\frac12\,\delta^{ab}$.
The total EMT is conserved
\be\label{Eq:EMT-cons}
	\partial^\mu\hat T_{\mu\nu} = 0, \quad \quad
	\hat T_{\mu\nu} = \sum_q\hat T_{\mu\nu}^q+\hat T_{\mu\nu}^g \; .
\ee
On ``classical level'' and for massless quarks QCD is invariant 
under conformal (``scale'') transformations: the associated current 
$j^\mu=x_\nu\hat{T}^{\mu\nu}$ satisfies 
$\partial_\mu j^\mu=\hat{T}_{\!\mu}^{\,\mu}$ and is 
conserved for light quarks in the chiral limit due to 
$\hat{T}_{\!\mu}^{\,\mu} = \sum_qm_q\bar\psi_q\psi_q$. Quantum 
corrections break the conformal symmetry due to the trace anomaly
\cite{Adler:1976zt,Nielsen:1977sy,Collins:1976yq,Adler:2004qt}
\be\label{Eq:ff-of-EMT-trace-1}
    	\hat{T}_{\!\mu}^{\,\mu} \equiv 
	\frac{\beta(g)}{2g}\;F^{a,\mu\nu}{F^{a,}}_{\!\!\mu\nu}
    	+(1+\gamma_m)\sum_qm_q\bar\psi_q\psi_q \;,\ee
where $\beta(g)$ is the $\beta$-function of QCD and $\gamma_m$ is 
the anomalous dimension of the mass operator. The trace anomaly
has the same form in QED.

{The vacuum matrix elements of the EMT are related to the quark 
and gluon vacuum condensates with important applications in the QCD 
sum rule approach \cite{Shifman:1978bx}. The EMT vacuum expectation 
values were also explored, e.g., to prove low energy theorems for 
correlators containing the gluon operator $F^2$ \cite{Migdal:1982jp}. 
The emergence of vacuum condensates reflects the rich and non-trivial 
structure of the QCD vacuum. In literature it is sometimes said that 
the vacuum is trivial in the light-front quantization approach, but 
this is strictly speaking not the case, see Ref.~\cite{Collins:2018aqt} 
for a review.

The main focus of this review are hadronic properties as encoded 
in the matrix elements of $\hat{T}^{\mu\nu}$ in one-particle~states.}
The matrix elements of the total EMT in one-particle 
states define the EMT form factors which are Lorentz scalars. Due to 
the EMT conservation (\ref{Eq:EMT-cons}) the form factors of the total 
EMT are renormalization scale invariant.

Since the individual quark and gluon operators, $\hat{T}_{\mu\nu}^q$ and 
$\hat{T}_{\mu\nu}^g$, are each separately gauge-invariant, we can
also define quark and gluon EMT form factors which are also 
Lorentz scalars. Since the separate quark and gluon EMT operators are
not conserved additional form factors appear in the decompositions of 
their matrix elements, and all individual quark and gluon form factors
acquire scale- and scheme-dependence.

\section{Definition of EMT form factors}
\label{Sec-2:EMT-def}

We use the covariant normalization
$\la p^\prime|\,p\ra=2p^0\,(2\pi)^3\delta^{(3)}(\bm{p^\prime}-\bm{p})$
of one-particle states, and introduce the kinematic variables 
$P= \frac12(p^\prime + p)$, $\Delta = p^\prime-p$, $t=\Delta^2$.
The EMT form factors of a spin-$\frac12$ {hadron in QCD}
are defined as  
\begin{align}
    \langle p^\prime,s^\prime| \hat T_{\mu\nu}^a(x) |p,s\rangle
    = \bar u{ }^\prime\biggl[
      A^a(t)\,\frac{\gamma_{\{\mu} P_{\nu\}}}{2}
    + B^a(t)\,\frac{i\,P_{\{\mu}\sigma_{\nu\}\rho}\Delta^\rho}{4m}
    + D^a(t)\,\frac{\Delta_\mu\Delta_\nu-g_{\mu\nu}\Delta^2}{4m}
    + {m}\,{\bar c}^a(t)\,g_{\mu\nu} \biggr]u\,e^{i(p^\prime-p)x}.
    \label{Eq:EMT-FFs-spin-12}
\end{align}
where the normalization of spinors is $\bar u(p,s)\, u(p,s) =2\,m$, and
we introduced the notation $a_{\{\mu} b_{\nu\}}=a_\mu b_\nu + a_\nu b_\mu$.
Exploring the Gordon identity 
$2m\bar u^\prime\gamma^\alpha u=
\bar u^\prime(2P^\alpha+i\sigma^{\alpha\kappa}\Delta_\kappa)u$
an alternative decomposition is obtained
\ba
    \la p^\prime,s^\prime| \hat T_{\mu\nu}^a(x) |p,s\rangle
    = \bar u^\prime\biggl[
      A^a(t)\,\frac{P_\mu P_\nu}{m}
    + J^a(t)\ \frac{i\,P_{\{\mu}\sigma_{\nu\}\rho}\Delta^\rho}{2m}
    + D^a(t)\,\frac{\Delta_\mu\Delta_\nu-g_{\mu\nu}\Delta^2}{4m}
    +{m}\,{\bar c}^a(t)g_{\mu\nu} \biggr]u\,\,e^{i(p^\prime-p)x} .
    \label{Eq:EMT-FFs-spin-12-alternative} \ea
The two representations are equivalent, and the form factors are
related as $A^a(t)+B^a(t) = 2\,J^a(t)$. 
{Form factors of the non-symmetric canonical EMT were discussed 
in \cite{Lorce:2015lna}. For a discussion of classical aspects of 
the EMT we refer to \cite{Blaschke:2016ohs,Forger:2003ut}.}
Let us remark that if the symmetries of QCD, parity and time 
reversal, are relaxed, then more form factors are possible. For
instance Dirac neutrinos have 5 EMT form factors 
\cite{Kobzarev:1962wt,Kobsarev:1970qm}, 
and Majorana neutrinos even 6 \cite{Ng:1993vh}.

Spin-0 hadrons like pion, $^4$He, etc have 3 EMT form factors {in QCD}
which can be defined as follows
\be\label{Eq:EMT-FFs-spin-0}
	\la p^{\,\prime\,}|\hat{T}^a_{\mu\nu}(x)|p\ra = 
	\biggl[2 P_\mu P_\nu\, A^a(t) + 
	\frac12(\Delta_\mu\Delta_\nu - g_{\mu\nu} \Delta^2)\,D^a(t)
	+ {2\ m^2}\,{\bar c}{ }^a(t)\,g_{\mu\nu} \biggr] \; e^{i(p^\prime-p)x}\,.
\ee
The individual quark and gluon form factors 
$A^a(t)$, $B^a(t)$ or $J^a(t)$, $D^a(t)$, ${\bar c}^a(t)$ depend 
on the renormalization scale which we do not indicate for brevity. 
Due to EMT conservation, Eq.~(\ref{Eq:EMT-cons}), the constraint
$\sum_a\bar{c}^a(t)=0$ holds, and the total form factors
$A(t)$, $B(t)$, $D(t)$ are renormalization scale invariant where we
defined $A(t)\equiv \sum_a A^a(t)$ with $a=g,\,u,\,d,\,\dots$ and 
analog for other form factors.

The total EMT form factors were introduced by Kobzarev and Okun 
\cite{Kobzarev:1962wt}  and by Pagels \cite{Pagels:1966zza} already 
in the 1960's for both spin-0 and spin-$\frac12$ hadrons
in somewhat different notation. 
	(We refer to Ref.~\cite{Hudson:2016gnq} for an overview of 
	the different notations used in literature.)
Hadrons with higher spins have additional EMT form factors
because, for instance in the spin-1 case the polarization vectors 
$\epsilon^{\prime\ast\mu}\epsilon^\nu$ can be used to generate further
symmetric Lorentz structures. 

\section{Constraints on the form factors}

For hadrons of any spin, there is a form factor $A(t)$ which 
accompanies the Lorentz structure proportional to $P^\mu P^\nu$.
Assuming adequate normalization this form factor satisfies at zero 
momentum transfer the constraint
\be\label{Eq:constraint-A}
	A(0)=1\,.
\ee
This is a consequence of translational invariance and reflects 
the fact that in the limit $\bm{p},\,\bm{p^\prime}\to 0$ only 
the 00-component remains in 
Eqs.~(\ref{Eq:EMT-FFs-spin-12}--\ref{Eq:EMT-FFs-spin-0}), and
$H=\int d^3x\;\hat{T}_{00}(x)$ is the Hamiltonian of the system with 
$H\,|\bm{p}\,\ra = m\,|\bm{p}\,\ra$ in the rest frame of the particle 
\cite{Pagels:1966zza}. The physical meaning of this constraint is that 
100\% of the hadron momentum is carried by its constituents in the infinite 
momentum frame, or in lightcone quantization \cite{Brodsky:1997de}.

If we define $A^Q(t)=\sum_qA^q(t)$ (and analog for other form factors below)
the quark and gluon form factors at zero-momentum transfer take in the 
limit of asymptotically large renormalization scale $\mu$ the following 
values \cite{Gross:1974cs,Politzer:1974sm}
\be\label{Eq:A-q-g-asymp}
	\lim\limits_{\mu\to\infty} \;A^Q(0) = \frac{N_f}{N_f+4C_F}, \quad
	\lim\limits_{\mu\to\infty} \;A^g(0) = \frac{4C_F}{N_f+4C_F}. 
\ee
Here $C_F = (N_c^2-1)/(2N_c)$ where $N_c=3$ is the number of colors 
in QCD. $N_f$ is the number of active quark flavors. The results
in Eq.~(\ref{Eq:A-q-g-asymp}) refer to the leading-order in the
QCD coupling $\alpha_s$, and show how quarks and gluons share 
the momentum of the hadron at asymptotically large scales. 
{The values of $A^Q(0)$ and $A^g(0)$} 
can be obtained from parameterizations of the parton 
distribution functions $f_1^a(x)$ extracted from deep-inelastic 
scattering experiments, 
\be
	A^Q(0) = \sum_q \int_0^1 d x \, x(f_1^q+f_1^{\bar q})(x),\quad \quad
	A^g(0) = \int_0^1 d x \, x f_1^g(x).
\ee
The quantity $f_1^a(x)\,{\rm d}x$ describes the probability to find 
(at a given normalization scale not indicated here for brevity)
a quark or gluon carrying the fraction $x$ of the hadron's momentum
in infinite-momentum frame/lightfront quantization in the interval 
$[x,\,x+ d x]$. Parameterizations of parton distribution functions 
are available for 
nucleon \cite{Gluck:1998xa,Martin:2009iq,Dulat:2015mca,Ball:2017nwa},
nuclei \cite{Hirai:2007sx,deFlorian:2011fp,Kovarik:2015cma}, 
pion \cite{Gluck:1999xe,Aicher:2010cb}, and kaon 
\cite{Badier:1980jq}. 
{For instance, in the nucleon quarks carry about $54\,\%$ and gluons 
about $46\,\%$ of the nucleon momentum at a scale of $\mu^2=4\,{\rm GeV}^2$ 
according to the leading order parameterizations of Ref.~\cite{Martin:2009iq}.}
The partonic interpretation is valid in leading order of $\alpha_s$,
and has restrictions for nuclei \cite{Brodsky:2002ue}.

The form factor $B(t)$ which exists for hadrons with $J>0$ 
satisfies at zero-momentum transfer the constraint
\be\label{Eq:constraint-B}
	B(0) = 0\,.
\ee
This constraint is referred to as the vanishing of the
``anomalous gravitomagnetic moment'' of a spin-$\frac12$ fermion and 
was first proven classically in \cite{Kobzarev:1962wt}
and discussed in quantum field theories in various contexts
\cite{Pagels:1966zza,Ji:1996ek,Teryaev:1999su,Brodsky:2000ii,Silenko:2006er,Teryaev:2016edw,Lowdon:2017idv}.
There is an interesting analogy to the anomalous magnetic moment 
of the electron: at one loop order in QED the anomalous magnetic
moment of the electron is given by the famous Schwinger term
$\alpha/2\pi$. At this order the fermionic contribution to the
anomalous gravitomagnetic moment of the electron is $\alpha/3\pi$.
This is compensated by the bosonic contribution $-\alpha/3\pi$
\cite{Brodsky:2000ii}, because photons also couple to gravity! 
It was shown that the anomalous gravitomagnetic moment
of a composite fermion vanishes order by order in the 
lightfront Fock expansion. This can be traced back to the
Lorentz boost properties of the {lightfront} Fock representation
\cite{Brodsky:2000ii}.

In the notation of Eq.~(\ref{Eq:EMT-FFs-spin-12-alternative}) 
the constraint (\ref{Eq:constraint-B}) is equivalent to
\be\label{Eq:constraint-J}
	J(0) = \frac12\,,
\ee
which reflects the fact that the contributions of quarks and gluons 
to the spin of the nucleon add up to $\frac12$ \cite{Ji:1996ek}. 
We will follow up on this below. At asymptotically large scales
the contributions of quarks and gluons to the nucleon spin is
analog to the momentum distribution in Eq.~(\ref{Eq:A-q-g-asymp}), 
namely \cite{Ji:1995cu}
\be\label{Eq:J-q-g-asymp}
	\lim\limits_{\mu\to\infty} \;J^Q(0) = \frac12\;\frac{N_f}{N_f+4C_F}, \quad
	\lim\limits_{\mu\to\infty} \;J^g(0) = \frac12\;\frac{4C_F}{N_f+4C_F}.
\ee
The constraints in 
Eqs.~(\ref{Eq:constraint-A}, \ref{Eq:constraint-B}, \ref{Eq:constraint-J})
were revisited recently in an axiomatic approach in Ref.~\cite{Lowdon:2017idv}.

The deeper reason for the existence of the constraints at zero-momentum 
transfer --- for the form factors $A(t)$ of all hadrons and the form
factors $B(t)$ or $J(t)$ for hadrons with $J\ge \frac12$ --- is owing
to the fact that these form factors are related to the generators 
of the Poincar\'e group which are ultimately related to the mass 
and spin of the particle.

In contrast to this the value of the form factor $D(t)$ at zero-momentum 
transfer is unconstrained. This value is often referred to as the 
$D$-term
$D\equiv D(0)$ \cite{Polyakov:1999gs}.
All hadrons independently of their spin $0,\,\frac12,\,1,\,\frac32,\,\dots$ 
possess a $D$-term. But this property is unknown and must be determined from 
experiment.\footnote{%
	Also the mass $m$ must be of course determined from experiment.
	However, once $m$ is known and an appropriate normalization 
	introduced, the values of $A(0)$ and $J(0)$ are fixed. This is 
	{\it not} the case for $D(0)$ showing that this is an 
	independent property.}
Below we will see that in contrast to $A(0)$ and $B(0)$ the $D$-term is not 
related to ``{\it external} properties'' of a particle like mass and spin, 
but to the stress tensor and {\it internal} forces.
The only general information about this property is that at asymptotically 
large scales the relative contributions of quarks and gluons to the $D$-term 
is the same as in Eqs.~(\ref{Eq:A-q-g-asymp},~\ref{Eq:J-q-g-asymp}) 
and can be expressed as \cite{Goeke:2001tz}
\be
	\lim\limits_{\mu\to\infty} \;D^Q(0) = D\;\frac{N_f}{N_f+4C_F}, \quad
	\lim\limits_{\mu\to\infty} \;D^g(0) = D\;\frac{4C_F}{N_f+4C_F}.
\ee

\section{Relation to observables}

\begin{figure}[h!]
\centering
\includegraphics[height=4.5cm]{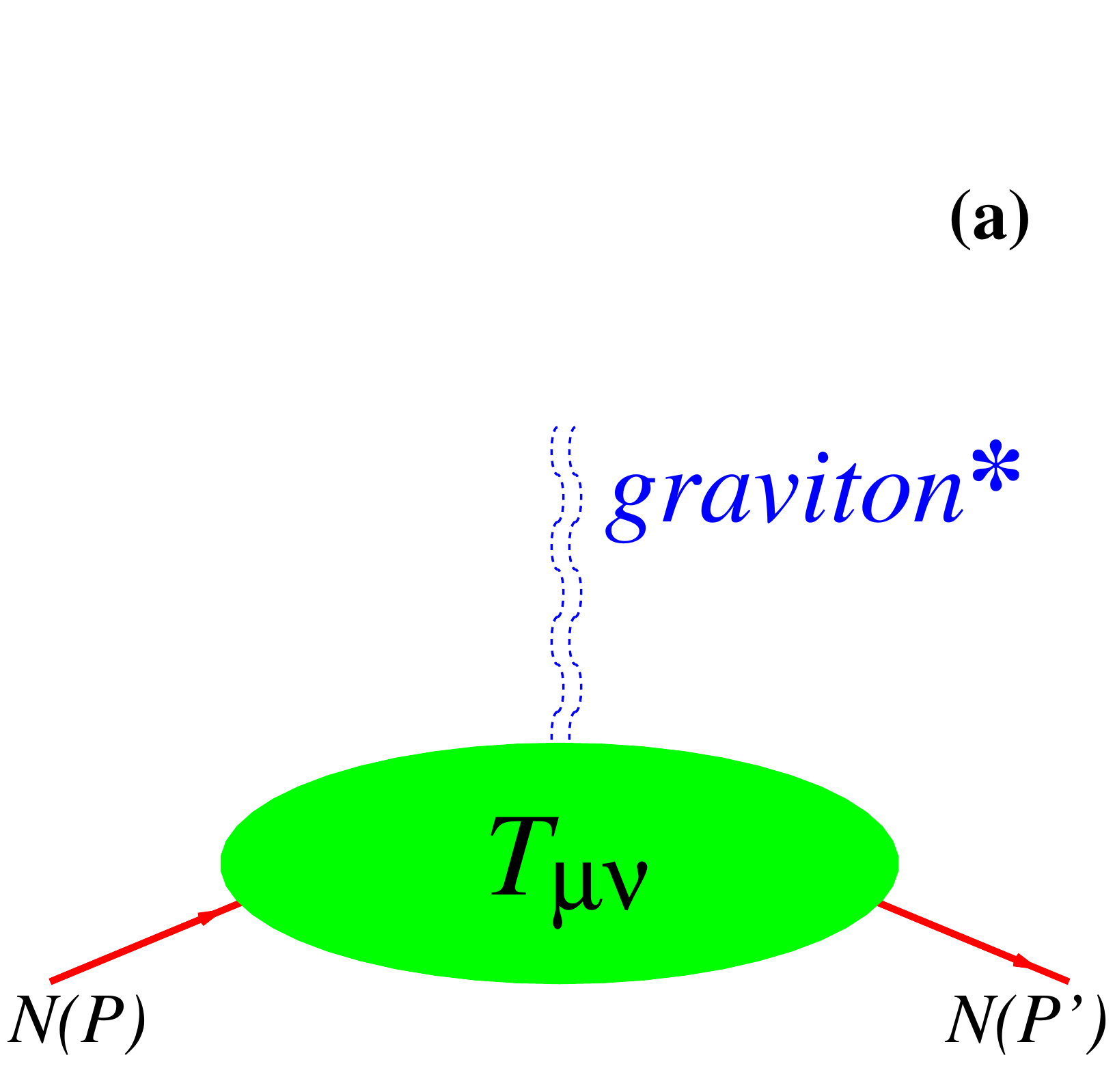}	\hspace{1.2cm}
\includegraphics[height=4.5cm]{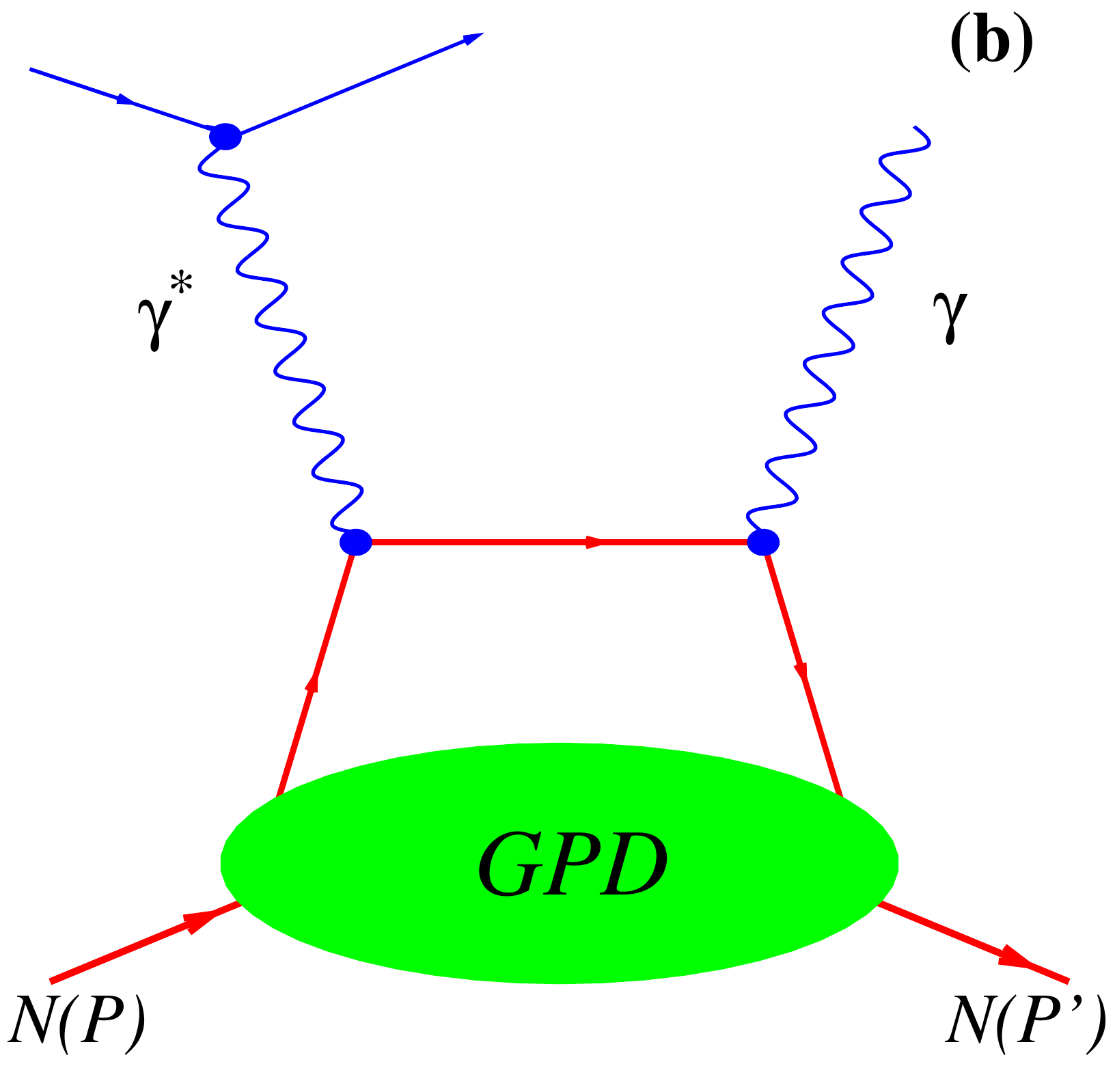} 	\hspace{1.2cm}
\includegraphics[height=4.5cm]{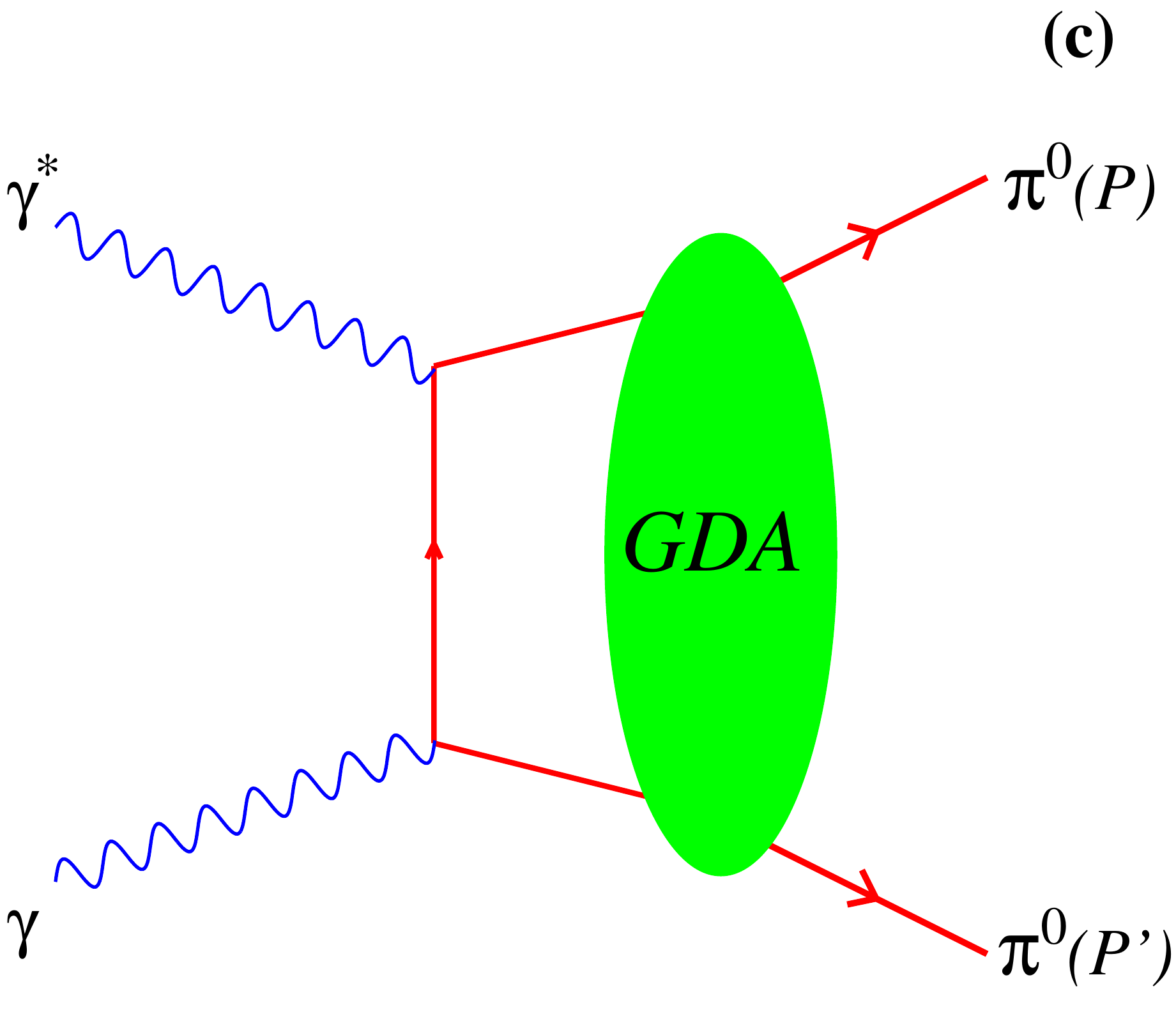}
\caption{\label{Fig-1:processes}
	(a) A natural but impractical probe of EMT form factors is 
	scattering off gravitons. (b) Hard-exclusive reactions 
	like deeply virtual Compton scattering (DVCS) provide 
	a realistic way to access EMT form factors through GPDs. 
	Here one of the relevant tree-level diagrams is shown.
	{(c) Information on the EMT structure of particles
	not available as targets, such as e.g.\ $\pi^0$,
	can also be accessed from studies of generalized distribution
	amplitudes (GDAs) which are an ``analytic continuation''
	of GPDs to the crossed channel. The shown reaction
	$\gamma^\ast\gamma\to \pi^0\pi^0$ (and analog for other hadrons)
	can be studied in $e^+e^-$ collisions.}}
\end{figure}

The most natural way to probe EMT form factors, scattering off 
gravitons in Fig.~\ref{Fig-1:processes}a, is also the least practical 
one. A practical opportunity to access EMT form factors emerged with 
the advent of GPDs 
\cite{Mueller:1998fv,Ji:1996ek,Radyushkin:1996nd,Radyushkin:1996ru,
Ji:1996nm,Collins:1996fb,Radyushkin:1997ki,Vanderhaeghen:1998uc,
Belitsky:2001ns,Goloskokov:2005sd,Kumericki:2007sa,Kumericki:2009uq,
Mueller:2013caa,Kumericki:2015lhb,Muller:2014wxa},
which describe hard-exclusive reactions, such as deeply virtual 
Compton scattering (DVCS) $eN\to e^\prime N^\prime \gamma$ sketched 
in Fig.~\ref{Fig-1:processes}b or hard exclusive meson production 
$eN\to e^\prime N^\prime M$. In the case of the nucleon,
the second Mellin moments of unpolarized GPDs yield 
the EMT form factors 
\begin{align}
	\int_{-1}^1{\rm d}x\;x\, H^a(x,\xi,t) = A^a(t) + \xi^2 D^a(t) \,, \quad \quad
        \int_{-1}^1{\rm d}x\;x\, E^a(x,\xi,t) = B^a(t) - \xi^2 D^a(t) \,.
    	\label{Eq:GPD-Mellin}
\end{align}
The GPDs can be viewed as ``amplitudes'' for removing from the nucleon
a parton carrying the fraction $x-\xi$ of the average momentum $P$ and 
putting back in the nucleon a parton carrying the fraction $x+\xi$,
while the nucleon receives the momentum transfer $\Delta$. For $\xi=0$
the momentum transfer is purely transverse and the Fourier transform
$H^a(x,b_\perp)=\int d^2\Delta_\perp/(2\pi)^2 \,
\exp(-i\bm{\Delta_\perp b_\perp})\,H^a(x,0,-\bm{\Delta^2_\perp})$
describes the probability to find a parton carrying the momentum 
fraction $x$ of the hadron and located at the distance $b_\perp$ from
the hadrons (transverse) center-of-mass on the lightcone. This
allows one to do femtoscale tomography of the nucleon
\cite{Burkardt:2000za,Burkardt:2002hr,Ralston:2001xs}.

Adding up the two equations in (\ref{Eq:GPD-Mellin}) and extrapolating
$t\to 0$ provides the key to the nucleon's spin decomposition: the fraction 
of the nucleon spin due to the angular momentum of the parton of 
type $a=g,\,u,\,d\,\dots$ is obtained from the Ji sum rule \cite{Ji:1996ek}
\be
	\lim\limits_{t\to0}\int_{-1}^1{\rm d}x\;x\, 
	\biggl(H^a(x,\xi,t) + E^a(x,\xi,t)\biggr) = 2J^a(0) \,.
    	\label{Eq:Ji-sum-rule}
\ee
These fractions are currently unknown. An unambiguous decomposition into 
the ``spin contribution'' and the ``orbital angular momentum contribution'' 
to $J^a(0)$ is not possible, though different schemes can be introduced.
We refer to the review article \cite{Leader:2013jra} for a discussion
of the controversy of the nucleon spin decomposition.

Note that GPDs are not direct observables and they cannot be fully 
restored from the DVCS observables, see the detailed discussion in 
Ref.~\cite{Polyakov:2007rv} based on the dual parameterizations of 
GPDs  \cite{Polyakov:2002wz}.
In particular, it was shown that the Mellin moments of GPDs 
(\ref{Eq:GPD-Mellin}) are not directly observable. However, the 
$D$-term (in contrast to $J^a(t)$ or $A^a(t)$) can be extracted 
directly from the DVCS observables through the logarithmic $Q^2$ 
dependence of the subtraction constant in the dispersion relations 
for the DVCS amplitudes 
\cite{Polyakov:2002yz,Teryaev:2005uj,Anikin:2007yh,Diehl:2007jb,Radyushkin:2011dh}, 
see the discussions in Section~\ref{Sec:expnucleon}.

First experimental results related to nucleon GPDs from studies of hard 
exclusive reactions came from the HERMES and HERA experiments at DESY 
and Jefferson Lab \cite{Airapetian:2001yk,Stepanyan:2001sm,
Ellinghaus:2002bq,Chekanov:2003ya,Aktas:2005ty,Airapetian:2006zr,
Camacho:2006qlk,Mazouz:2007aa,Airapetian:2007aa,Girod:2007aa,
Airapetian:2008aa,Airapetian:2009ac,Airapetian:2009bm,Airapetian:2009cga,
Airapetian:2009aa,Airapetian:2010ab,Airapetian:2010aa,Airapetian:2011uq,
Airapetian:2012mq,Airapetian:2012pg,Jo:2015ema}.
Studies of hard-exclusive reactions are an important part of 
ongoing experimental programs at Jefferson Lab and COMPASS at CERN.

A spin-0 hadron like pion has only one GPD, namely $H^a(x,\xi,t)$,
which is per se not directly accessible in experiment. However,
hard-exclusive production of two pions provides the possibility 
to access information on the two-pion distribution amplitude which
is an analog of a GPD in the crossed channel,
{see Fig.~\ref{Fig-1:processes}c.} This allows one to extract 
information on the EMT form factors of the pion (and similarly 
other hadrons) in the time-like region.

{For completeness let us remark that a relation of the fixed poles 
in the angular momentum plane in virtual Compton scattering 
\cite{Cheng:1970vg,Brodsky:1971zh,Brodsky:1972vv} to the $D$-term
was discussed in \cite{Brodsky:2008qu}.
However, it was shown that the $J = 0$ fixed pole universality 
hypothesis of Ref.~\cite{Brodsky:2008qu} is an external assumption 
and might never be proven theoretically \cite{Muller:2015vha}.}

\section{The last global unknown property of a hadron}

The $D$-term is sometimes referred to as the ``last unknown global property.'' 
To explain what this means we recall that the structure of 
hadrons, the bound states of strong interactions, is most conveniently
probed by exploring the other fundamental forces: electromagnetic, weak, 
and (in principle) gravitational interactions. The particles couple to 
these interactions via the fundamental currents 
$J^\mu_{\rm em}$, $J^\mu{\!\!\!\!}_{\rm weak}$, $T^{\mu\nu}_{\rm grav}$
which are conserved (in case of weak interactions we deal with 
partial conservation of the axial current, PCAC).
The matrix elements of these currents are described in terms of form 
factors which contain a wealth of information on the probed particle. 
The undoubtedly most fundamental information corresponds to the form 
factors at zero momentum transfer. For the nucleon, these are the 
``global properties:'' 
electric charge $Q$, magnetic moment $\mu$, axial coupling constant 
$g_A$, induced pseudo-scalar coupling constant $g_p$, mass $M$, spin 
$J$, and the $D$-term $D$. These properties, being related to 
external conserved currents, are scale- and scheme-independent 
in QCD. All global properties are in principle on equal footing 
and well-known, see Table~\ref{Table-global-properties}, 
with one exception: the $D$-term.

\begin{table}[h!]
\vspace{5mm}
\centering
\begin{tabular}{rllrrrl} 
		\hline
{\bf em:}	& $\partial_\mu J^{\mu^{\phantom X}}_{\rm em}=0$ 
		& $\la N^\prime|J^\mu_{{\bf em}}|N\ra$ 
		& $\longrightarrow$  
		& $Q$ 
		& $=$ & $1.602176487(40)\times10^{-19}$C\\
		&&&
		& $\mu$ 
		& $=$ & $2.792847356(23) \mu_N$ \\ \hline
{\bf weak:}	& PCAC 
		& $\la N^\prime|J^{\mu^{\phantom X}}_{{\bf weak}}|N\ra$ 
		& $\longrightarrow$  
		& $g_A$        	
		& $=$ & $1.2694(28)$ \\
		&&&
		& $g_p$ 
		& $=$ & $8.06(55)$
		\\ \hline
{\bf gravity:} & $\partial_\mu T^{\mu\nu}_{{\bf grav}}=0$ 
		& $\la N^\prime|T^{\mu\nu^{\phantom X}}_{{\bf grav}}|N\ra$ 
		& $\longrightarrow$  
		& $m$		
		& $=$ & $938.272013(23)\,$MeV$/c^2$ \\
		&&&
		& $J$ 		
		& $=$ & $\frac12$\\
		&&&
		& {\boldmath{$D$}} & $=$ & {\bf ?}
\\ \hline
\end{tabular}
\caption{\label{Table-global-properties}
The global properties of the proton defined in terms of matrix elements 
of the conserved currents associated with respectively electromagnetic, 
weak, and gravitational interaction. Notice the weak currents include 
the partially conserved axial current, and $g_A$ or $g_p$ are strictly 
speaking defined in terms of transition matrix elements in the neutron 
$\beta$-decay or muon-capture. The values of the properties are from 
the particle data book \cite{Patrignani:2016xqp} and \cite{Andreev:2012fj} 
(for $g_p$) except for the unknown $D$-term.}
\end{table}

In some cases (e.g. free particles, Goldstone bosons) the value 
of the $D$-term is fixed by general principles (see discussions below). 
For other particles the $D$-term is not fixed and it reflects the internal 
dynamics of the system through the distribution of forces. In strongly 
interacting systems the $D$-term is sensitive to correlations in the system.
For example, the baryon $D$-term behaves as $\sim N_c^2$ whereas all other 
global observables (mass, magnetic moments, axial charge, etc.) behave
at most as $\sim N_c$ in the large $N_c$ limit. For a large nucleus the 
$D$-term shows also anomalously fast increase with the atomic mass
number $D\sim A^{7/3}$.

\section{The static EMT and stress tensor}

The physical content of the information encoded in the EMT
form factors is revealed in the so-called Breit frame.
In this frame the momenta of the incoming and outgoing hadron are 
given by $p=(P-\frac12\Delta)$ and $p^\prime=(P+\frac12\Delta)$, such 
that $P=(E,0,0,0)$ and $\Delta=(0,\bm{\Delta})$ while $t=-\bm{\Delta}^2$.
Let us first discuss the spin $\frac12$ case which is relevant
for the nucleon.

In the Breit frame the components of the EMT in 
Eq.~(\ref{Eq:EMT-FFs-spin-12-alternative}) can be expressed as 
\sub{\begin{align}
\langle p^\prime,s^\prime| \hat T^{00}_{a}(0) |p,s\rangle
      	= 2\,m\,E\,\biggl[
        & A^{a}(t) - \frac{t}{4m^2}\bigl[A^a(t)-2J^a(t)+D^a(t)\bigr]
      	+ {\bar c}^{a}(t)\biggr]\delta_{ss^\prime}  
	\label{Eq:EMT-Breit-T00}\\
\langle p^\prime,s^\prime| \hat T^{ik}_{a}(0) |p,s\rangle
    	= 2\,m\,E\, \biggl[
	& D^{a}(t)\,\frac{\Delta^i\Delta^k-\delta^{ik}\bm{\Delta}^2}{4m^2}
    	- {\bar c}^{a}(t)\,\delta^{ik} \biggr] \delta_{ss^\prime}  
	\label{Eq:EMT-Breit-Tik}\\
\langle p^\prime,s^\prime| \hat T^{0k}_{a}(0) |p,s\rangle
	= 2\,m \,E\biggl[
	& J^a(t)\;\frac{(-i\,\bm{\Delta}\times\bm{\sigma}_{s^\prime s})^k}{2\,m}
	\biggr]
	\label{Eq:EMT-Breit-T0k}
\end{align}}
Notice that in the Breit frame 
$\bar u{ }^\prime u \equiv\bar{u}(p^\prime,s^\prime)u(p,s)=2E\,\delta_{ss^\prime}$ 
and the energy $E$ is given by $E=\sqrt{m^2+{\bm{\Delta}{ }^2}/4}$.
Here $\sigma^j_{s^\prime s}=\chi^\dag_{s'}\sigma^j\chi^{ }_{s}$ with
the nucleon Pauli spinors $\chi^{ }_{s'}$ and $\chi^{ }_{s}$ in the 
respective rest frames normalized as 
$\chi^\dag_{s'}\chi^{ }_{s} =\delta_{ss^\prime} $.
The matrix elements of $T^{0k}_a$ vanish if the polarization is 
along $\bm{\Delta}$. 

We can now define the static EMT $T^{\mu\nu}(\bm{r},\bm{s})$ of a 
hadron by Fourier transforming the matrix elements of the EMT in 
Eqs.~(\ref{Eq:EMT-Breit-T00}--\ref{Eq:EMT-Breit-Tik})
with respect to~$\bm{\Delta}$ as follows \cite{Polyakov:2002yz}
\be\label{EQ:staticEMT}
	T^a_{\mu\nu}(\bm{r},\bm{s})=\int \frac{d^3 \bm{\Delta}}{(2\pi)^3 2E}\ 
	e^{{-i} \bm{r\Delta}} \langle p'| \hat{T}^a_{\mu\nu}(0)|p\rangle,
\ee
where $\bm{s}$ denotes the polarization vector of the states 
$|p,s\ra$ and $|p^\prime,s^\prime\ra$ in the respective rest frames.

Let us first discuss the 00-component of (\ref{EQ:staticEMT}),
i.e.\ the energy density.
Due to the presence of the EMT-nonconserving term $\bar c^a(t)$ the 
energy density $T_{00}(r)$ can only be defined for the total system 
(recall that $\sum_a\bar c^a(t)=0$, see Sec.~\ref{Sec-2:EMT-def}). 
The total energy density in a hadron is thus given by 
\cite{Polyakov:2002yz}
\be\label{EQ:staticEMT-T00}
	T_{00}(r)=m \int \frac{d^3 \bm{\Delta}}{(2\pi)^3}\ 
	e^{{-i}\bm{r\Delta}} \,\biggl[
        A(t) - \frac{t}{4m^2}\bigl[A(t)-2J(t)+D(t)\bigr]\biggr]\,.
\ee
The energy density is a function of $r=|\bm{r}|$.
It is independent of polarization, and normalized as 
\be
	\int d^3r \,T_{00}(r) = m\,
\ee
due to the constraint (\ref{Eq:constraint-A}).
The expression (\ref{EQ:staticEMT-T00}) is analog 
to the electric charge distribution which can be mapped out by means of 
electron scattering expriments. In an analog way, (hypothetical) scattering 
off gravitons would allow one to access information on the spatial 
distribution of the energy inside a hadron. We stress that 
in this way we can only access the energy density of the total system. 
The decomposition of the nucleon mass in terms of contributions 
from quarks and gluons was extensively discussed in 
Refs.~\cite{Ji:1994av,Ji:1995sv,Lorce:2017xzd} 
{and in the reviews \cite{Liu:2015xha,Gao:2015aax}.}

The $ij$-components of static EMT define the stress tensor 
\cite{Polyakov:2002yz}.
The total quark + gluon stress tensor can be decomposed in a 
traceless part associated with shear forces $s(r)$ and a trace 
associated with the pressure $p(r)$, 
\begin{align}\label{Eq:stress-tensor-p-s}
	T^{ij}(\bm{r}) = \biggl(
	\frac{r^ir^j}{r^2}-\frac13\,\delta^{ij}\biggr) s(r)
	+ \delta^{ij}\,p(r)\,.
\end{align}
The shear forces are ``good observables'' and 
exist also separately for quarks and gluons (although 
$T_{\mu\nu}^q$ and $T_{\mu\nu}^g$ are not conserved separately,
the EMT-nonconserving $g_{\mu\nu}\,\bar{c}^a(t)$ in
(\ref{Eq:EMT-FFs-spin-12}) drop out from the traceless part of the 
stress tensor),
{i.e.\ we can define (scale-dependent) partial contributions $s^a(r)$ 
for $a=g,\,u,\,d,\dots$ to the shear forces. This also allows}
one to define the quark and gluon contributions to the $D$-term  
\cite{Polyakov:2002yz},
\be\label{Eq:stress-tensor-D-term}
	D^a = - \frac{2}{5}\,m \int{\rm d}^3r\;T^a_{ij}(\bm{r})\,
	\biggl(r^ir^j-\frac13\,r^2\delta^{ij}\biggr) 
	= {- \frac{4}{15}\,m \int{\rm d}^3r\;r^2s^a(r)} \,.
\ee
In contrast to this $p(r)$ is defined only for the total system, and 
has no relation to the separate $D^q$ and $D^g$ \cite{Polyakov:2002yz}.

The pressure $p(r)$ and shear forces $s(r)$ can be computed from 
$D(t)$ as follows:
\be
\label{Eq:relationSPD}
	s(r)= -\frac{1}{4 m}\ r \frac{d}{dr} \frac{1}{r} \frac{d}{dr}
	{\widetilde{D}(r)}, \quad
	p(r)=\frac{1}{6 m} \frac{1}{r^2}\frac{d}{dr} r^2\frac{d}{dr}
 	{\widetilde{D}(r)}, \quad
	{\widetilde{D}(r)=}
	\int {\frac{d^3\Delta}{(2\pi)^3}}\ e^{{-i} \bm{\Delta r}}\ D(-\bm{\Delta}^2).
\ee
From these equations one sees immediately that the von Laue stability 
condition (\ref{Eq:von-Laue}) is satisfied automatically 
provided $\widetilde{D}(r)$ is not too singular at the origin 
and satisfies $\lim_{r\to 0} r^2 \widetilde{D}'(r)=0$.

For the individual contributions of the constituents ($a=g,\,u,\,d,\dots$)
to the pressure and shear forces, $p^a(r)$ and $s^a(r)$,\footnote{For 
	brevity we suppress the dependence of these quantities on the 
	QCD normalisation scale.} 
we have:
\be
\label{Eq:relationSPDindividual}
	s^a(r)= -\frac{1}{4 m}\ r \frac{d}{dr} \frac{1}{r} \frac{d}{dr}
	{\widetilde{D^a}(r)}, \quad
	p^a(r)=\frac{1}{6 m} \frac{1}{r^2}\frac{d}{dr} r^2\frac{d}{dr} 
	{\widetilde{D^a}(r)} 
	- m \int {\frac{d^3\Delta}{(2\pi)^3}}\ e^{{-i} \bm{\Delta r}}\ \bar c^a(-\bm{\Delta}^2).
\ee
Comparing this equation with (\ref{Eq:relationSPD}) we see that the shear 
force distribution for individual components can be obtained from the 
$D^a(t)$ form factor, whereas the pressure distribution requires 
additionally the knowledge of the form factors $\bar c^a(t)$.
{The form factors $\bar c^a(t)$ can be related to the hadron matrix element
of the mixed quark-gluon operator $\bar\psi(x) \gamma_\nu F^{\mu \nu}(x) \psi(x)$ \cite{Kolesnichenko:1984,Braun:2004vf,Tanaka:2018wea}
and to the Mellin moments of the twist-4 GPDs \cite{Leader:2012ar,Leader:2013jra,Tanaka:2018wea}.}

The $0k$-components of the EMT are related to the spatial distribution 
of the nucleon spin.
The EMT non-conserving terms drop out in Eq.~(\ref{Eq:EMT-Breit-T0k}) 
since they are polarization independent, such that we can define the
individual contributions of quarks and gluons as follows
\be\label{Eq:static-EMT-T0k-1}
	J_a^i(\bm{r},\bm{s})=\epsilon^{ijk}r^jT_a^{0k}(\bm{r},\bm{s})\;.
\ee
Insterting the expression (\ref{Eq:EMT-Breit-T0k}) 
into Eq.~(\ref{Eq:static-EMT-T0k-1}) yields
\ba\label{Eq:static-EMT-T0k-2}
	J_a^i(\bm{r},\bm{s})
	&=& s^j\int \frac{d^3 \bm{\Delta}}{(2\pi)^3}\ e^{-i \bm{r\Delta}}\Biggl[
	\biggl(J^a(t)
	+\frac23\,t \;\frac{dJ^a(t)}{dt}\biggr)\delta^{ij}
	+\biggl(\Delta^i\Delta^j-\frac13\,{\bm{\Delta^2}}\;\delta^{ij}\biggr)
	\;\frac{dJ^a(t)}{dt}\Biggr] 
\ea
The first term in the square brackets corresponds 
to a monopole term \cite{Polyakov:2002yz}, and the second term 
corresponds to a quadrupole term of Ref.~\cite{Lorce:2017wkb}. 
In Ref.~\cite{Polyakov:2002yz} only the monopole term was considered 
as the 3D angular momentum density was implicitly defined as 
$J_a^i(\bm{r})=s^i \langle J_a^k(\bm{r},\bm{s}) s^k \rangle$, where 
$\langle\dots\rangle$ denotes averaging over the direction of the 
vector $\bm{s}$.
Integrating (\ref{Eq:static-EMT-T0k-2}) and summing over
quarks and gluons yields
\ba\label{Eq:static-EMT-T0k-3}
	\sum\limits_a\int d^3r\;J_a^i(\bm{r},\bm{s})
	= s^i J(0) 
\ea
where we used that the individual contributions $J^a(0)$ add 
up to $\sum_aJ^a(0)=J(0)$ with $J(0)=\frac12$ according 
to Eq.~(\ref{Eq:constraint-J}).
The possibility to access the form factors $J^a(t)$ from studies 
of hard exclusive reactions through GPDs \cite{Ji:1996ek} will give
important insights into the spin decomposition of the nucleon. 
We refer to Ref.~\cite{Leader:2013jra} for a review on the
nucleon spin decomposition, and to Ref.~\cite{Lorce:2017wkb}
for a discussion of the anugular momentum density in the
3D Breit-frame representation and 2D lightfront representation.

\section{EMT densities in spin-0 hadrons}
\label{Sec:spin-0}

{The EMT densities of spin-0 hadrons can be defined in the same 
way as in Eq.~(\ref{EQ:staticEMT}) with the obvious difference that 
$T^{0k}(\bm{r})=0$ and the form factor $J(t)$ is absent, which 
reflects the spin-0 character. The
expressions for $T^{00}(r)$ and~$T^{ij}(\bm{r})$ look almost exactly as in 
the fermionic case, in Eqs.~(\ref{EQ:staticEMT-T00},~\ref{Eq:relationSPD}),
 --- with one subtle difference. In the Breit frame the nucleon spinors 
$\bar u{ }^\prime u =2E\,\delta_{ss^\prime}$ contribute the factor $2E$ which 
cancels out the factor $1/(2E)$ in the definition of the static EMT in 
Eq.~(\ref{EQ:staticEMT}). In the spin-0 case no such compensation occurs,
and the EMT densities read \cite{Hudson:2017xug}
\sub{\begin{align}
	\label{Eq:static-EMT-T00-spin0}
	T_{00}(r)
	= 2m^2 &
	\int \frac{d^3\Delta}{2E(2\pi)^3}\ e^{{-i}\bm{r\Delta}} \,
	\biggl[A(t) - \frac{t}{4m^2}\bigl[A(t)+D(t)\bigr]\biggr]\,,
	\quad (\mbox{spin 0 case}) \\
	\label{Eq:static-EMT-Tij-spin0}
    	T_{ij}(\bm{r}\,) 
	= \frac{1}{2} &
	\int\frac{ d^3\Delta}{2E(2\pi)^3}\;e^{-i\bm{r}\bm{\Delta}}\;
    	\biggl[\Delta_i\Delta_j - \delta_{ij} \bm{\Delta}^2 \biggr]\,D(t) \,.
\end{align}}

We remark that the pressure and shear forces can be extracted from
Eq.~(\ref{Eq:static-EMT-Tij-spin0}) as follows
\sub{\begin{align}
	\label{Eq:static-EMT-p-spin0}
    	p(r) &= 
	\frac{1}{3}\int\frac{ d^3\Delta}{2E(2\pi)^3}
	\;e^{-i\bm{\Delta}\bm{r}}\;P_0(\cos\theta)\biggl[t\,D(t)\biggr] \,,\\
	\label{Eq:static-EMT-s-spin0}
    	s(r) &= 
	\frac{3}{4}\int\frac{ d^3\Delta}{2E(2\pi)^3}
	\;e^{-i\bm{\Delta}\bm{r}}\;P_2(\cos\theta)\biggl[t\,D(t)\biggr] \,,
\end{align}}
which illustrates their relations to respectively, the trace and the
traceless part of the stress tensor. With the replacement $2E\to 2m$
the Eqs.~(\ref{Eq:static-EMT-p-spin0},~\ref{Eq:static-EMT-s-spin0})
hold also for the nucleon and are equivalent to (\ref{Eq:relationSPD}).

Effectively, the expressions for the polarization-independent 00- and
$ij$-components of the static EMT in spin $\frac12$ vs spin 0 case differ}
by  rescaling the form factors $A(t)$ and $D(t)$  by the factor $m/E$ with 
$E=\sqrt{m^2+\bm{\Delta}^2/4}$.
Such modifications correspond to relativistic corrections to the static 
EMT densities (\ref{EQ:staticEMT}). (Notice that $E/m$ is the Lorentz 
gamma-factor for the boost from rest frame system to the Breit one.) 
There is no unique field-theoretic way to derive such modifications, 
see discussions in 
Refs.~\cite{Licht:1970pe,Kelly:2002if,Ji:1991ff,Holzwarth:1996xq}. 
For the case of heavy particles ($m R_h\gg1$ where $R_h$ is the hadron size)
the relativistic corrections modify the EMT densities at short distances of 
order $1/m$ which result in the relative corrections of order $1/(mR_h)^2$ to
the radii of EMT densities.\footnote{The $D$-term is not modified 
	by the relativistic corrections as it is defined at zero 
	momentum transfer.}
In most physically interesting cases such corrections are very small 
(see the estimates in Ref.~\cite{Hudson:2017xug}) and we shall 
systematically neglect them. A special case is $\pi$-meson. The pion is 
the (pseudo)Goldstone boson {of spontaneous chiral symmetry breaking},
so its mass is parametrically small and we have $m_\pi R_\pi \ll 1$, i.e.\ 
the relativistic corrections are big and the static EMT densities are not 
properly defined. However, for the pion one can apply the methods of chiral 
effective field theory, see discussion in Section~\ref{Sec:Goldstone}.

Let us remark that the stress tensor $T^{ik}(\bm{r})$ has the general 
form (\ref{Eq:stress-tensor-p-s}) in spin-0 and $\frac 12$ case.
For higher spin particles additional tensor structures $S_i S_j$, 
$(S_i r_j+S_j r_j) (\bm{S\cdot r})$ are possible. Also the shear forces 
and the pressure have additional dependence on $(\bm{S\cdot r})^2$. 
The case of higher spin particles will be considered elsewhere.

\section{Consequences from EMT conservation}
\label{Sec:EMT-conservation-consequences}

{The EMT conservation, $\partial^\mu\hat{T}_{\mu\nu}=0$, implies for the
static EMT $\nabla^i T_{ij}=0$. This yields the differential equation 
\be\label{Eq:diff-eq-s-p}
	\frac23\,s^\prime(r)+\frac2r\,s(r)+p^\prime(r)=0	\,,
\ee
i.e.\ the shear forces and pressure are not independent 
functions but connected to each other due to EMT conservation.
Another consequence of the EMT conservation is the von Laue condition 
\cite{von-Laue}, which shows how the internal forces balance inside 
a composed particle,
\be\label{Eq:von-Laue}
	\int_0^\infty {\rm d}r\,r^2p(r)=0  \,.
\ee 
This relation implies that the pressure must have at least one
node. In all model studies so far it was found that the pressure
is positive in the inner region, and negative in the outer region.
In our convention the positive sign means repulsion towards outside,
the negative sign means attraction directed towards inside.

The von Laue condition (\ref{Eq:von-Laue}) is a necessary condition
for stability, but not sufficient. 
We shall see below that also the pressure distribution inside 
an unstable particle like the $\Delta$-resonance satisfies the von Laue 
condition \cite{Perevalova:2016dln}. In models the von Laue condition is 
often equivalent to the virial theorem which expresses the fact that one
deals with a minimum of the action {\cite{Goeke:2007fp}}. A sufficient 
stability condition is stronger than (\ref{Eq:von-Laue}) and requires 
that one deals with the global mimimum of the action for the given 
quantum numbers.

Using Eq.~(\ref{Eq:diff-eq-s-p}) we can express the $D$-term in two
equivalent ways in terms of $s(r)$ [cf.\ Eq.~(\ref{Eq:stress-tensor-D-term})] 
and $p(r)$ as 
\be\label{Eq:D-equivalent-expressions}
         D =-\,\frac{4m}{15}\int d^3r\;r^2\, s(r)
           = m \int d^3 r\;r^2\, p(r)\;.
\ee
}

Using (\ref{Eq:relationSPD}) we can derive a number of interesting 
relations. For example, it is obvious that {(cf.\ Ref.~\cite{Goeke:2007fp}, 
App.~B)}
\be\label{Eq:generelized-Kelvin-relation}
\int_0^\infty dr\ \frac{2 s(r)}{r} =p(0).
\ee
This relation can be viewed as the generalization of well known Kelvin 
relation between the pressure ($p_0$) in a liquid spherical drop, its 
surface tension ($\gamma$) and the radius ($R_{\rm drop}$) of the drop 
$2\gamma/R_{\rm drop} =p_0$ \cite{Kelvin}. Using this analogy we define 
the average surface tension of the hadron as $\gamma=\int_0^\infty dr\ s(r)$ 
and the surface tension energy of a hadron as $\int d^3r\ s(r)$. For the 
liquid drop we have $s(r)=\gamma\delta(r-R_{\rm drop})$ and thus these 
definitions correspond to $\gamma$ and to $4\pi \gamma R_{\rm drop}^2 $ 
respectively. 
 
Another interesting relation is obtained by integrating the second equation 
in (\ref{Eq:relationSPD}) over a ball of radius $R$. Doing this we get:
\be
\label{Eq:finite-Laue-integral}
 	\int_{|\bm{r}|\le R} d^3r\ p(r)
	=\frac{4\pi R^3}{3}\left[\frac 23 s(R)+p(R)\right].
\ee
Below we shall see that the combination $\left[\frac 23 s(R)+p(R)\right]$ 
corresponds to the distribution of the normal component of the force and
must be positive {to guarantee} the stability of the system. This 
implies that the integral of the pressure over the ball of a {finite}
radius $R$ is also positive and the von Laue condition {(\ref{Eq:von-Laue})}
is realised if $s(r)$ and $p(r)$ drop at the infinity faster than $1/r^3$. 
This is typically the case, e.g.\ in the nucleon $s(r)$ and $p(r)$ 
vanish at large $r$ like $1/r^6$ in the chiral limit \cite{Goeke:2007fp},
and faster for finite $m_\pi$ as we shall see below.
 
{Note that the nontrivial shear forces distribution $s(r)$ 
is responsible for the structure formation in a hadron. Indeed, 
from Eqs.~(\ref{Eq:stress-tensor-p-s},\ref{Eq:diff-eq-s-p}) it 
follows that $s(r)=0$ corresponds to isotropic matter with a 
constant pressure.\footnote{{A constant pressure is also
	obtained if $s(r)\propto\frac{1}{r^3}$ which could hold 
	in a finite region $0<r<\infty$. It would be interesting
	to explore e.g.\ soliton models with piece-wise defined
	potentials where in a certain region(s) of $r$ this situation
	could be realized.}}
Anisotropy and non-trivial shape of the pressure distribution (hadron shape) 
appear due to non-trivial shear forces distribution $s(r)$, the latter is 
also called  pressure anisotropy \cite{TrawinskiLCC}. Interestingly 
the pressure anisotropy (shear forces distribution) plays an essential 
role in astrophysics \cite{TrawinskiLCC,CedricPK}, see the review 
\cite{Herrera:1997plx} on the role of pressure 
{anisotropy}
for self-gravitating systems in astrophysics and cosmology. 
} 
\section{Energy density and pressure in the center}

If the form factors $A(t)$ and $D(t)$ are known, then the energy density
and pressure in the nucleon center can be computed directly as:
\sub{
\begin{align}\label{eq:T00(0)}
	T_{00}(0)=\frac{m}{4\pi^2}\  
    &	\int_{-\infty}^0 dt\ \sqrt{-t}\ \biggl[
        A(t) - \frac{t}{4m^2}D(t)\biggr]\,, \\
	\label{eq:p(0)}
	p(0)=\frac{1}{24\pi^2 m} 
    &	\int_{-\infty}^0 dt\ \sqrt{-t}\ \; t D(t).
\end{align}}
These integrals are convergent if 
{$A(t)$ drops faster than $\sim |t|^{-3/2}$ and}
$D(t)$ drops faster than $\sim |t|^{-5/2}$ at large $t$. 
However, we have to remember that the integrals above are subject 
to relativistic corrections, see discussion in Sec.~\ref{Sec:spin-0}.
If the integrals (\ref{eq:T00(0)},~\ref{eq:p(0)}) are convergent the relative 
size of the relativistic corrections to $T_{00}(0)$ and $p(0)$ is of the order 
$1/(m R_h)$ and is small for the nucleon both parametrically 
(it is $\sim 1/N_c$ in the large $N_c$ limit) and numerically. 
If the integrals (\ref{eq:T00(0)},~\ref{eq:p(0)}) are divergent the present 
formalism is not able to make predictions for the values of $T_{00}(0)$
and $p(0)$.
Notice that in Eq.~(\ref{eq:T00(0)}) we consistently neglected 
the contribution $\frac{t}{4m^2}[A(t)-2J(t)]$, cf.\ 
Eq.~(\ref{EQ:staticEMT-T00}), which constitutes a relativistic 
correction of the type discussed above. 
{However, we keep the term  $\frac{t}{4m^2}D(t)$ in this equation 
as the $D$-term can be specially enhanced, e.g.\ in the large $N_c$ limit}
{for baryons (where $t\sim N_c^0$, $m\sim N_c$, $D\sim N_c^2$, 
see Sec.~\ref{Sec:XI-rE-rhoE}).}

\section{The mean square radius of the energy density}
\label{Sec:XI-rE-rhoE}

The energy density in a mechanical system satisfies the inequality
$T_{00}(r) > 0$. This allows us to introduce the notion of the 
mean square radius of the energy density as follows
\begin{equation}
\label{eq:r2E}
	\langle r^2\rangle_E 
	=  \frac{\int d^3r\  r^2\ T_{00}(r)}
		{\int d^3r\  T_{00}(r)}
	 \, .
\end{equation}
This definition gives the mean square radius of the energy density 
in terms of the slope of the form factor $A(t)$ and the $D$-term:
\be\label{Eq:radius-T00}
	\langle r^2\rangle_E =6 A'(0)-\frac{3 D}{2 m^2}.
\ee
Note that the second term in this equation is not a relativistic correction 
for baryons as $D\sim N_c^2$ and the baryon mass $m\sim N_c$ 
(we will discuss the large-$N_c$ limit below).
However, for a heavy meson like a quarkonium, this term could 
be neglected as a relativistic correction.

Due to the trace anomaly (\ref{Eq:ff-of-EMT-trace-1}) we can express 
(in the chiral limit of QCD) the mean square radius of the gluon operator 
$F^2\equiv {F^{a,}}_{\!\!\mu\nu}F^{a,\mu\nu}$ in terms of EMT form factors
\be
	\langle r^2\rangle_{F^2}= 6 A'(0)-\frac{9 D}{2 m^2}
	=\langle r^2\rangle_E-\frac{3 D}{ m^2}.
\ee
Below in Sections~\ref{Sec-9:mech-radius} and \ref{Sec:norm_tang_forces} 
we shall see that the $D$-term is always negative, which implies that 
{$\langle r^{\,2}\rangle_{\!F^2} > \langle r^2\rangle_E$.}

\section{The mechanical mean square radius}
\label{Sec-9:mech-radius}

The normal component of the total force exhibited by the system on 
an infinitesimal piece of area  $dS^j$ at the distance~$r$ has the form  
$F^i(\vec r)=T^{ij}(\vec r)\,dS^j=\left[\frac23s(r)+p(r)\right]\,dS^i$
{where $dS^j = dS \, r^j/r$}.
In Ref.~\cite{Perevalova:2016dln} we argued that for the mechanical stability 
of the system the corresponding force must be directed outwards. Therefore
the local criterion for the mechanical stability can be formulated as the 
inequality 
\be\label{Eq:local-stability-criterion}
	\frac 23 s(r) +p(r) > 0.
\ee
This inequality implies that the $D$-term for any stable system must be 
negative \cite{Perevalova:2016dln}, 
\be
	D < 0 \, .
\ee
The positive combination $\left[\frac 23 s(r) +p(r)\right]$ has the meaning 
of the normal force distribution in the system. This allows us to introduce 
the notion of the mechanical radius for hadrons:
\begin{equation}
\label{eq:mechanicalradius}
	\langle r^2\rangle_{\rm mech} 
	=  \frac{\int d^3r\  r^2\ \left[\frac 23 s(r) +p(r)\right]}
		{\int d^3r\ \left[\frac 23 s(r) +p(r)\right]}
	= \frac{6 D}{\int_{-\infty}^0 dt\ D(t)} \, ,
\end{equation}
where in the numerator we used Eqs.~(\ref{Eq:D-equivalent-expressions}) 
as $\int d^3r\  r^2\ \left[\frac 23 s(r) +p(r)\right]=-3D/2m$, while
in the denominator we used the von Laue condition (\ref{Eq:von-Laue})
and the relation of the surface tension energy of the system 
$\int d^3r\ s(r)$ to $D(t)$ given by
\begin{equation}
\label{eq:surfacetensionEnergy}
	\int d^3r\ s(r)=-\frac{3}{8 m} \int_{-\infty}^0 dt\ D(t)	 \, ,
\end{equation}
We can conjecture that the surface tension energy must be smaller than
the total energy of the system, i.e.\ $\int d^3r\ s(r)\le m$. This
would imply at least two things. First, the integral $\int_{-\infty}^0
dt\ D(t)$ must converge which implies that $D(t)$ must drop at large
$|t|$ faster than $\sim 1/t$. Second, the mechanical radius is bound
from below by $\langle r^2\rangle_{\rm mech} \ge -9 D/(4 m^2)$.  
We note however that our conjecture is still lacking a rigorous proof.

We see that $D$-term also determines the mechanical 
radius of the hadrons. Notice the unusual definition: 
unlike e.g.\ the mean square charge radius, the mechanical 
mean square radius is not related to the slope of a form factor.
Also note that one obtains a non-zero value of the mechanical radius 
if $D(t)$ drops faster than $1/t$ at large (negative) values of $t$. 
For the multipole Ansatz $D(t)=D/(1-t/\Lambda^2)^n$  
the resulting mechanical mean square radius is 
$\langle r^2\rangle_{\rm mech} = 6 (n-1)/\Lambda^2$.

In Ref.~\cite{Kumano:2017lhr}, where an extraction of EMT form 
factors of $\pi^0$ in the time-like region was reported, also a definition
of a ``mechanical radius'' was proposed, however, in terms of the slope 
of the form factor $D(t)$. This differs from our definition 
(\ref{eq:mechanicalradius}), and is not an appropriate measure 
of the true mechanical radius of a hadron \cite{Polyakov:2018guq}.

\section{Normal and tangential forces in hadrons}
\label{Sec:norm_tang_forces}

Let us consider the slice of a hadron defined in spherical coordinates 
by the equation $\theta=\pi/2$. Other choices of the slice are equivalent
to this for spherically symmetric hadrons, i.e.\ hadrons with spins 0 and 1/2.
At any point on the slice the force (pressure) acting on the infinitesimal 
area element $d\bm{S}=dS_r\bm{e}_r+dS_\theta\bm{e}_\theta+dS_\phi\bm{e}_\phi$  
has the following spherical components:
\be\label{Eq:force-spherical components}
	\frac{dF_r}{dS_r}=\frac 23 s(r)+p(r), \quad 
	\frac{dF_\theta}{dS_\theta}=\frac{dF_\phi}{dS_\phi}=-\frac 13 s(r)+p(r).
\ee
The normal ($dF_r$) and tangential ($dF_\phi,\, dF_\theta$) forces are the 
eigenvalues of the stress tensor $T_{ij}$ with $\bm{e}_r$, $\bm{e}_\theta$, 
$\bm{e}_\phi$ being the corresponding eigenvectors, which define the 
local principal axes. The positive (negative) eigenvalues correspond to
``streching'' (``squeezing'') along the corresponding principal axes. 
For spherically symmetric hadrons (spin-0, spin-1/2) the tangential 
forces {$dF_\phi, dF_\theta$} are equal {to each other}, for higher 
spin hadron, generically, they are different and spin dependent. In a
stable spherically symmetric system the normal component of the force 
$dF_r{/dS_r}=\frac 23 s(r) +p(r)$ must correspond to ``streching'' forces 
(to be positive, {see Sec.~\ref{Sec-9:mech-radius}}) otherwise 
the system would collapse in ``squeezing'' direction. 
The tangential forces {change sign with the distance $r$}
because possible ``squeezing'' is averaged to zero for spherically 
symmetric systems. It would be interesting to consider the local 
stability conditions for higher spin hadrons.

The positivity of the normal component of the force $dF_r$ allowed 
us to define the notion of the mechanical radius
in Sec.~\ref{Sec-9:mech-radius}. The tangential components 
of the force $dF_\theta$ and $dF_\phi$ also allow us to define  additional 
stability conditions with a nice, physically intuitive  
interpretation as the mechanical stability of {lower} dimensional 
subsystems of the whole 3D system. The $\theta$-component 
of the force acts perpendicular to the $\theta=\pi/2$ hadron's slice. 
Let us consider the $\theta$ component of the force acting on the 
infinitesimaly narrow ring of the width $dr$. The corresponding force is:
\be\label{Eq:tang-force-ring}
	dF =\left[-\frac 13 s(r)+p(r)\right] 2\pi r dr.
\ee 
As the pressure is the isotropic part of the stress tensor, the above 
expression can be written for the 2D subsystem (slice of the hadron 
$\theta=\pi/2$) as follows:
\be
	({\rm 2D\ pressure)} \times d{\rm (2D\ volume)}.
\ee 
It is a general relation between the force experienced 
by the $(n-1)D$ subsystem from the {viewpoint} 
of the $nD$ system: $dF^{(nD)}=p^{(n-1)D} dV^{(n-1)D}$, 
where $p^{(n-1)D}$ is the pressure in $(n-1)D$ subsystem and 
$dV^{(n-1)D}$ is the volume element in it. This equation tells 
that the force acting on the subsystem can be considered as 
a free energy from point of view of the {lower}-dimensional 
subsystem, see, e.g.\ the discussions in Chapter~II of \cite{LLv7}.
We shall discuss {these relations elsewhere in more detail} \cite{XYZ-new}.

The von Laue stability  condition for the 2D subsystem 
($\int dV^{(2D) } p^{(2D) }=0$) has the form:
\be
\label{Eq:tang-stability}
	2\pi \int_0^\infty dr\ r \left[-\frac 13 s(r)+p(r)\right] =0.
\ee 
This is a new tangential stability condition.

If we further consider the 
1D subsystem (e.g.\ a ray $\phi=0$) of the 2D slice $\theta=\pi/2$ we can 
compute the 2D force acting on the line element\footnote{A line (1D) element 
	plays the role of a ``volume element'' for a 1D system and is the 
	``surface element'' from the point of view of the 2D system.} 
$dr$ of 1D ray ($\phi=0, \theta=\pi/2$) as:
\be
\label{Eq:tang-stability-1D-xx}
dF^{(\rm 2D)}=\left[-\frac 43 s(r) +p(r)\right] dr,
\ee
For the derivation of this equation 
see the discussion after Eq.~(\ref{Eq:stress-tensor-p-s-nD}).
Again the equation (\ref{Eq:tang-stability-1D-xx}) has the 
structure (1D pressure)$\times d$(1D volume). 
The corresponding 1D von Laue stability condition reads:
\be
\label{Eq:1Dstability}
\int_0^\infty dr \left[-\frac 43 s(r)+p(r)\right] =0.
\ee
Using Eq.~(\ref{Eq:relationSPD}) we can relate the normal and 
tangential pressures to the $D$-form factor:
\be
\label{Eq:relationNTD}
	\frac 23 s(r)+p(r)
	= \frac{1}{{2}m} \frac 1r \frac{d}{dr} \widetilde D(r), \quad 
	-\frac 13 s(r)+p(r)
	= \frac{1}{4 m} \frac{1}{r} \frac{d}{dr} r\frac{d}{dr} 
	\widetilde D(r), \quad
	-\frac 43 s(r)+p(r)
	= \frac{1}{2 m} \frac{d^2}{dr^2}  
	\widetilde D(r).
\ee
From these expressions we see immediately that the tangential stability 
conditions (\ref{Eq:tang-stability}) and (\ref{Eq:1Dstability})
are satisfied automatically.  
Interestingly we can write these expressions and Eq.~(\ref{Eq:relationSPD}) 
as
\be
\label{Eq:pLaplace}
	p^{{\rm (nD)}}(r)=\frac{1}{2 n m}  \partial^2_{\rm nD} \widetilde D(r),
\ee
where $\partial^2_{{\rm nD}}$ is the $n$-dimensional ($n$D) Laplace operator 
and the pressures in the $n$-dimensional subsystems are:
\be
	p^{{\rm (3D)}}(r)=p(r), \quad 
	p^{{\rm (2D)}}(r)=-\frac 13 s(r)+p(r),\quad 
	p^{{\rm (1D)}}(r)=-\frac 43 s(r)+p(r).
\ee
The von Laue stability condition in $n$-dimensions has the form:
\be
	\int d^nr\ p^{{\rm (nD)}}(r) =0,
\ee
which is automatically satisfied due to Eq.~(\ref{Eq:pLaplace}).

Generically for the $n$-dimensional spherically symmetric subsystem 
the conserved stress tensor has the form:
\begin{align}\label{Eq:stress-tensor-p-s-nD}
	T^{\rm (nD)}_{ij}(\bm{r}) = \biggl(
	\frac{r_ir_j}{r^2}-\frac1n\,\delta_{ij}\biggr) s^{\rm (nD)}(r)
	+ \delta_{ij}\,p^{\rm (nD)}(r)\,.
\end{align}
The naive restriction of $T^{\rm (nD)}_{ij}(\bm{r})$ to the $(n-1)$D 
subspace leads to a non-conserved stress tensor. To construct a 
conserved stress tensor one has to take into account the forces 
acting on the $(n-1)$D subsystem from outside. The corresponding  
conserved stress tensor for the  $(n-1)$D subsystem,  
i.e.\ a stress tensor satisfying {
$\nabla^a T^{\rm (n-1)D}_{ab}(\bm{r})=0$,
has the form:
\begin{align}\label{Eq:stress-tensor-p-s-n-1D}
	T^{\rm (n-1)D}_{ab}(\bm{r}) = 
	T^{\rm (nD)}_{ab}(\bm{r}) -
	\left(\delta_{ab}-
	\frac{r_a r_b}{r^2}\right)\ \frac{s^{\rm (nD)}(r)}{n-2}\,,
\end{align}
where $a,b=1,\ldots, n-1$ and $\bm{r}\in (n-1)$D~subspace. }
The last term in this equation takes into account the forces experienced 
by $(n-1)$D subsystem from outside.
From this one obtains the following relations between pressures 
and shear forces of subsystems with different dimensions as:
\be
	p^{\rm (n-1)D}(r)=-\frac 1n\ s^{\rm (nD)}(r) +p^{\rm (nD)}(r), \quad 
	s^{\rm (n-1)D}(r)=\frac{n-1}{n-2}\ s^{\rm (nD)}(r).
\ee
Interesting is that we can consider a hadron as a 3D subsystem 
of some $n$-dimensional mechanical system with pressure and shear 
force distributions\footnote{The general relations  are
	$p^{\rm (nD)}(r)=p^{\rm (kD)}(r)+\frac{(k-1)(n-k)}{k\cdot n}s^{\rm (kD)}(r)$
	and $s^{\rm (nD)}(r)=\frac{k-1}{n-1} s^{\rm (kD)}(r)$.}:
\be
	p^{\rm (nD)}(r)=p(r)+\frac{2(n-3)}{3 n} s(r), \quad 
	s^{\rm (nD)}(r)=\frac{2}{n-1} s(r),
 \ee 
where $p(r)$ and $s(r)$ are pressure and shear forces distribution in 
a 3D hadron. {Such a picture can have interesting connections to AdS/QCD.}

From the positivity of the normal forces  {and the first equation in 
(\ref{Eq:relationNTD})}  we obtain an important constraint 
for the 3D Fourier transform of the $D$-form factor, 
$\widetilde{D}(r)$ defined in Eq.~(\ref{Eq:relationSPD}):
it must be monotonicaly increasing function of $r$. For hadrons we expect 
that $\widetilde{D}(r)\to 0$ for $r\to \infty$, thus $\widetilde{D}(r)$ must 
be negative! So we obtain that not only $D$-term must be negative, but also 
the whole form-factor in the coordinate space is negative.

The tangential stability condition (\ref{Eq:tang-stability}) implies that 
the tangential forces ($dF_\phi, dF_\theta$) must at least once change their 
signs. This implies that the tangential forces inside a hadron must switch 
their nature from ``streching" to ``sqeezing". Such pattern is illustrated 
on Fig.~\ref{Fig:visualisation} of Section~\ref{Sec:Nucleon} where 
{a model prediction for the}
distribution of the tangential forces inside the nucleon is shown. 
One clearly sees that the tangential forces are ``stretching" inside the 
sphere of radius $\sim 0.5$~fm and are ``squeezing" outside it.

We obtained here the von Laue (\ref{Eq:von-Laue}) and tangential 
(\ref{Eq:tang-stability},\ref{Eq:1Dstability}) stability conditions from physical considerations 
about forces. Technically they can be traced back to the EMT conservation 
${\nabla^i T^{ij}(\bm{r})}=0$. 
Using the conservation of EMT we can derive a set of interesting relations 
which are presented in Appendix~\ref{App:SR-EMT-densities}.
In particular, we can obtain the non-linear integral relation among EMT 
densities (\ref{Eq:spectacular}) which can be rewritten equivalently as:
\be\label{Eq:spectacular-rewritten}
	\int  d^3r\biggl[
	{\rm tr}(T(\bm{r})^2)-\frac 12 {\rm tr}(T(\bm{r}))^2\biggr]=0,
\ee
where $T(\bm{r})$ is the matrix corresponding to $T_{ij}(\bm{r})$.	
Such kind of relations can be very useful for analysis of dispersion 
relations for EMT form factors.

\section{\boldmath $D$-term in free theories: 
Distinguishing bosons and fermions}
\label{Sec:boson-vs-fermion}

Before discussing the phenomenologically interesting {cases} 
it is instructive to inspect free field theories. 
Remarkably, the particle property $D$-term can ``distinguish'' between 
non-interacting pointlike (``elementary'') bosons and fermions in the 
following sense. 
A free spinless boson has a non-zero intrinsic $D$-term $D=-1$.
In sharp contrast to this, the $D$-term of a free spin-$\frac12$ 
fermion is zero, see \cite{Hudson:2017xug,Hudson:2017oul} and 
references therein.

This unexpected finding deserves a comment. One can give a pointlike 
boson a ``finite, extended, internal structure'' (by ``smearing out'' 
its energy density $T_{00}(r)=m\,\delta^{(3)}(\bm{r})$ with e.g.\ a 
narrow Gaussian, and analog for other densities) such that the property 
$D=-1$ is preserved. This yields automatically the characteristic shapes 
for $p(r)$ and $s(r)$ found in dynamical model calculations 
\cite{Hudson:2017xug}.
Interesting field theories of extended (solitonic $Q$-ball type) solutions 
with $D=-1$ can be constructed where such ``smearing out'' is implemented 
dynamically \cite{Hudson:2017xug}. In general interactions modify 
the free theory value $D=-1$, {see next section. But the point is: 
a spin-zero boson has an intrinsic $D$-term.}

The situation is fundamentally distinct for fermions: here 
interactions do not modify the $D$-term, they {\it generate} it. 
A non-zero fermionic $D$-term is {\it purely} of dynamical origin 
\cite{Hudson:2017oul}. Recalling that all known matter is fermionic,
this indicates the importance to study the physics of the $D$-term.

\section{\boldmath $D$-term in weakly interacting theories}

In this section we review the $D$-term in theories with an interaction 
so weak that a perturbative treatment is justified, starting with 
the $\Phi^4$ theory, see \cite{Callaway:1988ya} for a review, 
defined by ${\cal L} = \frac12\,(\partial_\mu\Phi)(\partial^\mu\Phi)- V(\Phi)$
with $V(\Phi) = \frac12\,m^2\Phi^2 + \frac{\lambda}{4!}\,\Phi^4$.
The EMT of this theory was studied in Ref.~\cite{Callan:1970ze},
see also \cite{Coleman:1970je,Freedman:1974gs,Freedman:1974ze,
Lowenstein:1971vf,Schroer:1971ud,Collins:1976vm}.
For scalar fields the canonical EMT obtained from the Noether 
theorem is symmetric,
$\hat{T}^{\mu\nu}(x)=(\partial^\mu\Phi)(\partial^\nu\Phi)-g^{\mu\nu}{\cal L}$,
and yields $D=-1$ in free Klein-Gordon theory 
\cite{Pagels:1966zza,Hudson:2017xug}. This symmetric EMT follows 
from Eq.~(\ref{Eq:EMT-from-gravity}) with a ``mimimal coupling'' 
of the theory to gravity $S_{\rm grav,min} = \int d^4x\;\sqrt{-g}[
\frac12\,g^{\mu\nu}(\partial_\mu\Phi)(\partial_\nu\Phi)-V(\Phi)]$.
However, already on classical level this EMT is not conformally 
invariant, not even for $m\to0$. This can be remedied by working with 
a non-minimal coupling term of the scalar field to the curvature
\be\label{Eq:S-grav-II}
	S_{\rm grav} = \int d^4x\;\sqrt{-g}\biggl(
	\frac12\,g^{\mu\nu}(\partial_\mu\Phi)(\partial_\nu\Phi)- V(\Phi)
	-\frac12\,h\,R\,\Phi^2\biggr) , \quad
	h = \frac14\biggl(\frac{n-2}{n-1}\biggr),
\ee
where $R$ is the Riemann scalar, $n$ denotes the number of space-time 
dimensions. The effect of this non-minimal coupling is to add an 
``improvement term:'' 
$\hat{T}^{\mu\nu} \to \hat{T}^{\mu\nu} + \hat{\Theta}^{\mu\nu}_{\rm improve}$
with $\hat{\Theta}^{\mu\nu}_{\rm improve} = -h
(\partial^\mu\partial^\nu-g^{\mu\nu}\square)\,\phi(x)^2$ \cite{Callan:1970ze}. 
It is often said one can add to the EMT operator ``any quantity whose 
divergence is zero and which does not contribute to Ward identities'' 
\cite{Collins:1976vm}. In fact, adding this $\hat{\Theta}^{\mu\nu}_{\rm improve}$
preserves $\partial_\mu\hat{T}^{\mu\nu}=0$, has the important virtue of 
making $\hat{T}^{\mu\nu}$ a finite renormalized operator at 1, 2, 3 loop 
level \cite{Collins:1976vm}, and does not affect the $00$-- and 
$0k$--components such that mass and spin of the particle are not altered. 
However, it modifies the stress tensor components, and the $D$-term is 
changed from its free field theory value $D_{\rm free}=-1$ to 
$D_{\rm interact}=-1+4h = - \frac13$ in $n=4$ space-time dimensions.

This is a remarkable result \cite{Hudson:2017xug}. As the $D$-term is an 
observable it must be uniquely defined. Thus, total derivatives cannot 
be arbitrarily added to the EMT. Rather it is essential to establish a 
unique definition of the EMT and verify (in QCD) whether the
matrix elements of this EMT operator are probed in experiment. 

Notice that the gravitational background field was used here only to derive 
the EMT operator: after the variation in Eq.~(\ref{Eq:EMT-from-gravity}) 
the metric $g^{\mu\nu}$ was set to flat space. However, the renormalizability 
of the $\Phi^4$ theory can be studied also in weakly curved gravitational 
background fields: the same improvement term is needed there to make 
$\hat{T}^{\mu\nu}$ finite \cite{Brown:1980qq}.  
Since no quantum theory of gravity is known, it is of course also not 
known whether the term $\hat{\Theta}^{\mu\nu}_{\rm improve}$ would ensure 
renormalizability if quantum gravity effects were included. 
At this point one might be tempted to think that gravity is far too
weak to be of relevance in particle physics. However, the lesson we
learn is that even infinitesimally small interactions in $\Phi^4$ 
theory can impact the $D$-term. So why not infinitesimally small 
gravitational interactions?
We note that the $D$-term emerges to be strongly sensitive to interactions. 
One must consistently include all forces, perhaps even gravity, to determine 
the true improvement term and the true value of the $D$-term  
\cite{Hudson:2017xug}.

At this point it is instructive to mention also studies in the 
$\Phi^3$ theory defined by 
${\cal L} = \frac12\,(\partial_\mu\Phi)(\partial^\mu\Phi) - V(\Phi)$
with $V(\Phi) = \frac12\,m^2\Phi^2 + \frac{\lambda}{3!}\,\Phi^3$.
This is not a realistic theory with a potential not even bound
from below. However, it is a popular toy model in many situations
including studies of GPDs \cite{Radyushkin:1996nd} where GPDs 
$\propto{\cal O}(\lambda^2)$ can be computed in one-loop order 
with {\it no} $D$-term \cite{Pobylitsa:2002vw}.
Historically a lot of intuition about GPDs, and their analytic properties
was gained in \cite{Radyushkin:1996nd} on the basis of diagrams
from the $\Phi^3$ theory, see \cite{Pobylitsa:2002vw} for further
applications {of the $\Phi^3$ theory}. Such diagrams give rise to GPDs 
$\propto{\cal O}(\lambda^2)$ {where} $\lambda=\lambda(\mu)$
has a renormalization scale dependence. Hence a GPD 
{being proportional to} ${\cal O}(\lambda^2)$ cannot 
contribute to the renormalization scale independent $D$-term.
Only the tree level diagram can do this, see previous section.
But such diagrams are not interesting from the diagramatic
point of view, and bear no insights on analytic properties of GPDs. 
Consequently, they were not considered and the {$D$-term} 
was overlooked in \cite{Radyushkin:1996nd}. This point was clarified 
in \cite{Polyakov:1999gs}. Untill today the modelling of GPDs is most 
conveniently done in the so-called ``double distribution Ansatz'' 
developed on the basis of the perturbative calculations in $\phi^3$ 
theory in \cite{Radyushkin:1996nd}  {\it supplemented} by the $D$-term 
\cite{Polyakov:1999gs}. This story underlines the deep relation
of the $D$-term to non-perturbative physics. 

The weakest interaction in nature is gravity. Although a theory of 
quantum gravity is not yet known, the leading quantum corrections 
can be computed from the known low energy structure of the theory 
\cite{Donoghue:1993eb}. These calculations are challenging
\cite{Donoghue:2001qc,Khriplovich:2002bt,BjerrumBohr:2002kt,Khriplovich:2004cx}.
The loop corrections to the Reissner-Nordstr\"om and Kerr-Newman metrics
\cite{Khriplovich:2002bt,BjerrumBohr:2002kt,Khriplovich:2004cx}
show how (QED, gravity) interactions generate quantum long-range
contributions to the stress tensor and other EMT densities. A 
consistent description of the $D$-term requires the inclusion
of all contributions: also the short-distance contributions which 
cancel exactly the long-distance ones in the von Laue condition. 
The results of these works therefore do not allow us to gain insights 
on how much quantum gravity corrections contribute to the $D$-terms
of elementary (charged) fermions. However, these results give
us a feeling of gravitational infrared corrections at large
distances.

\section{\boldmath Illustration of forces in the $Q$-ball toy model}

{Before discussing hadrons} it is instructive to review 
{another strongly interacting theory}, namely $Q$-balls 
\cite{Mai:2012yc,Mai:2012cx,Cantara:2015sna}.
$Q$-balls are non-topological solitons in theories with global symmetries,
e.g.\ U(1) \cite{Friedberg:1976me,Coleman:1985ki,Lee:1991ax}.
They might have played a role in the early universe and are dark matter 
candidates \cite{Kusenko:1997si,Enqvist:1997si,Kasuya:1999wu}.
$Q$-balls exhibit a variety of families of solutions including stable, 
metastable, unstable solitons and radial excitations.
In Ref.~\cite{Mai:2012yc,Mai:2012cx,Cantara:2015sna} the EMT of 
$Q$-balls was studied in the complex scalar theory 
${\cal L}=\partial_\mu\Phi^\ast\partial^\mu\Phi-V$ 
with $V=A\,\Phi^\ast\Phi-B\,(\Phi^\ast\Phi)^2+C\,(\Phi^\ast\Phi)^3$ where 
$A,\,B,\,C$ are positive constants. This theory is not renormalizable 
and understood as an effective theory \cite{Coleman:1985ki}. 
The soliton solutions are of the type 
$\Phi(t,\bm{r}\,)=e^{i\omega t}\phi(r)$ where 
$\omega_{\rm min}<\omega<\omega_{\rm max}$ with 
$\omega_{\rm min}^2=2A(1-B^2/4AC)$ and $\omega_{\rm max}^2=2A$. 
For $\omega\to\omega_{\rm min}$ one deals with absolutely stable 
solitons \cite{Coleman:1985ki}.
For $\omega\to\omega_{\rm max}$ one deals with $Q$-clouds, extremely 
unstable solutions which delocalize, and disociate into an infinitely 
dilute system of free quanta \cite{Alford:1987vs}.

\begin{figure}[t!]

\vspace{-3mm}

\centering
\includegraphics[height=3cm]{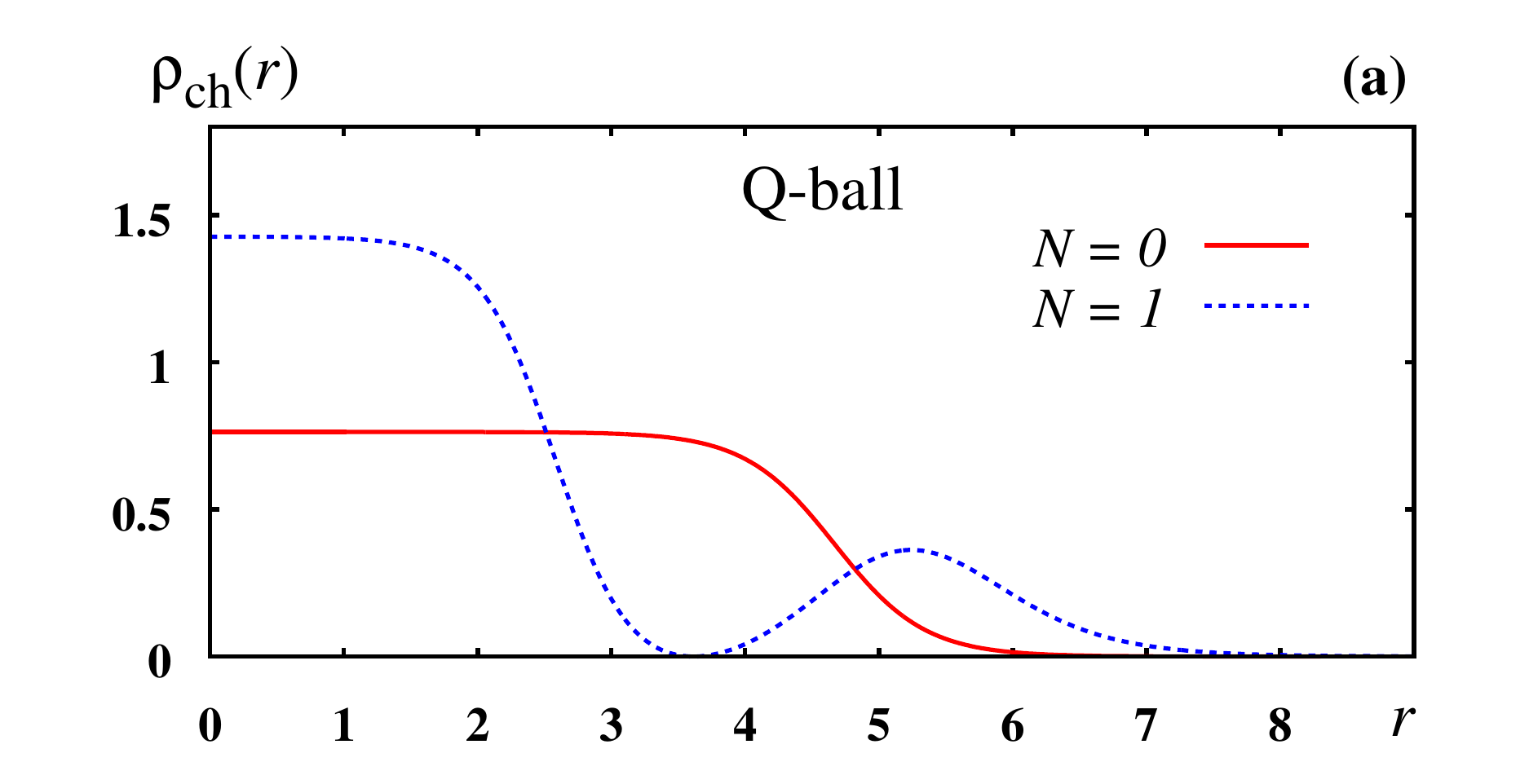}
\includegraphics[height=3cm]{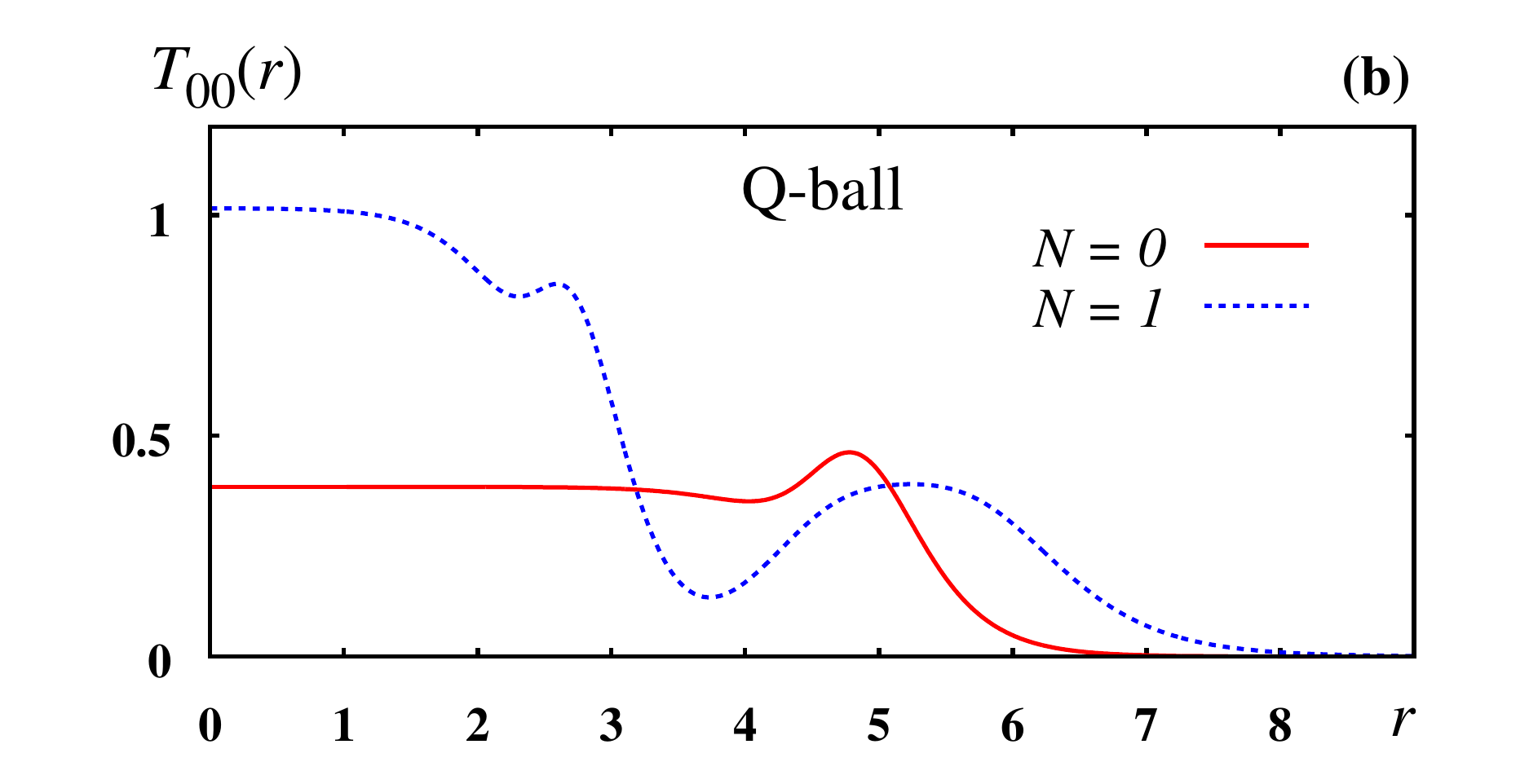}
\includegraphics[height=3cm]{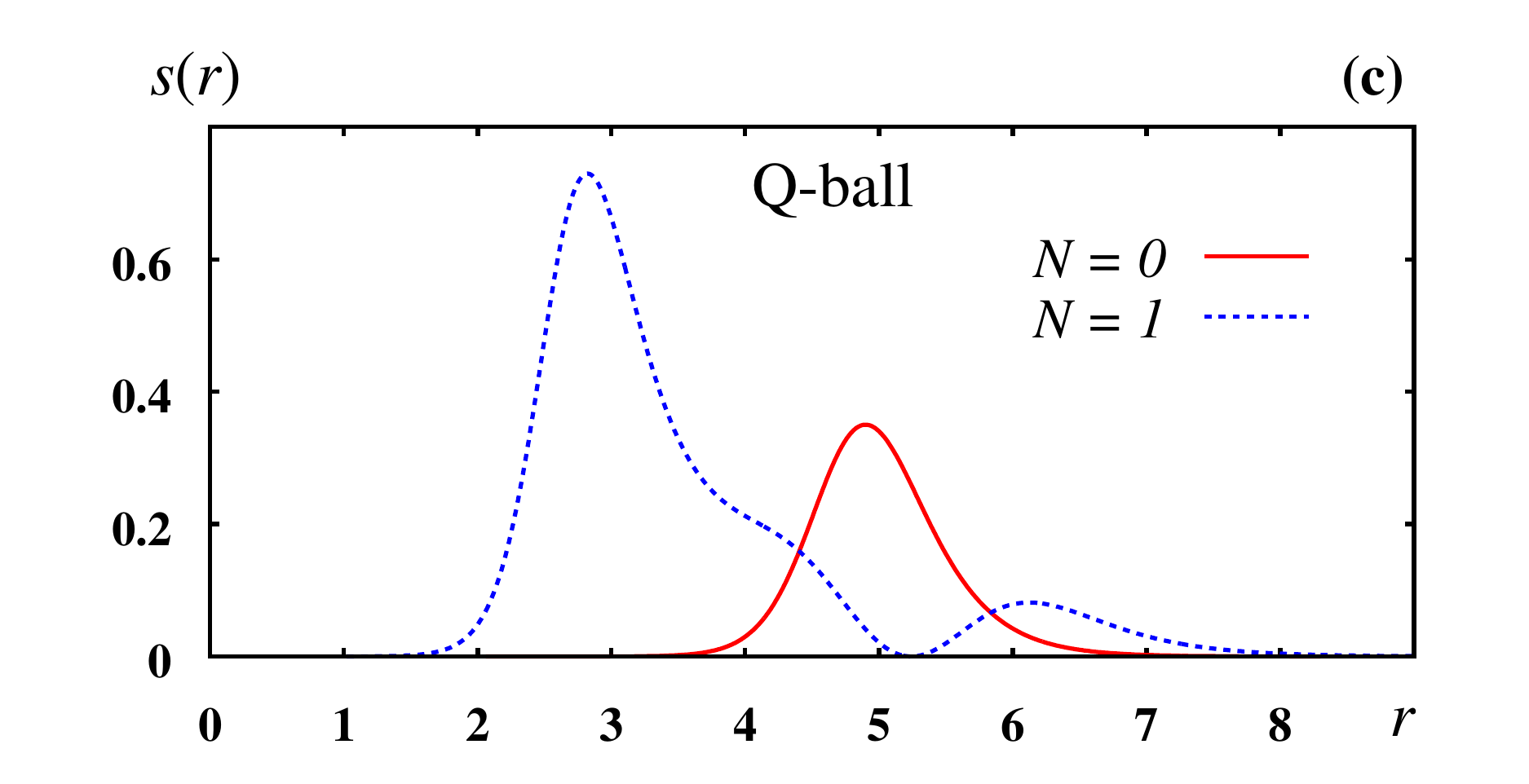}

\includegraphics[height=3cm]{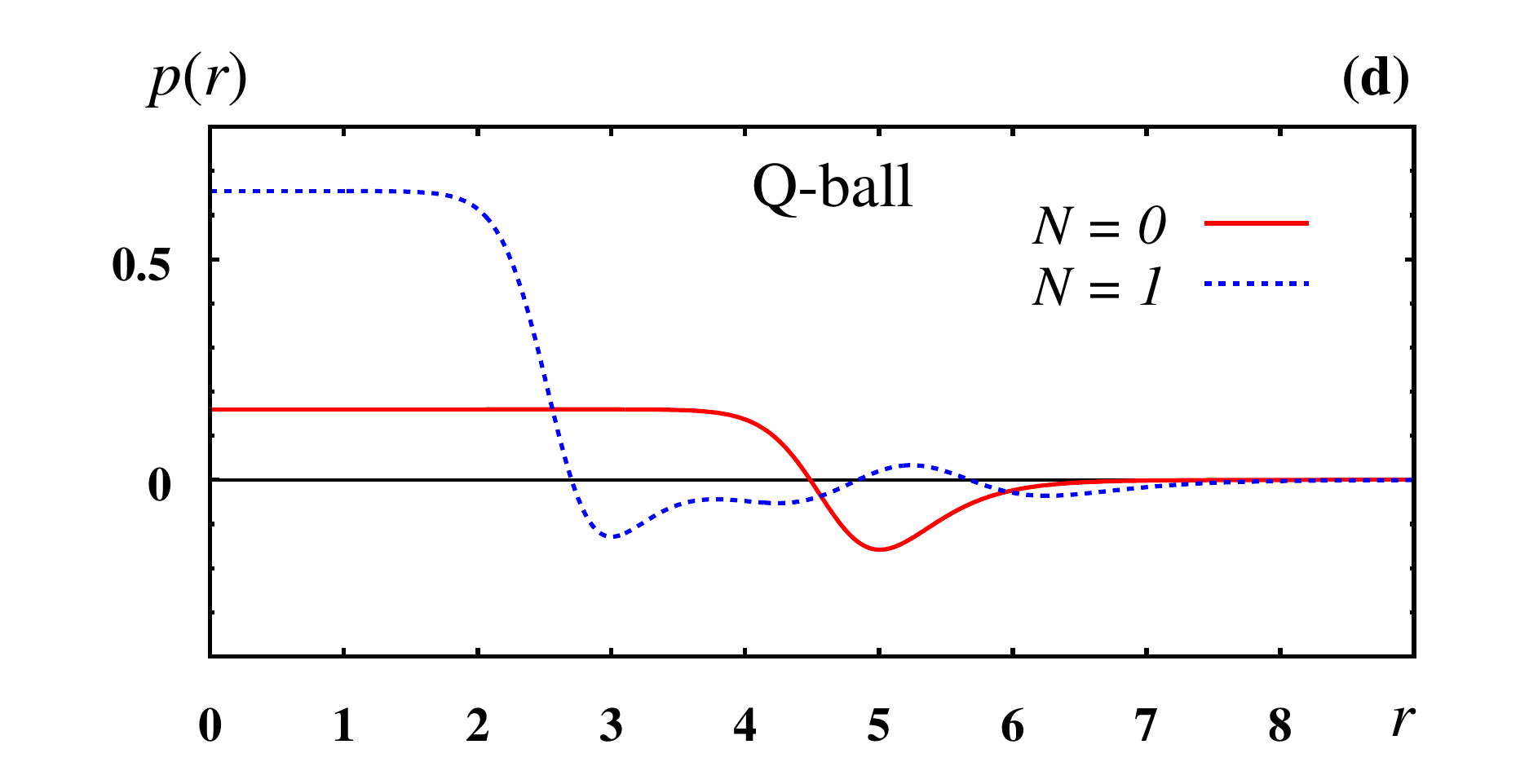}
\includegraphics[height=3cm]{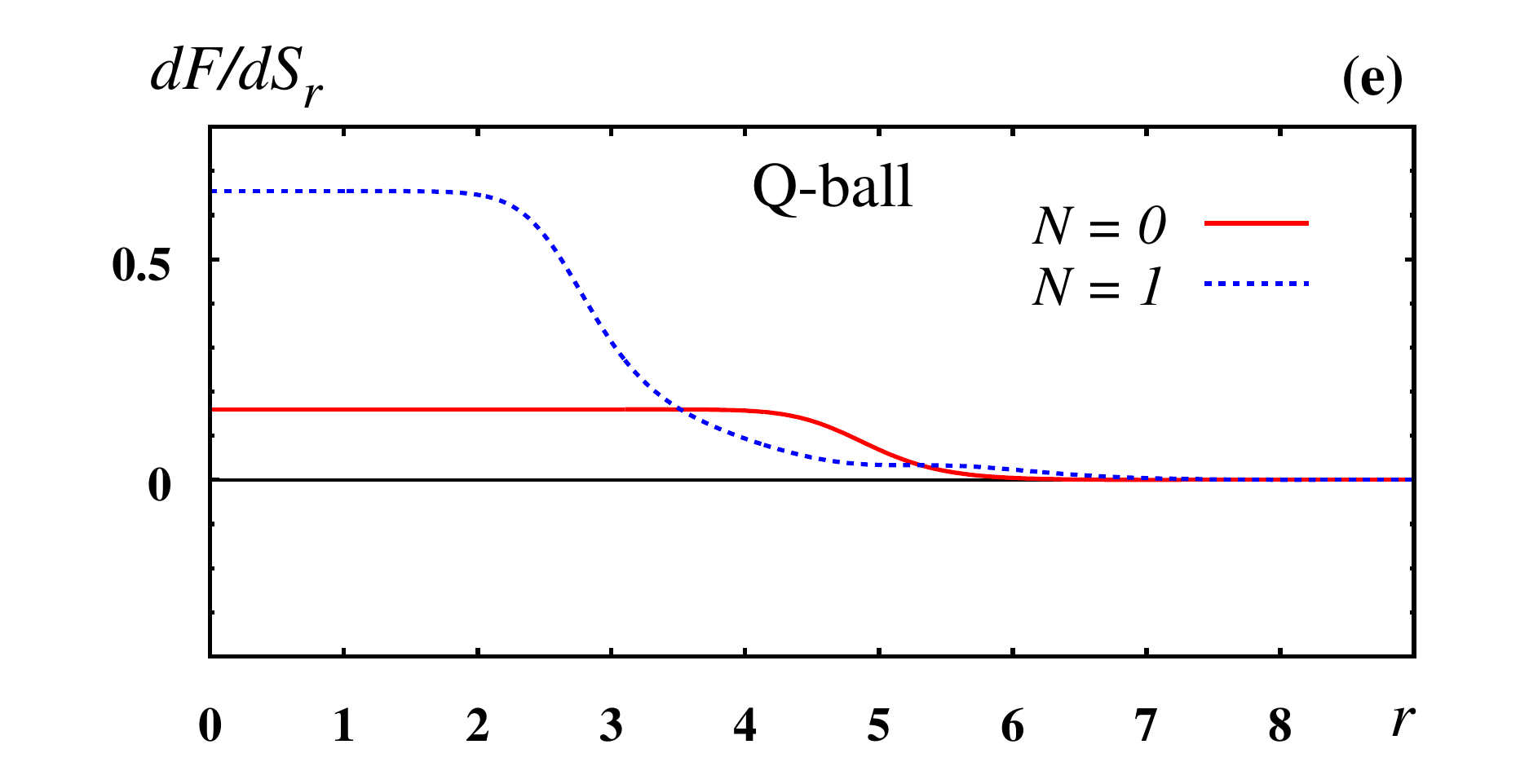}
\includegraphics[height=3cm]{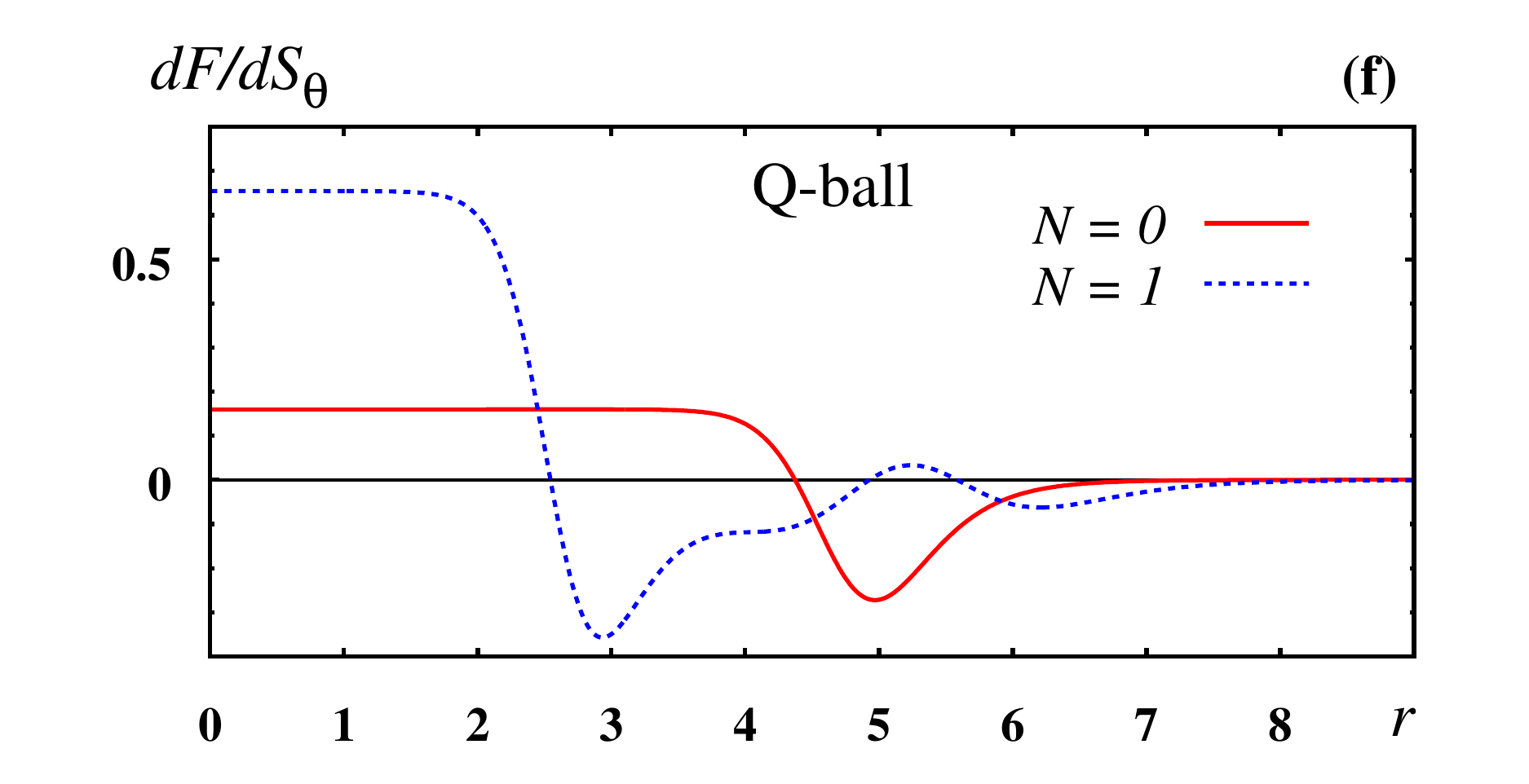}

\vspace{-1mm}

\caption{\label{Fig:Qball}
	Charge distribution $\rho_{\rm ch}(r)$, 
	energy density $T_{00}(r)$, shear forces $s(r)$,
	pressure $p(r)$, normal force $dF/dS_r$ and tangential
	force $dF/dS_\theta$ inside a stable $Q$-ball \cite{Mai:2012yc}
	($N=0$, solid line) and inside the first radial excitation
	($N=1$, dotted line) \cite{Mai:2012cx}.
	Both solutions $N=0\,, \,1$ have the same charge $Q$.
	All quantities are in units of an appropriate of 
	power of GeV, $r$ is in ${\rm GeV}^{-1}$.}
\end{figure}

In Fig.~\ref{Fig:Qball} we show results from \cite{Mai:2012yc,Mai:2012cx}
obtained with the parameters 
$[A,\,B,\,C]=[1.1\,{\rm GeV}^2,\,2.0,\,1.0\,{\rm GeV}^{-2}]$
for the ground state $N=0$ and first excited state $N=1$ with the
charge $Q=342$. The absolutely stable ground state has
$m=286\,{\rm GeV}$ and $D=-2.5\times10^5$. The mechanical radius, 
mean square radius of the energy density, and charge radius are given by
$[\la r^2\ra^{1/2}_{\rm ch},\,\la r^2\ra^{1/2}_{\rm mech},\,\la r^2\ra^{1/2}_E]=
[3.8,\,4.0,\,4.4]\,{\rm GeV}^{-1}$. 
The $N=1$ state has $m=461\,{\rm GeV}$ and $D=-4.8\times10^5$ with radii 
$[\la r^2\ra^{1/2}_{\rm ch},\,\la r^2\ra^{1/2}_{\rm mech},\,\la r^2\ra^{1/2}_E]=
[4.9,\,3.6,\,5.1]\,{\rm GeV}^{-1}$. 

The charge distribution of the ground state in Fig.~\ref{Fig:Qball}a
is uniform in the interior and drops to zero over a relatively narrow 
``egde-region.'' $T_{00}(r)$ exhibits a similar behavior except for the 
characteristic peak at the ``edge'' of the system due to the surface 
tension, see Fig.~\ref{Fig:Qball}b. The shear forces indicate most
clearly the position of the ``edge,'' Fig.~\ref{Fig:Qball}c. The 
pressure is positive and constant in the interior, crosses the
zero in the edge region and stays negative for large $r$, see
Fig.~\ref{Fig:Qball}d.
The normal forces are positive, Fig.~\ref{Fig:Qball}e, and comply with 
the local stability requirement (\ref{Eq:local-stability-criterion}).
The tangential forces are positive in the inner region, change sign 
in the edge region, and remain negative at large $r$, see
Fig.~\ref{Fig:Qball}f.

For the $N=1$ excited state the situation is much different:
a portion of the charge is carried by an outer shell 
which is separated from the interior, see Fig.~\ref{Fig:Qball}a. 
The shell structure is visibile also in $T_{00}(r)$ and $s(r)$, 
see Figs.~\ref{Fig:Qball}b, c. In general, the $N^{\rm th}$ 
excited state exhibits $N$ shells, and $p(r)$ has $(2N+1)$ 
zeros: always with the pattern of being positive in the center
and negative in the asymptotic large-$r$ region. 
The excited solution is much heavier, but not significantly
larger: as a consequence the energy density and the magnitude
of the foces in its interior are much larger compared to the
ground state. Remarkably, the mechanical radius of the 
excited solution is smaller than that of the ground state
solution. This is due to the fact that the normal forces
diminish with $r$ much earlier for $N=1$ as compared to $N=0$.

This shows that the mechanical radius of a system gives a much
different view as compared to the charge radius or energy density radius. 
In general, the mass and charge of the $N^{\rm th}$ excited 
state grow like $N^3$ and the radii grow like $N$.
But the $D$-term exhibits the strongest growth with
$N^8$ \cite{Mai:2012cx}.

In the strict $Q$-ball limit 
$\varepsilon_{\rm max}^2=\omega_{\rm max}^2-\omega^2\to 0$
\cite{Coleman:1985ki} the properties scale as 
$\gamma\propto\varepsilon_{\rm max}^0$,
radii $\propto\varepsilon_{\rm max}^{-1}$,
$m$ and $Q\propto\varepsilon_{\rm max}^{-3}$, while
$D\propto\varepsilon_{\rm max}^7$ \cite{Mai:2012yc}. 
In the opposite $Q$-cloud limit 
$\varepsilon_{\rm min}^2=\omega^2-\omega_{\rm min}^2\to 0$
\cite{Alford:1987vs} the properties behave as
$\gamma\propto\varepsilon_{\rm min}^{3}$,
radii $\propto\varepsilon_{\rm min}^{-1}$,
$m$ and $Q\propto\varepsilon_{\rm min}^{-1}$, while
the strongest behavior is exhibited by 
$D\propto\varepsilon_{\rm min}^{-2}$ \cite{Cantara:2015sna}. 
The $D$-term emerges as the property which is most 
strongly sensitive to the dynamics, and always negative.

If even such extremely unstable systems as $Q$-clouds 
have negative $D$-terms, the question emerges whether a physical
system exists with a positive $D$-term. No such system with a 
positive $D$-term is known so far.

The $Q$-ball system was very useful to educate our intuition,
and to prepare the discussion of hadronic EMT properties.

\section{D-term and strong forces inside various hadrons}
\label{Sec:D-term-hadrons}

In this Section we review calculations of $D$-terms of
hadrons including nuclei, Goldstone bosons, nucleon, vector mesons,
deuteron and the $\Delta$-resonance.

\subsection{Nuclei in liquid drop model}

A liquid drop is the simplest mechanical system which can provide us 
with physical intuition. It can also serve as a model for large atomic 
nucleus. The pressure and shear forces distributions in the liquid drop 
are the following \cite{Polyakov:2002yz}:
\be\label{Eq:PS_drop}
	p(r)=p_0 \theta(r-R) -\frac{p_0 R}{3} \delta(r-R), \quad 
	s(r)=\gamma \delta(r-R),
\ee
where $R$ is the radius of the drop, $p_0$ is the pressure inside the drop, 
and $\gamma$ is the surface tension coefficient related to $p_0$ and $R$
by the Kelvin relation: $p_0=2 \gamma/R$ \cite{Kelvin}. Obviously the von 
Laue and tangential stability conditions,
Eqs.~(\ref{Eq:von-Laue},~\ref{Eq:tang-stability}), are satisfied automatically.
One also sees immediately from Eq.~(\ref{Eq:PS_drop}) that the distribution 
of the normal pressure:
\begin{wrapfigure}[13]{R}{5cm}
\centering
\includegraphics[width=3.3cm]{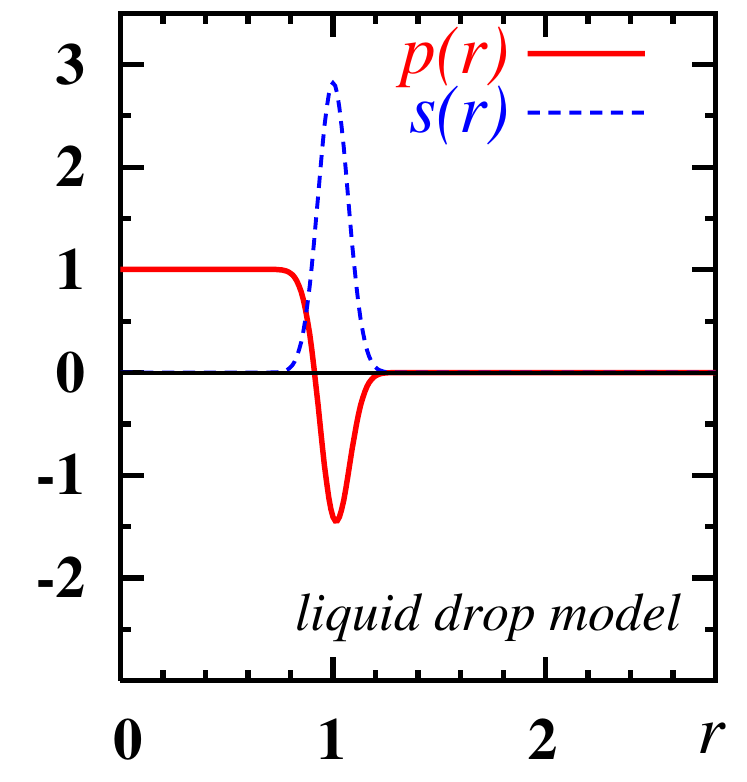}
\caption{\label{FIG-01:liquid-drop}
	The pressure and shear forces of nuclei 
	(in units of $p_0$) as functions of $r$ (in units of 
	nuclear radius $R_A$) in the liquid drop model. }
\end{wrapfigure}

\noindent 
\be
	\frac{dF_r}{d S_r}=\frac 23 s(r)+p(r) =p_0 \theta(r-R),
\ee
is positive (as required by the local stability of the system) and has 
step-like form. This form is what one expects intuitively for a liquid 
drop. Using the definition of the mechanical radius 
(\ref{eq:mechanicalradius}) we obtain the intuitively clear result 
\be	
	\langle r^2\rangle_{\rm mech}=\frac 35 R^2 \, .
\ee
The $D$-term for a liquid drop is
\be\label{Eq:Dterm-nucleus}
	D =-\frac45\,\biggl(\frac{4\pi}{3}\biggr)
	\,m \gamma\, R^4 \;.
\ee
A more realistic description of a large nucleus is obtained by
considering finite-skin effects which
make the $\Theta$- and $\delta$-functions in $p(r)$ and $s(r)$ smooth, 
see Fig.~\ref{FIG-01:liquid-drop}. The finite-skin effects make the 
$D$-term more negative \cite{Polyakov:2002yz}.

It is remarkable that the liquid drop model applied to a large nucleus 
gives $D_{\rm nucleus}\propto A^{7/3}$ since the nuclear masses and radii
grow like $m_A\propto A$ and $R_A\propto A^{1/3}$ with the mass number,
{while $\gamma\propto A^0$} \cite{Polyakov:2002yz}. Numerical
calculations in more sophisticated nuclear models support this 
prediction. In \cite{Guzey:2005ba} in the Walecka model the
$D$-terms of selected isotopes with $J^\pi=0^+$ were studied,
see Table~\ref{Table-2}a. For oxygen $^{16}$O and heavier nuclei
it was found $D \propto A^{2.26}$ \cite{Guzey:2005ba} 
in good agreement with \cite{Polyakov:2002yz}. 
For completeness we remark that in a model based on a 
non-relativistic nuclear spectral function a different 
$A$-behavior was found \cite{Liuti:2005qj}.

\begin{table}[b!]

\begin{center} 

\vspace{5mm}

\ \hspace{3mm}
\begin{tabular}{ccccccc}
	\hline\hline 
\hspace{-9mm} {\bf (a)} \ \ isotope 
	& $^{12}$C 
	& $^{16}$O 
	& $^{40}$Ca 
	&  $^{90}$Zr 
	& $^{208}$Pb  $\displaystyle\phantom{\frac00}$\\ \hline
$D$ 	& -6.2 	\  
        & -115  \
	& -1220 \
	& -6606	\
	& -39356   $\displaystyle\phantom{\frac00}$\\ \hline\hline
	\end{tabular}
\hspace{1.5cm}
	\begin{tabular}{ccccccc}
	\hline\hline
\hspace{-9mm} {\bf (b)} \ \ Goldstone boson \hspace{-5mm}
	& pion 
	& kaon 
	& $\eta$-meson $\displaystyle\phantom{\frac00}$\\ \hline
$D$ 	& \ $-0.97 \pm 0.01$ \	  
	& \ $-0.77 \pm 0.15$ \
	& \ $-0.69 \pm 0.19$ $\displaystyle\phantom{\frac00}$\\ \hline\hline
	\end{tabular}
	\end{center}
\caption{\label{Table-2}
	(a) 
	$D$-terms of selected nuclear isotopes with spin-parity quantum numbers
	$J^P=0^+$ \cite{Guzey:2005ba}. For $^{16}$O and heavier nuclei it is 
	approximately $D\approx -0.246\, A^{2.26}$ as predicted in the 
	liquid drop model \cite{Polyakov:2002yz}.
	(The convention used in Ref.~\cite{Guzey:2005ba} is $d_A=\frac45D$.)
	(b)
	$D$-terms of (pseudo) Goldstone bosons of chiral symmetry breaking
	$\pi$, $K$, $\eta$ with $J^P=0^-$ based on one-loop chiral theory
	\cite{Donoghue:1991qv} with estimated uncertainties 
	\cite{Hudson:2017xug}.
	In the chiral limit the $D$-term of a Goldstone boson is $D=-1$
	\cite{Novikov:1980fa,Voloshin:1980zf}.}
\end{table}

\subsection{Goldstone bosons}
\label{Sec:Goldstone}

The $D$-terms of  Goldstone bosons of spontaneous chiral symmetry breaking 
are given in the soft pion limit~by
\be\label{Eq:D-Goldstone}
	D = - 1 \;.
\ee
This result was obtained in 
\cite{Novikov:1980fa,Voloshin:1980zf,Voloshin:1982eb,Leutwyler:1989tn}
and rederived from a soft-pion theorem for pion GPDs in 
Ref.~\cite{Polyakov:1998ze}. 
{For an early study of EMT form factors of pseudoscalar mesons based on
 current-algebra techniques we refer to \cite{Raman:1971hs}.}
The leading (log enhanced) chiral correction to the $D$-term are given by
\cite{Donoghue:1991qv} 

\be\label{Eq:D-Goldstone-leading-log}
	D =- \left(1-\frac{m_\pi^2 \ln(\mu^2/m_\pi^2)}{48 \pi^2 f_\pi^2} 
	+{\cal O}(m_\pi^2)_\mu \right) 
\ee  
where $f_\pi\approx 93$~MeV is the pion decay constant. The
renormalization scale  $\mu$ can be chosen to be of the order of 
e.g.\ the $\rho$-meson mass, but the ${\cal O}(m_\pi^2)_\mu$ 
corrections in (\ref{Eq:D-Goldstone-leading-log}) depend on $\mu$ 
such that the total result for $D$ is scale independent. The formula 
(\ref{Eq:D-Goldstone-leading-log}) shows that the leading log 
chiral correction reduces the absolute value of the $D$-term.

Including the ${\cal O}(m_\pi^2)_\mu$ corrections in 
(\ref{Eq:D-Goldstone-leading-log}) computed in
\cite{Donoghue:1991qv} yields the results for the $D$-terms
of pions, kaons and $\eta$-mesons shown in Table~\ref{Table-2}b 
\cite{Hudson:2017xug}. 
The quoted uncertainties reflect the uncertainty of the low
energy constants entering to this order \cite{Donoghue:1991qv}
and include rough estimates of other chiral corrections (e.g.\ 
whether one uses $f_\pi\approx93\,{\rm MeV}$ at the physical value 
of the pion mass or $f_\pi\approx 88\,{\rm MeV}$ in the chiral limit). 
As expected, the deviations from the chiral limit value 
(\ref{Eq:D-Goldstone}) increase with mass and range from few percent 
for pions, to $20\,\%$ for kaons, to $30\,\%$ for $\eta$-mesons.
Electromagnetic and isospin-breaking corrections to pion 
EMT form factors were studied in \cite{Kubis:1999db}.

The slopes of the pion EMT form factors were computed in \cite{Polyakov:1998ze} 
in the large $N_c$ limit using the instanton liquid picture of QCD vacuum with 
the following result:
\be\label{eq:slopesNc}
	A'(0)=-D'(0)=\frac{N_c}{48 \pi^2 f_\pi^2},
\ee
where the pion decay constant 
{behaves in the chiral limit as $f_\pi^2={\cal O}(N_c)$.} 
We see that in the large $N_c$ limit the slopes of $A(t)$ and $D(t)$ 
form factors are the same. Interesting is that this corresponds to the 
fourth order effective chiral action for the Goldstone bosons 
(described by the nonlinear chiral field $U(x)=\exp[i \pi^a\tau^a/f_\pi]$) 
in external gravitational field of the form:
\be
	S= -\frac{N_c}{96 \pi^2} \int d^4x\ \sqrt{-g}\   
	{\rm tr}\left(\partial_\mu U \partial_\nu U^\dagger\right)  
	\left[R^{\mu\nu} -\frac 12 g^{\mu\nu} R\right] ,
\ee 
in which Goldstone bosons {couple} not separately to the Ricci tensor 
($R_{\mu\nu}$) and the scalar curvature ($R$) but to their combination which
is the Einstein tensor {$R^{\mu\nu} -\frac 12 g^{\mu\nu} R$}. 
Probably there is some deep 
reason for {this} peculiarity which still remains to be revealed.

While in large $N_c$ limit the slopes of $A(t)$ and $D(t)$ coincide,
a drastic difference arises in sub-leading $1/N_c$ order due to pion 
loop corrections. These corrections lead to a non-analytical dependence 
of $D'(0)$ on the pion mass, whereas $A'(0)$ remains analytical in $m_\pi$. 
This non-analyticity leads to a divergence of the $D(t)$ slope in the 
chiral limit of the form \cite{Leutwyler:1989tn}:
\be\label{eq:slopesChiralDiv}
	-D'(0)_{\rm chiral\ loop} = 
	\frac{\ln\left(\mu^2/m_\pi^2\right)}{24 \pi^2 f_\pi^2}.
\ee  
For the numerical estimate of the $D(t)$ slope we can combine the results in 
Eqs.~(\ref{eq:slopesNc},\ref{eq:slopesChiralDiv}): 
\be\label{eq:slopesNum}
 	-D'(0)= 
	\frac{N_c}{48 \pi^2 f_\pi^2}
	+\frac{\ln\left(\mu^2/m_\pi^2\right)}{24\pi^2 f_\pi^2}
	=(0.73+1.66)~{\rm GeV}^{-2}=2.40~{\rm GeV}^{-2}.
 \ee
Here for the numerical estimate we use $\mu=m_\rho$ and physical pion mass 
of $m_\pi=0.140$~GeV. An important conclusion from the consideration of slopes 
of the EMT form factors in chiral theory is that for the pion $-D'(0)$ should 
be larger than $A'(0)$ due to the different behavior of these slopes in the 
chiral limit. 
 
It is also instructive to compare Eq.~(\ref{eq:slopesNum}) with the analogous 
estimate (see section~6.1 of Ref.~\cite{Polyakov:1998ze}) for the slope of pion 
charge form factor $F_{\rm e.m.}(t)$:
\be\label{eq:slopeEM}
 	F_{\rm e.m.}'(0)=\frac{N_c}{24 \pi^2 f_\pi^2}+
	\frac{\ln\left(\mu^2/m_\pi^2\right)}{96 \pi^2 f_\pi^2}
	=(1.46 +0.42)~{\rm GeV}^{-2}=1.88~{\rm GeV}^{-2}.
\ee
First, the obtained numerical value is in good agreement with the experimental 
value of $F_{\rm e.m.}'(0)=(1.86\pm0.03)$~GeV$^{-2}$ \cite{Amendolia:1986wj},
which indicates that such an estimate gives sensible results for pion form 
factors. Second, it is very instructive to compare the expressions 
(\ref{eq:slopesNum}) and (\ref{eq:slopeEM}) -- one sees that the 
chiral loop corrections to the slope of the pion $D(t)$ form factor 
are four times larger than the to slope of the pion charge form factor, 
whereas the large $N_c$ (``core") contribution is two times smaller -- 
the slope of the pion $D(t)$ form factor is dominated by chiral logs.
This demonstrates that the $D$-term is very sensitive to physics of 
spontaneous breakdown of the chiral symmetry in QCD and study of the 
$D$-term can provide us with new effective tools for probing the 
mechanisms of chiral symmetry breaking in QCD.

The low energy effective chiral Lagrangian in curved space-time {and the}
gravitational form factors of the pion were also studied in quark model
frameworks \cite{Megias:2004uj,Megias:2005fj,Broniowski:2008hx},
AdS/QCD models in \cite{Jugeau:2009mn}, and covariant and 
light-front constituent models \cite{Frederico:2009fk}. 
{The result (\ref{eq:slopesNc}) was rederived in Ref.~\cite{Megias:2004uj}
in the large $N_c$ limit in quark spectral models, where it was noted that
the equality of the slopes of $A(t)$ and $D(t)$ was independent
of the particular realization of the spectral model.}
The von Laue condition for the pion was studied in \cite{Son:2014sna}.
A study of pion EMT form factors in lattice QCD was reported in 
Ref.~\cite{Brommel:2005ee}.

\subsection{Nucleon}
\label{Sec:Nucleon}

The first model studies of the nucleon $D$-term were performed 
in the bag model \cite{Ji:1997gm}, chiral quark soliton model 
\cite{Petrov:1998kf}, see also \cite{Schweitzer:2002nm,Ossmann:2004bp,
Goeke:2007fp,Goeke:2007fq,Wakamatsu:2007uc}, and Skyrme model 
\cite{Cebulla:2007ei,Jung:2013bya}.
The quark contributions to the $D$-term were also studied in the QCD 
multi-color limit $N_c\to\infty$ \cite{Goeke:2001tz}, lattice QCD 
\cite{Hagler:2003jd,Gockeler:2003jfa,Hagler:2007xi},  
dispersion relations \cite{Pasquini:2014vua}, and  quark models
\cite{Pasquini:2007xz,Hwang:2007tb,Abidin:2008hn,Brodsky:2008pf,Abidin:2009hr,
Chakrabarti:2015lba,Mondal:2015fok,Mondal:2016xsm,Kumar:2017dbf}.

\begin{figure}[b!]
\centering
\includegraphics[height=4.5cm]{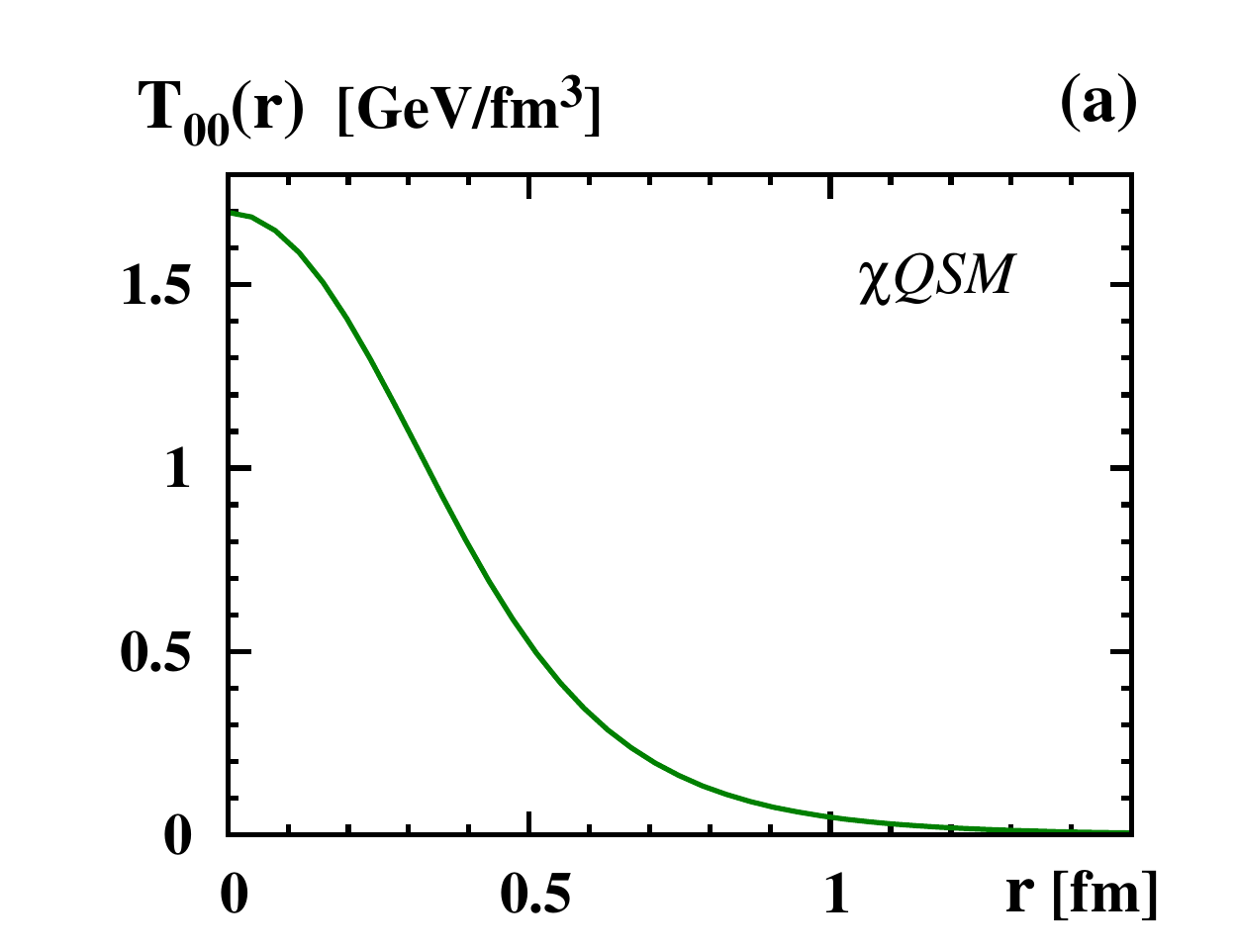}
\includegraphics[height=4.5cm]{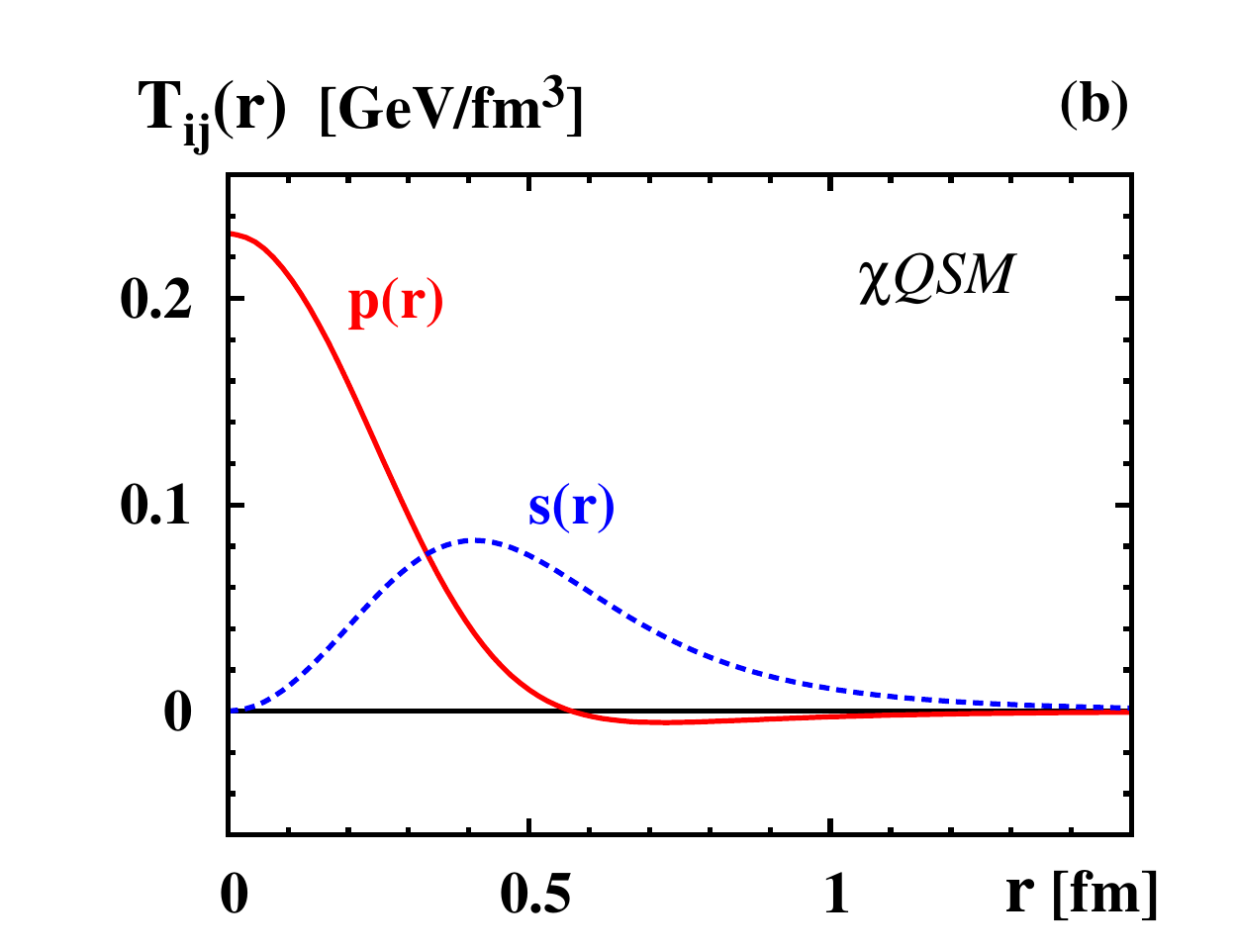}\\
\includegraphics[height=4.5cm]{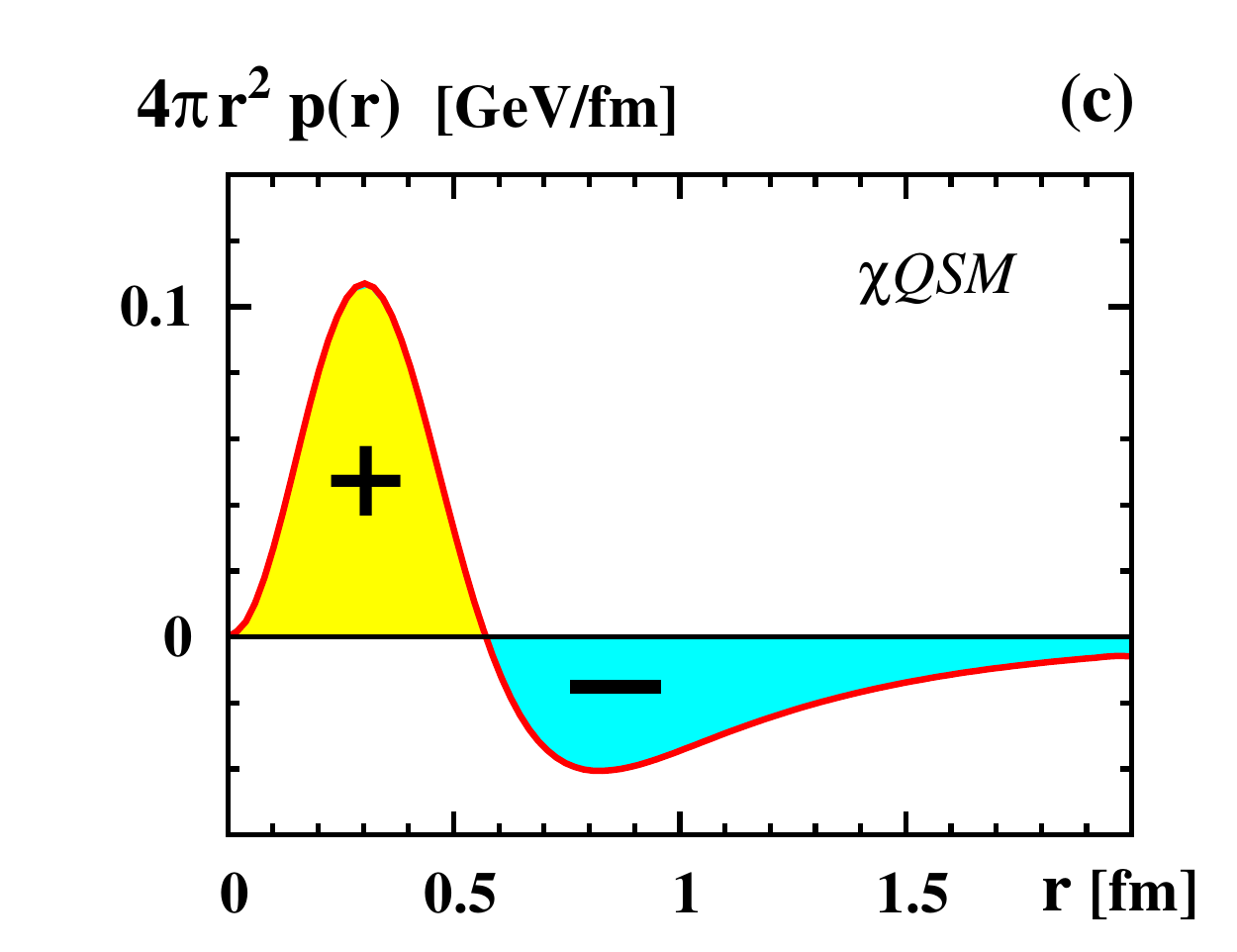}
\includegraphics[height=4.5cm]{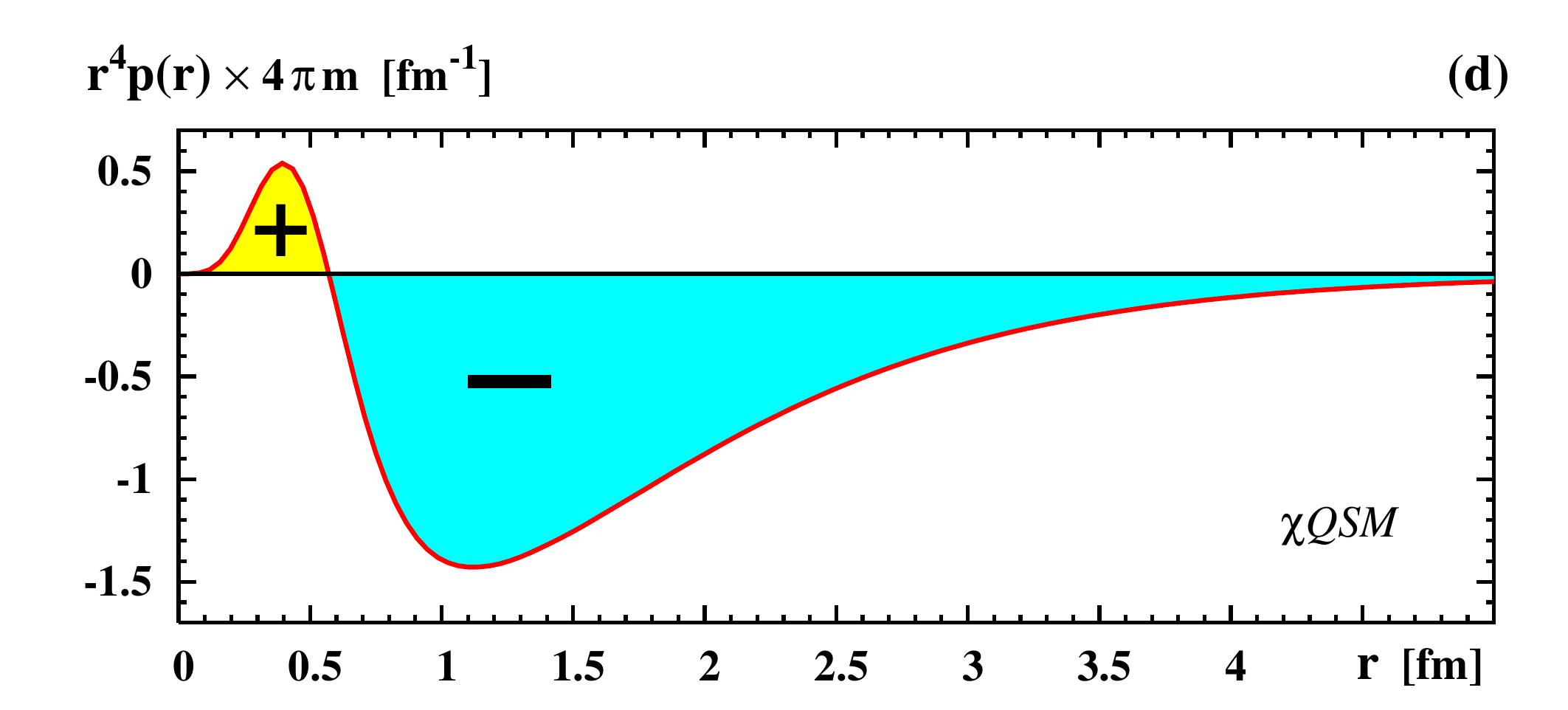}
\caption{\label{FIG-02:pressure-CQSM}
	EMT densities of the nucleon from the chiral quark soliton 
	\cite{Goeke:2007fp}. (a) Energy density $T_{00}(r)$, (b) 
	densities $p(r)$ and $s(r)$ of the stress tensor $T_{ij}(r)$,
        and (c) $4\pi r^2p(r)$ where the shaded areas above and below 
	the $x$-axis are exactly equal to each other which demonstrates 
	how the von Laue condition (\ref{Eq:von-Laue}) is realized. 
	(d) The integrand of the $D$-term is proportional to
	$r^4p(r)$ and yields $D<0$ upon integration. The negative
	sign of $D$ emerges as a natural consequence of
	the ``stability pattern'' \cite{Goeke:2007fp}.}

\centering
\includegraphics[height=4.5cm]{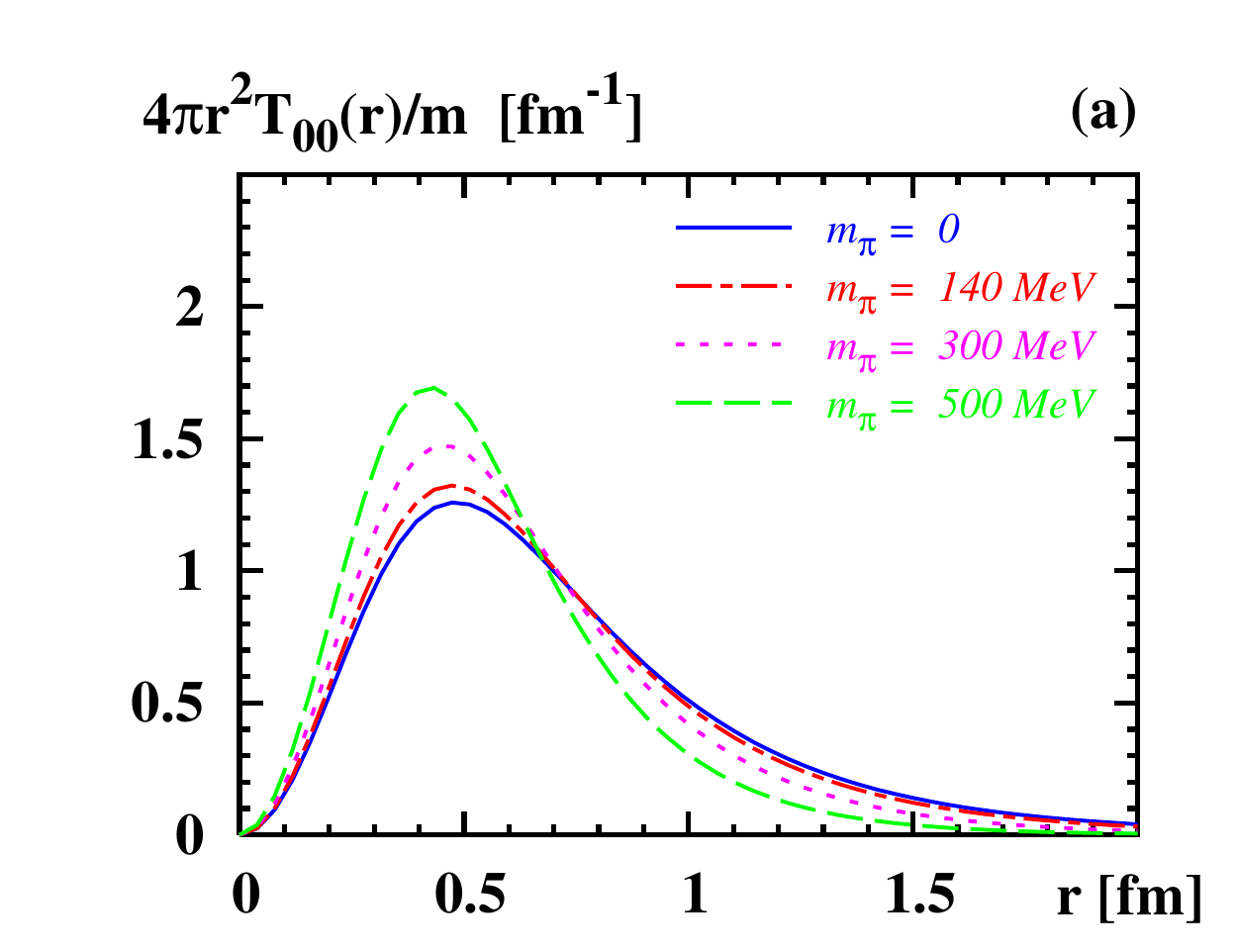}%
\includegraphics[height=4.5cm]{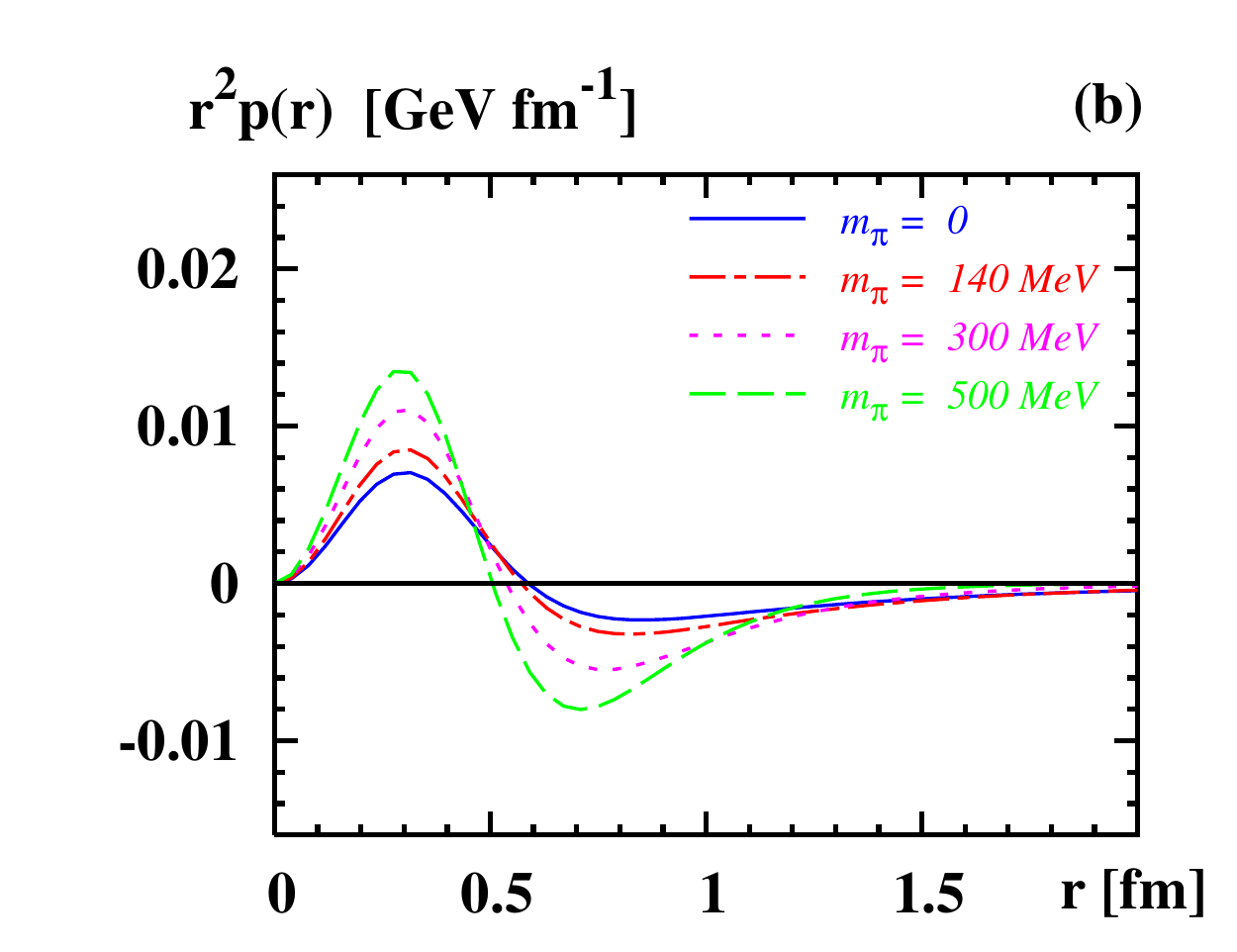}%
\includegraphics[height=4.5cm]{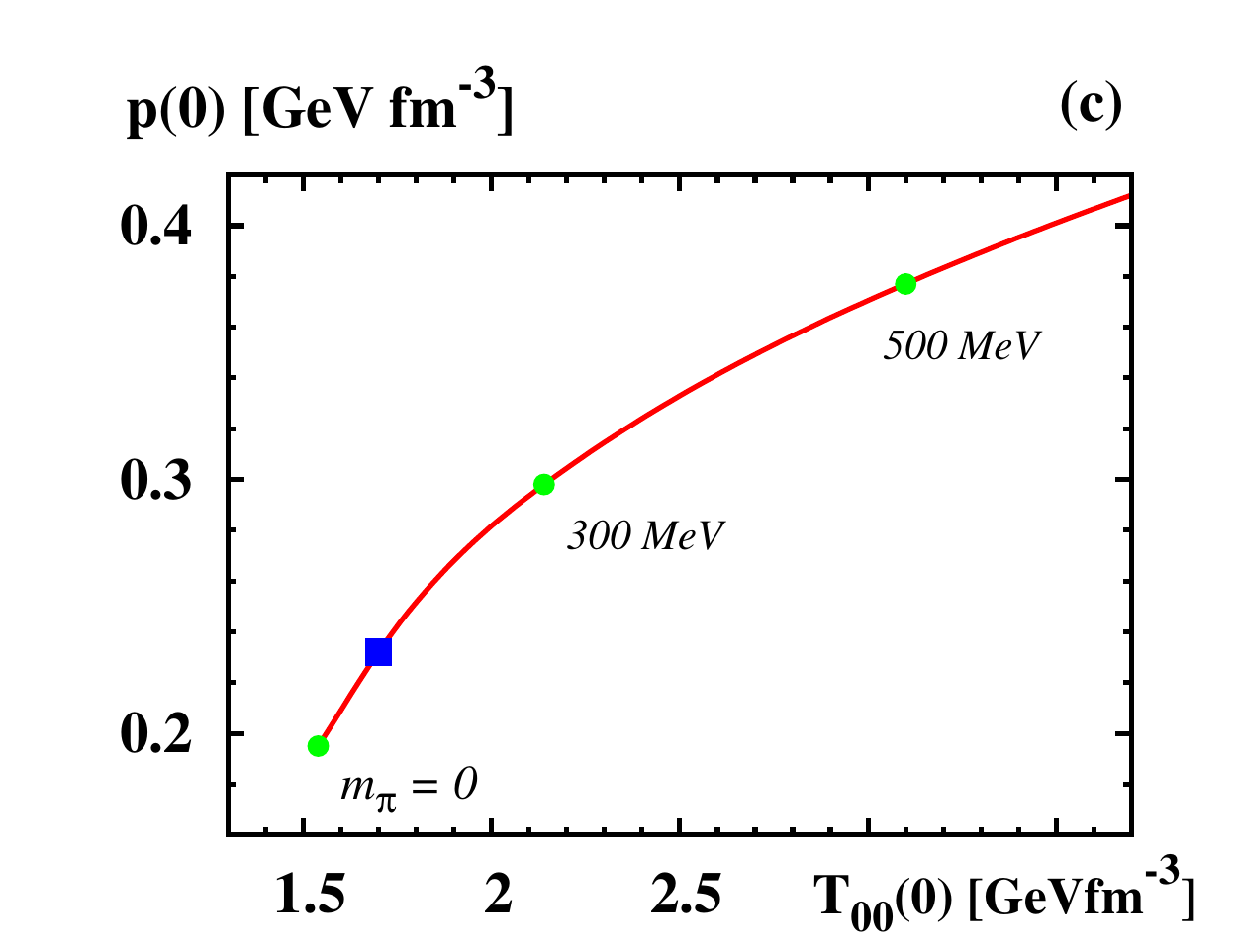}
\caption{\label{Fig-05:mpi-dep-CQSM}
	The pion mass dependence of the nucleon EMT densities from 
	chiral quark soliton \cite{Goeke:2007fq} from the chiral limit
	up to $m_\pi$ of the order of magnitude of the kaon mass. 
	(a) Energy density normalized as $4\pi r^2T_{00}(r)/m$ such 
	that the curves integrate to unity, and (b) $r^2p(r)$ which 
	integrates to zero.
        (c) The pressure in the center as function of the energy density
	in the center.}
\end{figure}

The bag and chiral quark soliton model were used in 
Ref.~\cite{Hudson:2017oul} to illustrate how interactions can 
generate the $D$-term of a fermion.
In the bag model {a non-zero $D$-term emerges when interactions 
are introduced in the shape of the bag boundary condition which is
imposed to simulate confinement and bind the otherwise free quarks.}
In the chiral quark soliton model the $D$-term {vanishes when the
chiral interactions are ``switched off'' and the free theory is restored
in a limiting procedure}.
{The bag model with massless quarks gives a small value $D=-1.1$ 
\cite{Ji:1997gm,Hudson:2017oul}.} The chiral models predict a more 
sizable $D$-term in the range \cite{Schweitzer:2002nm,Ossmann:2004bp, 
Goeke:2007fp,Goeke:2007fq,Wakamatsu:2007uc,Cebulla:2007ei,Jung:2013bya}
\be\label{Eq:D-chiral-models}
	-4 \lesssim D \lesssim -2\;.
\ee

In the large $N_c$ limit the $D$-term of the nucleon exhibits 
the flavor hierarchy \cite{Goeke:2001tz}
\be\label{Eq:D-large-Nc}
	|D^u(t)+D^d(t)| \sim N_c^2 \;\; \gg \;\; |D^u(t)-D^d(t)|\sim N_c\,.
\ee
This result is supported by numerical calculations in chiral 
quark soliton model \cite{Wakamatsu:2007uc} and lattice QCD 
\cite{Hagler:2003jd,Gockeler:2003jfa,Hagler:2007xi} .

Fig.~\ref{FIG-02:pressure-CQSM} shows the EMT densities from the 
chiral quark soliton model ($\chi$QSM) \cite{Goeke:2007fp}. 
In the center $T_{00}(0) = 1.7\,{\rm GeV/fm}^3$ which is approximately
13 times the nuclear matter density while $p(0) = 0.23 {\rm GeV/fm}^3$, 
which corresponds to $3.7\cdot 10^{29}$ atmospheric pressures.
The positive pressure in the center means repulsion, and negative 
$p(r)$ for $r \gtrsim 0.6\,{\rm fm}$ means attraction. Repulsive and 
attractive forces balance each other exactly according to the von 
Laue condition (\ref{Eq:von-Laue}).

The von Laue condition can be rigorously proven in the $\chi$QSM 
\cite{Goeke:2007fp} by exploring a theorem known as ``virial theorem:'' 
the soliton mass is a functional of the soliton profile. One may
consider a special class of variations of the profile function generated 
by the dilatational transformations $r\to \lambda\,r$.
This yields an energy functional $m(\lambda)$ which has a minimum
at $\lambda=1$. The von Laue condition can now be expressed as
$\int{\rm d}^3r\,T_{ii}(r)= m'(\lambda)|_{\lambda=1}=0$ \cite{Goeke:2007fp}.
This shows that this condition is satisfied by any stationary solution: 
global minimum, local minimum, other extremum, saddle point of the 
action. This means the von Laue condition is necessary
but not sufficient for stability.

The von Laue condition can be proven in exactly the same way also in the 
Skyrme model \cite{Cebulla:2007ei} and bag model \cite{Neubelt-et-al}.
These models have in common that they describe the nucleon in terms 
of a static mean field, even though in these models the mean fields 
are realized in much different ways. The generic mean field picture 
of the nucleon is justfied in QCD in the large-$N_c$ limit 
\cite{Witten:1979kh,Witten:1983tx}. Thus, the connection of the 
von Laue condition and the virial theorem is of more general 
character than the respective models: it holds in the large-$N_c$ 
limit in QCD. It is not known whether a connection of the von 
Laue condition and extrema of the action can be established 
also in QCD with finite $N_c$.

It is interesting to investigate what happens when one increases
the value of the current quark masses 
(as it was routinely done until recently in lattice QCD studies).
In this case the hadron masses increase, while their sizes 
decrease. For the EMT densities it has
the following implications: the energy density in the center of the
nucleon increases and so does the pressure, see 
Fig.~\ref{Fig-05:mpi-dep-CQSM}. This implies a more
negative $D$-term \cite{Goeke:2007fq}.

Modifications of the $D$-term of the nucleon in nuclear matter were 
studied in \cite{Kim:2012ts,Jung:2014jja}.  As the density 
of the nuclear medium increases, the energy density in the center of 
the nucleon bound in the medium and the pressure both decrease. 
The size of the system, however, grows and the $D$-term becomes 
more negative \cite{Kim:2012ts,Jung:2014jja}.

Chiral perturbation theory cannot predict the value of the nucleon 
$D$-term, but it predicts its $m_\pi$-dependence and the small-$t$ 
behavior of $D(t)$ \cite{Belitsky:2002jp,Ando:2006sk,Diehl:2006ya}. 
The slope of $D(t)$ at zero-momentum transfer diverges in the chiral 
limit as $D^\prime(0) \sim 1/m_\pi$. This behavior is reproduced also
in chiral models \cite{Goeke:2007fp,Cebulla:2007ei}. 

In Section~\ref{Sec-9:mech-radius} the mechanical radius of a hadron 
was defined {\it not} in terms of the slope of $D(t)$. Applying the 
definition of the mechanical radius (\ref{eq:mechanicalradius}) to 
the nucleon case, one can see on general grounds that the corresponding 
mechanical radius (in contrast to  $D^\prime(0)$ and to the charge radius 
of the nucleon) is finite in the chiral limit ($m_\pi\to 0$). Therefore, 
one expects that the nucleon mechanical radius should be smaller than, 
say, the charge radius. Indeed, the chiral quark soliton model predicts 
the mechanical radius of the proton to be about $25\,\%$ smaller than 
its mean square charge radius: 
$\langle r^2\rangle_{\rm mech}\approx0.75\,\langle r^2\rangle_{\rm charge}$.

\begin{figure}[b!]
\centering
\includegraphics[height=11cm]{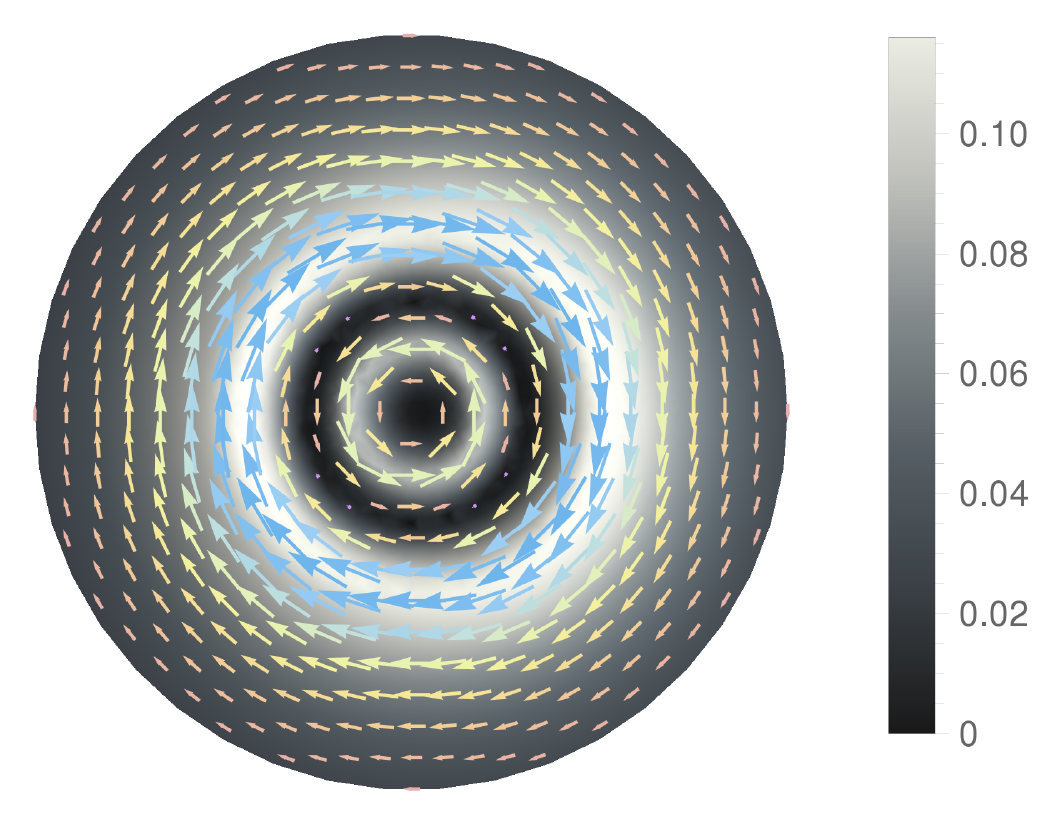}
\caption{\label{Fig:visualisation}
	Visualisation of the $\phi$-component of the tangential force 
	(the 2D vector vector field $4\pi r^2 T_{ij} {\bm e}^{\phi}_j$ )
	distribution in the nucleon from the {chiral quark soliton model.}  
	The radius of the disc on the figure is 1.5~fm, the colour legend
	gives the absolute value of the tangential force in GeV/fm.}
\end{figure}

It is instructive to see details of the strong forces distribution inside 
the nucleon. The radial (normal) forces in
Eq.~(\ref{Eq:force-spherical components}), are 
always ``stretching" (directed outwards the nucleon centre) and monotonically 
decrease with distance from the centre. The distribution of the tangential 
forces provides us with further fine details of how the strong forces keep the
nucleon together. From the stability condition (\ref{Eq:tang-stability}) it 
is clear that the tangential force must at least once change its direction. 
Studying these forces one can pose very intriguing questions about nature 
of strong forces -- how many times do the forces change from ``stretching'' 
to ``squeezing''? What does this number mean? What does distinguish the 
regions of ``stretching'' and ``squeezing''? What do we learn about the 
confinement mechanism from this? 

Presently we are not able to answer the above posed questions. 
Here we just report the results on the force distribution 
in the nucleon from models.
In Fig.~(\ref{Fig:visualisation}) we plot the vector field of the
$\phi$-component of the tangential force (the  2D vector vector field 
$4\pi r^2 T_{ij} {\bm e}^{\phi}_j$ ) inside the nucleon\footnote{See also 
	recent lattice {calculations} of the spatial distribution of forces 
	for the heavy quark $\bar Q Q$ pair in 
	Ref.~\cite{Yanagihara:2018qqg}. 
	The formalism provided here paves a way to perform 
	analogous studies on the lattice for hadrons.}
obtained from EMT densities from the 
{chiral quark soliton model \cite{Goeke:2007fp}.}  

One clearly sees that at a distance of $r\approx 0.5$~fm from the nucleon 
centre the tangential  force changes its direction, and turns from 
``stretching'' to ``squeezing''.
Thus, we see that there are two qualitatively different regions inside 
the nucleon -- they are distinguished by the chirality of the tangential 
forces. It would be interesting to understand at the microscopic level the 
physical reasons for {the emergence of} these two different regions. 

In the lattice QCD study \cite{Hagler:2007xi} a hybrid approach based
on domain wall valence quarks with $2+1$ flavors of improved staggered 
sea quarks was used. 
The range $0.1\,{\rm GeV}^2 < -t < 1.2 \,{\rm GeV}^2$ was covered for 
pion masses from $760\,{\rm MeV}$ down to $350\,{\rm MeV}$. Depending 
on the chiral extrapolation method the following values were obtained 
which do not include disconnected diagrams:
$D^Q=-1.07 \pm 0.25$ using covariant baryon chiral perturbation theory,
and $D^Q=-1.68\pm 0.22$ using heavy baryon chiral perturbation theory
at the physical value of the pion mass in $\overline{MS}$ scheme at
$\mu^2=4\,{\rm GeV}^2$.
The quark contribution to the $D$-term from dispersion relations 
 \cite{Pasquini:2014vua} refers to the same $\mu^2$ and is in the
range $-1.54\lesssim D^Q\lesssim -1.27$ in good agreement with
the lattice result. Considering that the results from chiral models
(\ref{Eq:D-chiral-models}) show the total $D$-term, the
dispersion relation and lattice result agree well with these
models \cite{Goeke:2007fq,Cebulla:2007ei}.

The nucleon EMT form factors $A(t)$ and $B(t)$ were also studied in 
approaches based on light front wave functions such as AdS/QCD models 
or spectator models 
\cite{Pasquini:2007xz,Hwang:2007tb,Abidin:2008hn,Brodsky:2008pf,Abidin:2009hr,
Chakrabarti:2015lba,Mondal:2015fok,Mondal:2016xsm,Kumar:2017dbf}. 
Such models are often based on a light-front Fock state expansion.
Typically the form factors $A(t)$ and $B(t)$ can be evaluated, which 
are simply related to the helicity non-flip and helicity flip matrix 
elements of the component $\hat{T}{ }_{++}$ of the EMT. 
Being related to the stress tensor $\hat{T}{ }_{ij}$ the form factor
$D(t)$ naturally ``mixes'' good and bad light-front components and 
is described in terms of transitions between different Fock state 
components in overlap representation. As a quantity intrinsically 
non-diagonal in a Fock space, it is difficult to study the $D$-term
in approaches based on light-front wave-functions. This is due to
the relation of the $D$-term to internal dynamics: a complete 
description of a hadron requires the inclusion of {\it all} Fock components.

\subsection{Size of the forces in the nucleon, and 
comparison with linear potential confinement forces}

Very frequently, e.g. in colour tube models, the confinement forces 
are related to the linear potential $V_{\rm conf}(r)=\sigma r$, where 
$\sigma\sim 1$GeV/fm is estimated from the slope of meson Regge 
trajectories. Recently the spatial distribution of the stress tensor 
for a heavy quark $\bar Q Q$ pair was directly measured on the lattice:
the typical size of the forces $\sim 1$GeV/fm {was confirmed}
\cite{Yanagihara:2018qqg}.
Such a linear interquark potential corresponds to a constant force 
between quarks $F=\sigma$. Our aim is to compare this force with 
{the forces encoded in the stress tensor}.

\begin{figure}[b!]
\centering
\includegraphics[height=5cm]{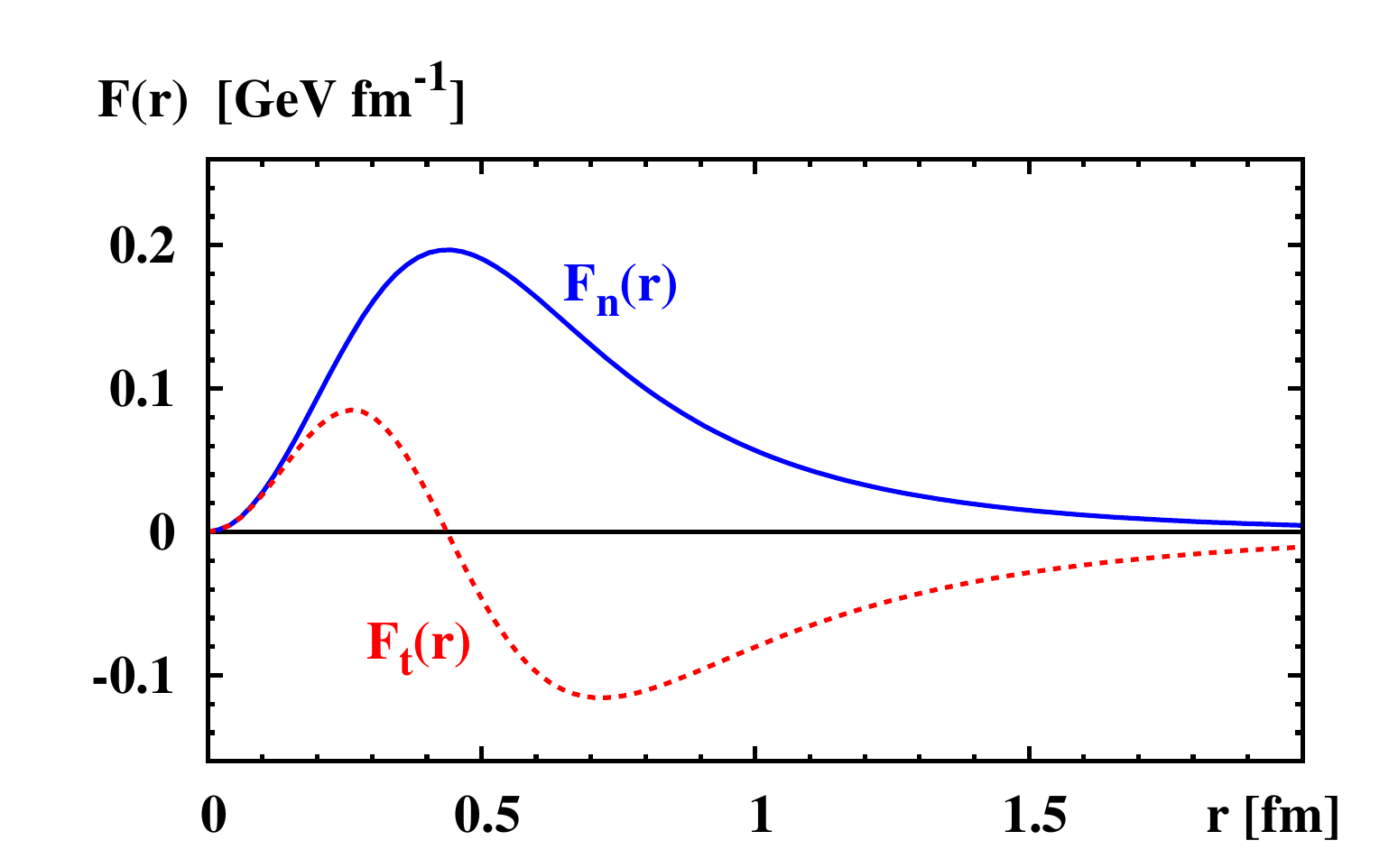}
\caption{\label{Fig:norm_forces}
The normal force $F_n=4\pi r^2 [ \frac 23 s(r)+p(r)]$ (solid) and 
tangential force $F_t=4\pi r^2 [-\frac 13 s(r)+p(r)]$ (doted) 
experienced by a spherical shell of radius $r$ in the nucleon 
computed in the $\chi$QSM. }
\end{figure}

The spherical shell of radius $r$ in the nucleon experiences the 
normal force $F_n=4\pi r^2 [\frac 23 s(r)+p(r)]$ and tangential force
$F_t=4\pi r^2 [-\frac 13 s(r)+p(r)]$. 
We use the chiral quark-soliton model ($\chi$QSM) results of 
Ref.~\cite{Goeke:2007fp} to compute the corresponding forces. 
The result is shown on Fig.~\ref{Fig:norm_forces}, we see that 
the maximally achieved strength is five times smaller 
than the confining forces in a colour tube model.

\subsection{Spin-1 hadrons}

{Light vector mesons were studied in Ref.~\cite{Abidin:2008ku} 
using light-front wave-functions obtained from an AdS/QCD model.
For the $\rho$-meson the mean square radius of the energy density was 
found to be {$\la r^2\ra_E = 0.21\,{\rm fm}^2$.} This is significantly
smaller then the mean square charge radius of $\rho^+$ determined 
to be {$\la r^2\ra_{\rm ch} = 0.53\,{\rm fm}^2$} in the same approach 
\cite{Grigoryan:2007vg}.

The GPDs for the deuteron were introduced in \cite{Berger:2001zb} 
and studied in details in Ref.~\cite{Cano:2003ju}. 
The EMT form factors of the deuteron were studied in 
Ref.~\cite{Mondal:2017lph} using a deuteron wave function
from a softwall AdS/QCD model.
The $D$-term was not computed neither for the vector mesons nor for 
deuteron. We are not aware of a calculation of the $D$-term of a 
spin-1 hadron. To best of our knowledge, the only {spin 1} particle 
whose $D$-term has been studied is the photon, which has a negative 
$D$-term \cite{Friot:2006mm,Gabdrakhmanov:2012aa}.

}

\subsection{$\Delta$-resonance}

\begin{figure}[h!]
\centering
\includegraphics[height=4.5cm]{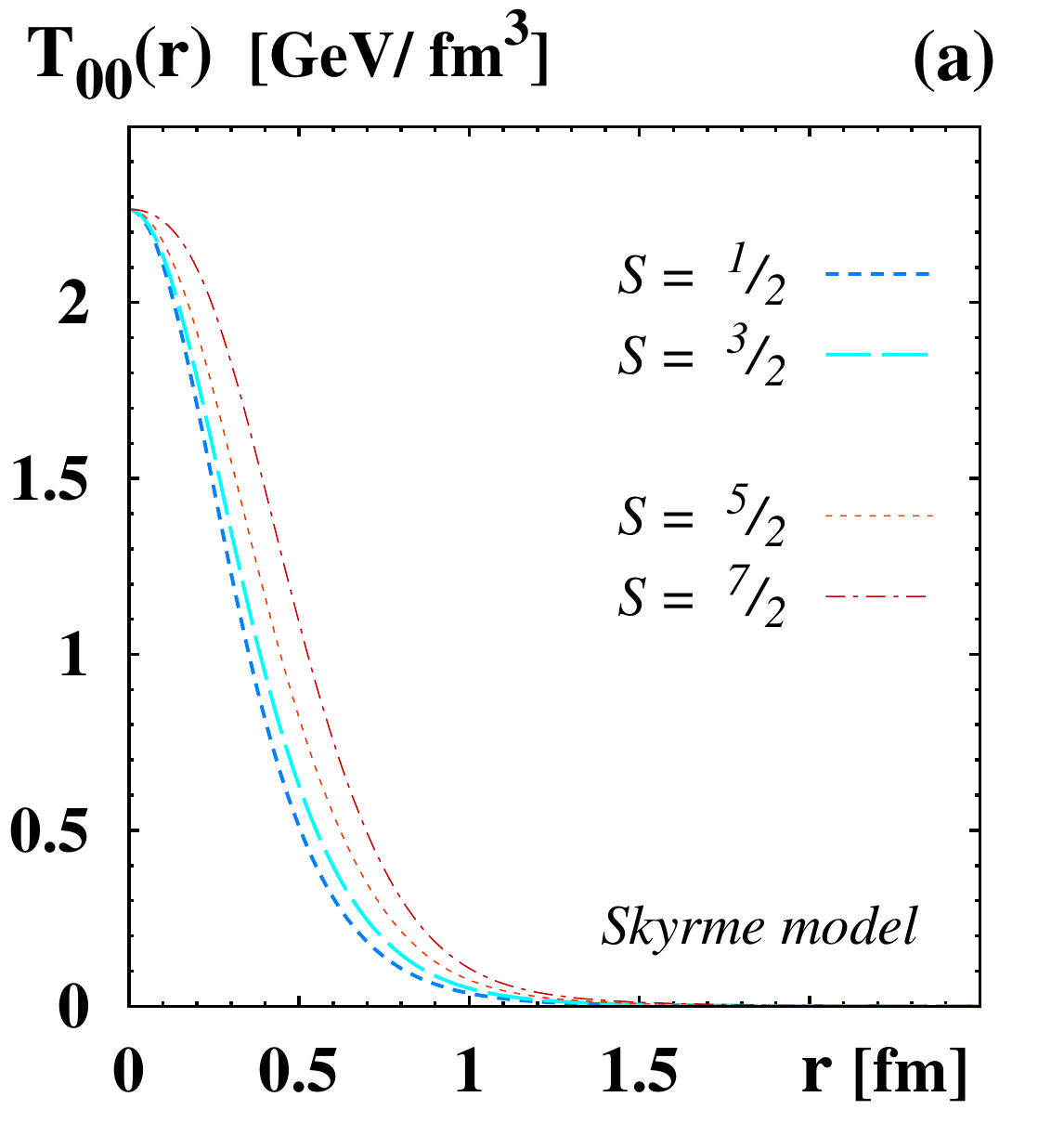}%
\includegraphics[height=4.5cm]{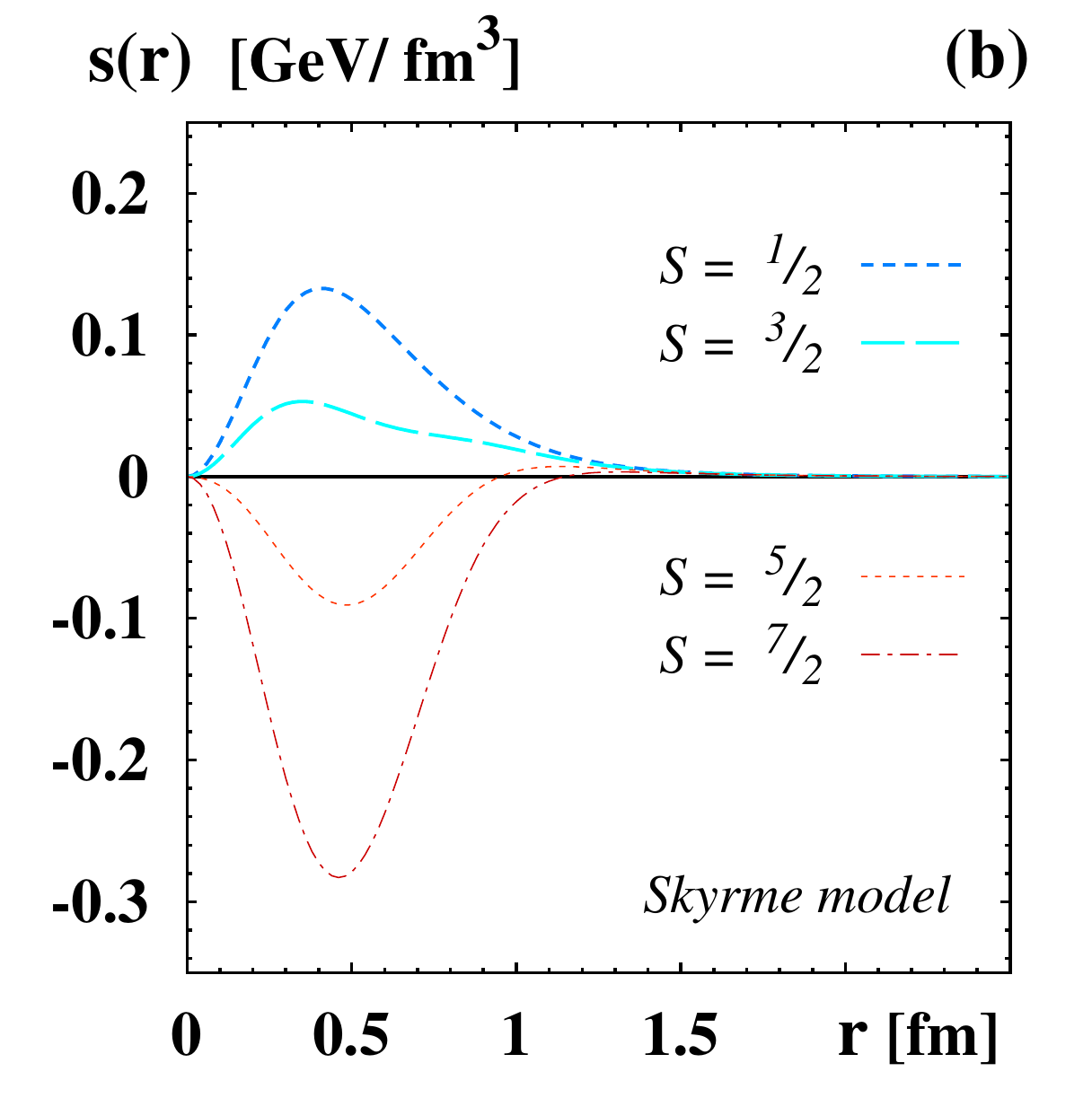}%
\includegraphics[height=4.5cm]{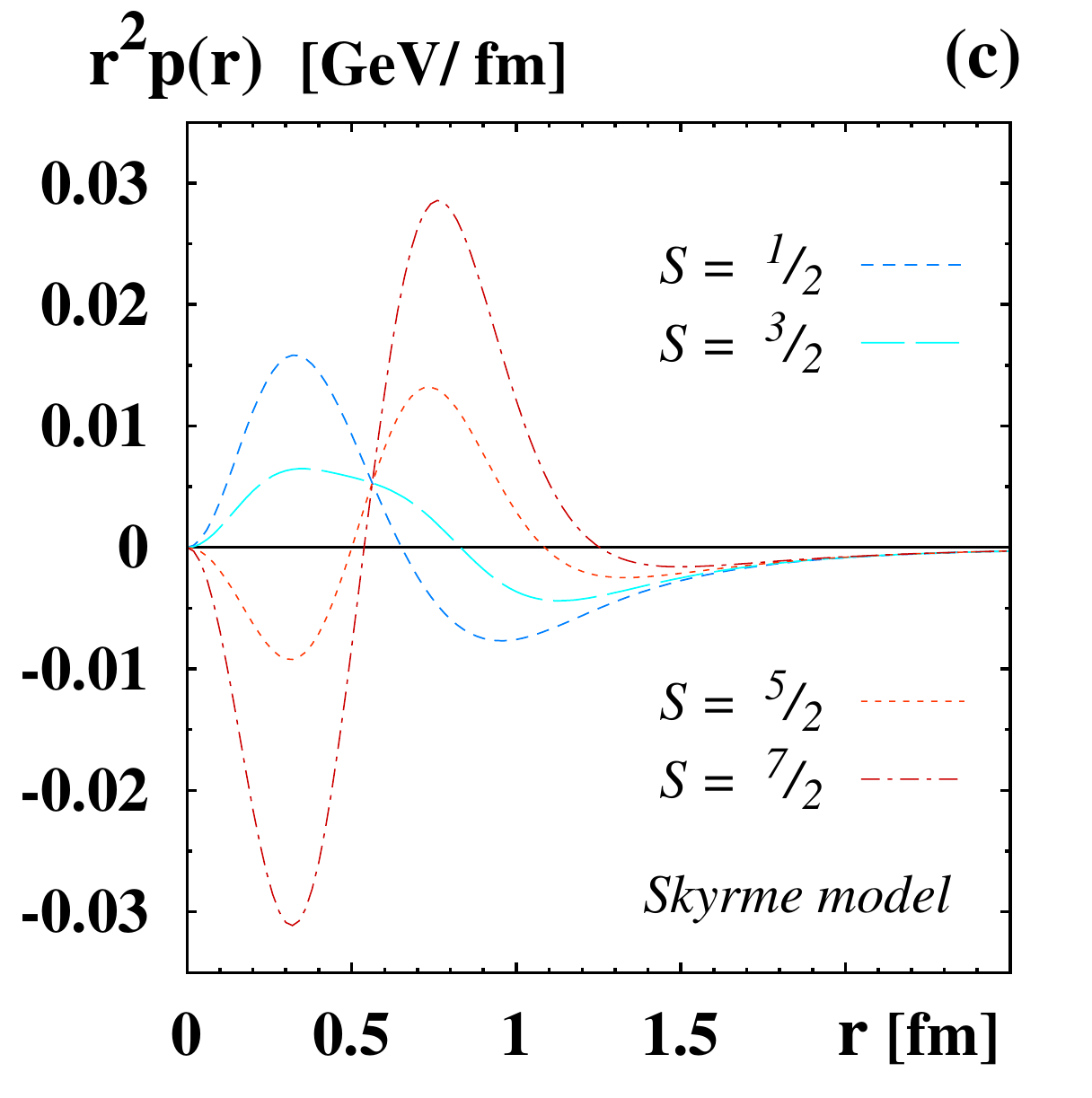}%
\includegraphics[height=4.5cm]{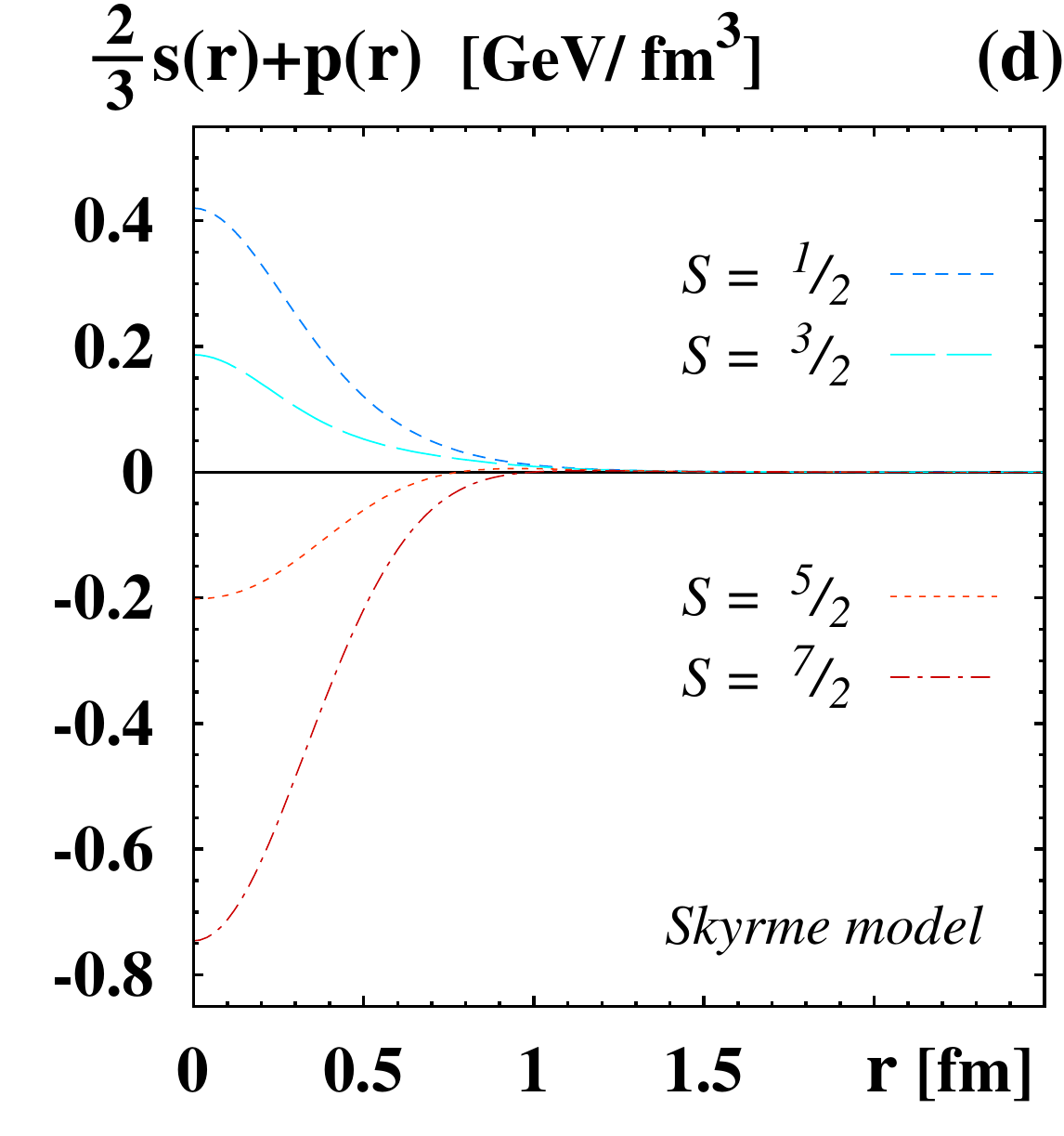}
\caption{\label{FIG-04:rigid-rotator}
	EMT densities in Skyrme model for the quantum numbers
	$S=I=\frac12,\;\frac32,\;\frac52,\;\frac72$.
	Energy density $T_{00}(r)$ indicates the states are heavier 
	with increasing spin, and does not reveal anything unusual.
	More insightful are $s(r)$ and $p(r)$  which flip signs and 
 	violate the stability criterion $\frac23s(r)+p(r)\ge 0$ in 
	Eq.~(\ref{Eq:local-stability-criterion}) for $S=I\ge\frac52$. 
	This calculation explains why 
	$S=I=\frac12,\;\frac32$ are physical states (nucleon, $\Delta$)
	while $S=I\ge\frac52$ are unphysical states and not observed in 
	nature \cite{Perevalova:2016dln}.}
\end{figure}

The $D$-term of the $\Delta$ was studied in the Skyrme model in
Ref.~\cite{Perevalova:2016dln}. Despite being an unstable state,
the $\Delta$-resonance has nevertheless a negative $D$-term.
In the Skyrme model one finds the intuitive result that the $\Delta$ 
appears to have a larger radius than the nucleon, and the internal 
forces are weaker, {see Fig.~\ref{FIG-04:rigid-rotator}.}
The $D$-term is negative, but its magnitude is about $25\,\%$ 
smaller as a consequence of the reduced forces.
Using the formalism of Ref.~\cite{Perevalova:2016dln} we can 
compute also the mechanical radius for $\Delta$-resonance,
its value is by about 25\% larger than the mechanical radius 
of the nucleon, that reflects that the normal forces in 
$\Delta$-resonance are more spread out.
{In \cite{Perevalova:2016dln} not only the $\Delta$-resonance
was studied, but also higher spin-isospin states which are
predicted in the Skyrme model but not observed in nature.
The study of the EMT densities of such states provides
an explanation why no such states are observed in nature.}

Soliton models based on the large-$N_c$ expansion
(chiral quark-soliton model, Skyrme model)
describe light baryons with spin and isospin quantum numbers 
$S=I=\frac12,\;\frac32,\;\frac52,\;\dots\,$ as different rotational 
states of the same soliton solution. In this ``rigid rotator approach''
$1/N_c$ corrections are considered by expanding the action in terms 
of the angular velocity of the rotating solitons. 
The quantum numbers $S=I=\frac12$ and $\frac32$ correspond to the 
nucleon and $\Delta$. But rigid rotator states with $S=I\ge\frac52$ 
are not observed in nature. For a long time this was considered an
unsatisfactory artifact of the approach, until it was shown that 
such states actually do not exist --- not even in the rigid rotator
framework \cite{Perevalova:2016dln}. Using the Skyrme model it was 
shown that the $1/N_c$ corrections are a small perturbation for the
nucleon with $S=I=\frac12$. The corrections are more sizable for 
$S=I=\frac32$ but the $\Delta$ stil complies with stability criteria.
But for the rigid rotator artifact states $S=I\ge\frac52$ the $1/N_c$ 
corrections become so destabilizing, that the basic stability criterion 
(\ref{Eq:local-stability-criterion}) is violated. These are therefore
unphysical states (which would have a positive $D$-term). The results 
from the Skyrme model are shown in Fig.~\ref{FIG-04:rigid-rotator}.

\section{EMT densities and applications to hidden-charm 
penta- and tetra-quarks}

The extraction of the $D$-term will not only give insights on
how the internal strong forces balance inside the nucleon. Knowledge 
of EMT form factors has also applications to the spectroscopy of 
some of the recently observed exotic hidden-charm hadrons,
pentaquark and tetraquark states.

In the LHCb experiment pentaquark states were observed 
\cite{Aaij:2015tga} which decay in $J/\psi$ and proton. One of these 
states, the narrow $P^+_c(4450)$ with {a width}
$\Gamma\sim 40\,{\rm MeV}$, can be 
described exploring that quarkonia are small compared to the nucleon 
size. This justifyies a multipole expansion which shows that the 
baryon-quarkonium interaction is dominated by the emission of two 
virtual chromoelectric dipole gluons in a color singlet state. 
The effective interaction \cite{Voloshin:1979uv} can be 
expressed in terms of the quarkonium chromoelectric polarizability 
$\alpha$ and nucleon EMT densities as 
$V_{\rm eff}(r) = -\,\alpha\,\frac{4\pi^2}{b}\,(\frac{g}{g_s})^2
[\nu\,T_{00}(r)-3\,p(r)]$. Here $b =\frac{11}{3}N_c-\frac23 N_f$ 
is the leading coefficient of the Gell-Mann-Low function, 
$g_s$ ($g$) is the strong coupling constant at the scale 
associated with the nucleon ({quarkonium}) size, and the 
parameter $\nu$ was estimated to be $\nu\sim 1.5$ \cite{Eides:2015dtr}.

\begin{figure}
\centering
\includegraphics[height=7.2cm]{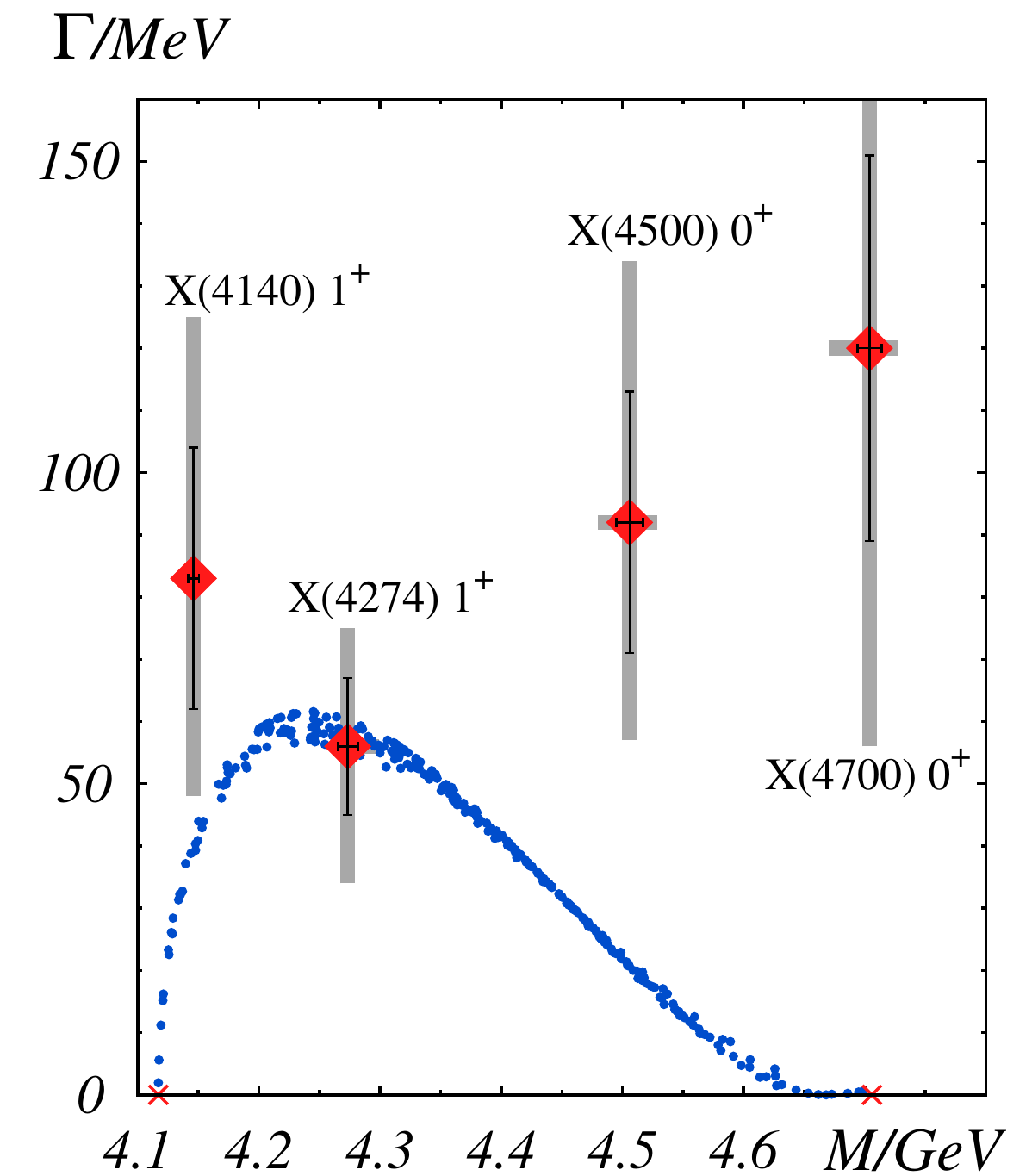} \ 
\caption{\label{FIG-01:observation-picture}
	The decay width $\Gamma$ vs mass $M$ of tetraquarks obtained 
	in Ref.~\cite{Panteleeva:2018ijz} from varying the parameters 
	which describe the unknown $\phi$-meson EMT form factors within 
	a wide range of values. 
	The crosses on the $M$-axis indicate the {kinematic} 
	bounds $m_J+m_\phi < M < m_\psi+m_\phi$. 
	For comparison we show the four tetraquarks in the $J/\psi$-$\phi$ 
	resonance region with their statistical (thin lines) and systematic 
	(shaded areas) uncertainties and spin parity assignments 
	\cite{Aaij:2016nsc}. The state $X(4274)$ emerges as a candidate 
	for the description as a hadrocharmonium. This method can be used 
	to identify other possible hadroquarkonia.}
\end{figure}

For realistic values of $\alpha(1S)$, the potential $V_{\rm eff}(r)$ 
is not strong enough to bind nucleon and $J/\Psi$ with the EMT 
densities from the chiral quark soliton. But {bound states} 
of the mass $4450\,{\rm MeV}$ exist in the $\psi(2S)$-nucleon channel 
for $\alpha(2S)\sim 17\,{\rm GeV}^{-3}$ \cite{Eides:2015dtr} which is 
close to perturbative QCD estimates of this chromoelectric 
polarizability \cite{Peskin:1979va,Bhanot:1979vb}. 
{The $J^P$ quantum numbers of these $s$-wave bound states are 
$\frac12{ }^-$ and $\frac32{ }^-$ which is among the possible 
spin-parity assignments of the LHCb partial wave analysis \cite{Aaij:2015tga}.
The 2 states are mass degenerate in the heavy quark limit. Their mass 
splitting was roughly estimated to be of ${\cal O}(20\,{\rm MeV})$
\cite{Eides:2015dtr}.}  The decay of $P^+_c(4450)$ 
is governed by the same $V_{\rm eff}(r)$ but with a much smaller
transition polarizability $\alpha(2S\to1S)\sim {\cal O}(1)$ 
\cite{Peskin:1979va,Bhanot:1979vb,Voloshin:2007dx} which
explains the relatively narrow width \cite{Eides:2015dtr}. 
The other putative pentaquark state seen by LHCb, $P^+_c(4380)$, 
is much broader with $\Gamma\sim 200\,{\rm MeV}$ and not 
described by this binding mechanism \cite{Eides:2015dtr}. 
These findings are confirmed {using Skyrme model predictions
for nucleon EMT densities} indicating they are largely model-insensitive 
\cite{Perevalova:2016dln}.
The approach {predicts also} bound states 
of $\psi(2S)$ with $\Delta$ \cite{Perevalova:2016dln} and 
hyperons \cite{Eides:2017xnt} which will allow us to test 
this theoretical framework.

One can generalise the formalism applied to heavy pentaquarks 
\cite{Eides:2017xnt} to the case of $\psi(2S)$-light meson bound states, 
the tetraquarks. The case of $\psi(2S)$-$\phi$ meson bound states was 
studied in Ref.~\cite{Panteleeva:2018ijz}. It was assumed there that the 
chromoelectric polarizability $\alpha(2S)\approx 17$~GeV$^{-3}$ is fixed 
by the pentaquark mass $P_c(4450)$. However, very little is known about 
{the $\phi$-meson EMT densities} necessary for the calculation of the 
effective $\psi(2S)$-$\phi$ potential. In Ref.~\cite{Panteleeva:2018ijz} 
the corresponding densities were therefore parametrised by flexible 
Ans\"atze whose parameters were varied in a 
{wide range of values for the $\phi$-meson square radii 
$0.05\  {\rm fm}^2 <\langle r^2\rangle_{E, \rm mech}<1\ {\rm fm}^2$ 
and $D$-term $-15<D<0$.} 
Not surprisingly a wide range was obtained for the masses $M$ of the 
corresponding tetraquarks and their partial $J/\psi$-$\phi$ decay 
widths $\Gamma$. 

Interestingly, it was found that the tetraquark masses and the partial 
decay widths are strongly correlated,\footnote{A similar correlation of 
	the mass and width was found in Ref.~\cite{Garcilazo:2018san} 
	in a different context.}
see Fig.~\ref{FIG-01:observation-picture}.  This is remarkable: even 
though we know nothing about the $\phi$-meson structure, {the formalism}
predicts that $M$ and $\Gamma$ of candidate $\phi$-$\psi(2S)$ tetraquarks 
are systematically correlated.
The Fig.~ \ref{FIG-01:observation-picture} shows also the
$J/\psi$-$\phi$ resonances observed by the LHCb collaboration 
\cite{Aaij:2016nsc}. The state $X(4274)$ observed in the 
$J/\psi$-$\phi$ channel has a width of $\Gamma=56\pm 11^{+8}_{-11}$~MeV 
\cite{Aaij:2016nsc} and fits exactly in the range predicted by the
calculations in Ref.~\cite{Panteleeva:2018ijz}.

\section{First experimental results}

{Recently first information on the $D$-terms of the proton and
the neutral pion became available from phenomenological analyses of 
experimental data. In this section we review what is currently known.}

\subsection{Nucleon}
\label{Sec:expnucleon}

The $D$-term was shown to be of importance for the 
phenomenological description of hard-exclusive reactions 
\cite{Kumericki:2007sa,Kumericki:2009uq,Mueller:2013caa,Kumericki:2015lhb},
see also the reviews \cite{Guidal:2013rya,Kumericki:2016ehc} and references 
there in.
{The $D$-term can be accessed in DVCS with help of fixed-$t$ dispersion 
relations \cite{Teryaev:2005uj,Anikin:2007yh,Diehl:2007jb,Radyushkin:2011dh}, 
for the LO DVCS Compton form factor ${\cal H}(\xi,t)=\int_{-1}^1 dx
(\frac{1}{\xi-x-i0}-\frac{1}{\xi+x-i0})\ H(x,\xi,t)$ one obtains
\be
\label{Eq:disprel}
	{\rm Re}{\cal H}(\xi,t)=\Delta(t) +\frac{ 1}{\pi} {\rm p.v.} 
	\int_0^1 d\xi' \ {\rm Im}{\cal H}(\xi',t)\left(\frac{1}{\xi-\xi'}
	-\frac{1}{\xi+\xi'}\right) \, .
\ee
The corresponding subtraction constant $\Delta(t)$ in the leading QCD order 
is related to the $D$-term in the following way:
\be
	\Delta(t)= 2 \int_{-1}^1 dz \frac{{\cal D}(z,t)}{1-z},
\ee 
with ${\cal D}(z,t)$ having the following expansion in the Gegenbauer 
polynomials $C_n^{3/2}(z)$:
\be
\label{Eq:gegen}
{\cal D}(z,t) =(1-z^2) \sum_{k=1}^\infty \left[e_u^2\ d^u_{2 k-1}(t)+e_d^2\  d^d_{2 k-1}(t)\right]\ C_{2 k-1}^{3/2}(z),
\ee
where $e_q$ is the electric charge of the quark with flavour $q$.
In the above equation we neglected contributions of strange and 
heavy quarks. The EMT form factor $D^q(t)=\frac45\,d_1^q(t)$. We remind 
that the quantities considered here ($d_1^q(t)$, ${\cal D}(z,t)$, etc.) 
depend on the QCD normalisation point $\mu^2$.
We do not write explicitly this dependence for brevity. {The QCD evolution equations for the quark and gluon $D$-term are the same as for the
second Mellin moments of the quark and gluon parton distributions.}  

\begin{figure}[b!]
\centering
\vspace{-4mm}
\includegraphics[height=10cm]{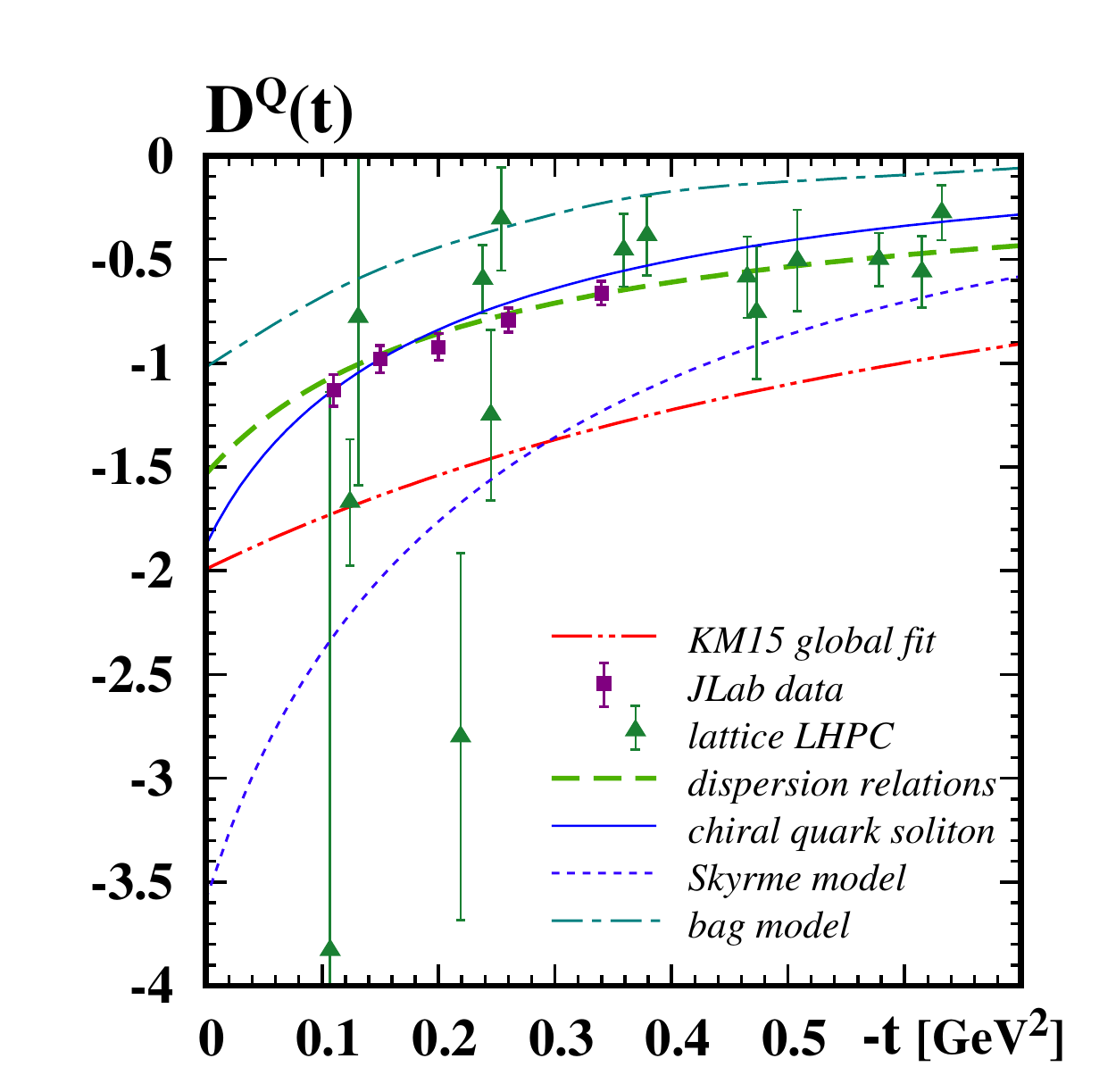} \ 
\vspace{-2mm}
\caption{\label{Fig:Nature}
The $D^Q(t)$ form factor obtained from the KM15 fit \cite{Kumericki:2015lhb} 
in comparison to $D^Q(t)$ from the Jefferson Lab analysis \cite{Nature},
calculations from dispersion relations \cite{Pasquini:2014vua}, lattice 
QCD \cite{Hagler:2007xi}, and results from the bag \cite{Ji:1997gm}, chiral 
quark soliton \cite{Goeke:2007fp} and Skyrme \cite{Cebulla:2007ei} model.
The JLab result \cite{Nature} refers to a normalisation point 
around $\mu^2=1.5$~GeV$^2$. The KM15 fit, dispersion relations and lattice 
results show the contribution of quarks to the $D$-term at the QCD 
scale of $4\,{\rm GeV}^2$. The bag, Skyrme, 
and chiral quark soliton (cf.\ footnote~\ref{Footnote-ChQSM})
models show the total $D$-term form factor 
which is renormalization scale independent.} 
\end{figure}

The first experimental access to the subtraction constant $\Delta(t,\mu^2)$ 
based on the most complete database of DVCS results was obtained in 
\cite{Kumericki:2015lhb} (KM15 fit) in the form: 
\be
\label{Eq:KM15}
\Delta(t,\mu^2)=- \frac{C}{\left(1-t/M_C^2\right)^2},
\ee
with parameters $C=2.768$ and $M_C=1.204$~GeV at the QCD normalisation 
point of $\mu^2=4$~GeV$^2$. The statistical uncertainty of the parameters 
are of order $20-30$\% \cite{KresimirPK}, but the authors of 
Ref.~\cite{Kumericki:2015lhb} refrained from publishing the precise value 
of the statistical error bars due to large systematic uncertainties 
(see the discussion of this point in relation to the $D$-term in 
Ref.~\cite{Kresimir}) \footnote{We are grateful to 
	Kresimir Kumeri\v cki for discussion of this point.}. 

We can relate the LO subtraction constant $\Delta(t,\mu)$ to the EMT 
form factor $D^Q(t,\mu^2)=D^d(t,\mu^2)+D^d(t,\mu^2)$
using the following simplifying assumptions:
\begin{itemize}
\item only the first coefficient $d_1^q(t)$ of the Gegenbauer expansion 
(\ref{Eq:gegen}) is taken into account. In the asymptotic limit of infinitely 
large renormalization scale $\mu$ all $d_i^q(t)$ for $i>1$ 
vanish, except for $d_1^q(t)$ which determines the
asymptotic form of GPDs \cite{Goeke:2001tz} and is related 
to the EMT form factor $D^q(t)=\frac45\,d_1^q(t)$;
\item dominance of the flavour singlet combination of the quark 
$D$-term $d_1^u\approx d_1^d \approx d^Q/2$. This can be justified by
in the limit of large number of colours, see Eq.~(\ref{Eq:D-large-Nc}).
\end{itemize}
Under these assumptions we obtain:
\be
\label{Eq:KM15_assumptions}
	D^Q(t)=\frac{4}{5}\;
	\frac{1}{2 (e_u^2+e_d^2)}\Delta(t)=\frac{18}{25}\ \Delta(t).
\ee
The KM15 fit Eq.~(\ref{Eq:KM15}) corresponds to the negative $D$-term of $D^Q=-2.0$ at $\mu^2=4$~GeV$^2$ with about 20\% statistical uncertainty 
and unestimated systematic one.
The result of the KM15 fit \cite{Kumericki:2015lhb} corresponding to 
Eqs.~(\ref{Eq:KM15},\ref{Eq:KM15_assumptions}) is shown in Fig.~\ref{Fig:Nature}
in comparison with theoretical predictions and other fits to DVCS data.}

Recently an analysis of the JLab data \cite{Girod:2007aa,Jo:2015ema}\footnote{These data are included in the experimental database of Ref.~\cite{Kumericki:2015lhb} } 
 was reported \cite{Nature} where an experimental information on 
the quark contribution to the $D$-term was {also} extracted. Additionally, the pressure distribution in the proton was presented in Ref.~\cite{Nature}.
Below we compare the theoretical predictions with the data on 
the form factor, and not with the pressure distribution of \cite{Nature}
as the latter was obtained under model assumptions which are 
still missing clear justification.

{In Ref.~\cite{Nature} the dispersion relations subtraction constant 
$\Delta(t)$ (see Eq.~(\ref{Eq:disprel}) for the definition) at the 
normalisation point of $\mu^2=1.5$~GeV$^2$ was presented on their 
Fig.~4 \cite{BEGPC}.
The main difference of the analysis in \cite{Nature} with that in 
\cite{Kumericki:2015lhb} is the much smaller systematic uncertainties 
in the former. This difference calls for a clarification.}

The $D^Q(t)$ form factor obtained from the analysis of \cite{Nature} with help 
of Eq.~(\ref{Eq:KM15_assumptions}) is also shown in Fig.~\ref{Fig:Nature}
where for comparison we include the results for the $D$-term form 
factor from dispersion relations \cite{Pasquini:2014vua}, lattice QCD 
\cite{Hagler:2007xi} and models \cite{Ji:1997gm,Goeke:2007fp,Cebulla:2007ei}.

The dispersion relation study of Ref.~\cite{Pasquini:2014vua} used information 
on pion parton distribution functions which fixes the overall 
normalization of the form factor: in Fig.~\ref{Fig:Nature} 
the result for $D^Q(t)$ is shown which is normalized as 
$D^Q=-1.56$. 
The results from the dispersion relations and lattice QCD show 
the quark contribution to $D^Q(t)$ and refer 
to the scale $\mu^2=4\,{\rm GeV}^2$ \cite{Pasquini:2014vua,Hagler:2007xi}.
The lattice data were obtained in a hybrid approach using domain wall 
valence quarks with $2+1$ flavors of improved staggered sea quarks not 
including disconnected diagrams. The ``dataset 6'' from \cite{Hagler:2007xi}
shown in Fig.~\ref{Fig:Nature} was taken on a $28^3\times32$ lattice with 
a lattice spacing $a=0.124\,{\rm fm}$ and a pion mass of 
$m_\pi=(352.3\pm 1.4)\,{\rm MeV}$. 
The results from bag \cite{Ji:1997gm}, Skyrme \cite{Cebulla:2007ei},
and chiral quark soliton\footnote{\label{Footnote-ChQSM} The 
   chiral quark soliton model is based on the instanton picture of the 
   QCD vacuum which is valid at a low scale set by the inverse instanton 
   size $\rho_{\rm av}^{-1}\simeq 0.6\,{\rm GeV}$, see e.g.\ the review by 
   D.~I.~Diakonov \cite{Diakonov:2002fq}. 
   Therefore the $D$-term calculation \cite{Goeke:2007fp} can also 
   be viewed as the quark contribution $D^Q(t)$ at the normalisation point 
   of $\mu^2\simeq \rho_{\rm av}^{-2}\simeq 0.4$~GeV$^2$, see the
   discussion in \cite{Diakonov:1996sr}. 
   The instanton calculus allows one to evaluate systematically hadronic 
   matrix elements of QCD operators in an expansion in powers of the small 
   parameter provided by the instanton packing fraction. The instanton 
   contribution to the gluon D-term is suppressed by this small parameter 
   and will be studied in future works.}
\cite{Goeke:2007fp} model show the total scale--independent $D(t)$.

Keeping all this in mind, Fig.~\ref{Fig:Nature} shows a
remarkable agreement. The MIT bag model \cite{Ji:1997gm}
seems to underestimate the magnitude of the $D$-term form factor,
the Skyrme model \cite{Cebulla:2007ei} seems to overestimate it 
(though, with different parameter fixing than in 
\cite{Cebulla:2007ei}, a better discription may be possible). 
The results from dispersion relations \cite{Pasquini:2014vua} and
chiral quark soliton model \cite{Goeke:2007fp} compare  very well 
to the experimental results from \cite{Nature}.

\subsection{Pion}

Recently in Ref.~\cite{Kumano:2017lhr} the first extraction of the pion 
EMT form factors from the BELLE data on $\gamma^*\gamma\to 2 \pi^0$
\cite{Masuda:2015yoh} was reported. The results for the quark part of 
EMT form factors ({$A^Q(t)$ and $D^Q(t)$} in our notation, 
see Eq.~(\ref{Eq:EMT-FFs-spin-0})) were presented. The 
results for the form factors at zero momentum transfer are:
\be
	A^Q(0) \approx 0.70, \quad D^Q(0)\approx -0.75.
\ee
These results are in agreement with the normalisation condition for the 
{\it full} form factor $A(0)=1$ and with the soft pion theorem $D=-1$, 
given that quarks carry only a fraction (about 70\% according to the 
result of Ref.~\cite{Kumano:2017lhr}) of the pion mass, and the gluon 
contribution to the $D$-term is not extracted. Also it is important that 
the analysis of \cite{Kumano:2017lhr} shows that the $D$-term is definitely 
negative as it should be for mechanical stability of the pion.
The result obtained in \cite{Kumano:2017lhr} for the slopes of the pion EMT 
form factors are:
\be
	\frac{1}{A^Q(0)}\frac{d}{dt}A^Q(0) = 1.33\sim2.02~{\rm GeV}^{-2}, \quad 
	\frac{1}{D^Q(0)}\frac{d}{dt}D^Q(0) = 8.92\sim10.35~{\rm GeV}^{-2}.
\ee
These results confirm the inequality $-D'(0)>A'(0)$ expected from chiral 
theory, however the numerical values are in sharp contrast with our estimate  
(\ref{eq:slopesNum}) based on the instanton picture of QCD vacuum combined 
with the chiral perturbation theory. It would be very important to understand 
which dynamical mechanism leads to anomalously large slopes of the pion EMT 
form factors obtained in analysis of Ref.~\cite{Kumano:2017lhr}.

\section{Conclusions}

We have reviewed aspects of the physics associated with the $D$-term 
and other EMT properties. The physics of EMT form factors is important 
for a variety of problems including the description of hadrons in strong 
gravitational fields, hard exclusive processes, hadronic decays of heavy 
quarkonia, and the description of certain exotic hadrons with hidden 
charm as hadroquarkonia.

The matrix elements of the EMT contain fundamental information on a
particle, namely the mass, spin, and $D$-term. While mass and spin
are related to the Casimir operators of the Poincar\'e group, the 
$D$-term is related to the stress tensor and internal forces inside 
a composed particle.
When interpreted in the Breit frame the Fourier transforms of the 
EMT form factors give insights on the 3D spatial densities describing 
the distributions of energy, pressure and shear forces. 

In free field theory the $D$-term of a spin-zero boson is negative,
but that of a spin $\frac12$ fermion is zero. This indicates an
interesting distinction of bosons and fermions. 

In interacting theories the $D$-term in general is not fixed, except
for the Goldstone bosons of chiral symmetry breaking for which the
$D$-term is determined by soft-pion theorems to be $D=-1$ in the chiral
limit. For other hadrons the $D$-term is not fixed, and reflects the 
internal dynamics of the system through the distribution of forces,
and is sensitive to correlations in the system. For example, the 
baryon $D$-term behaves as $\sim N_c^2$ whereas all other global 
observables (mass, magnetic moments, axial charge, etc.) behave at 
most as $\sim N_c$ in the large $N_c$ limit. For a large nucleus the 
$D$-term shows also anomalously fast increase with the atomic mass
number $D\sim A^{7/3}$.

The form factor $D(t)$ provides the key to introduce mechanical
properties. For instance, we have given a definition of the
mechanical radius of a hadron, discussed the concepts of normal
and tangential forces, and presented (on the basis of model results)
a picture of the forces inside the nucleon. Remarkably, the forces
change their directions in the inner region of the nucleon vs
the outer region. This change of sign is dicated by the conservation
of the EMT, but the physics behind it is presently not fully understood.

We have reviewed results from models, lattice QCD, dispersion relations
and the experimental information on the $D$-term of the nucleon
based on DVCS data, and the neutral pions based
on Belle data. The theoretical approaches and the first phenomenological
extractions agree that the $D$-term is negative, as predicted on the
basis of stability requirements.

Open questions include the issue of the mass decomposition of
the nucleon \cite{Ji:1994av,Ji:1995sv} and how it is related to 
the internal forces \cite{Lorce:2017xzd}.
It is a legitim but unanswered question whether the forces encoded in
the EMT can tell us interesting lessons about confinement. Another
interesting question is whether one can define further mechanical
properties of hadrons such as e.g.\ speed of sounds inside a hadron.
It would be also interesting to investigate the relation (if it
exists) to hydrodynamical transport properties such as viscosity 
in strongly interacting systems in heavy ion collisions. 
The analogy of the $D$-term to the vacuum cosmological constant
observed in \cite{Teryaev:2013qba} is a further topic worthwhile 
investigating. These topics will be discussed in future studies.

\ \\
\noindent{\bf Acknowledgments.}
We are grateful to M.~Cantara, J.~Hudson, 
J.~Panteleeva, B.~Pasquini, I.~Perevalova, and M.~Vanderhaeghen
for collaboration on various topics discussed in this review. We 
are thankful to 
Stan Brodsky, Kresimir Kumeri\v cki, C\'edric Lorc\'e and 
Oleg Teryaev for illuminating 
discussions and inspiration.
This work was supported by NSF (grant no.\ 1406298),
the Wilhelm Schuler Stiftung,
and by CRC110 (DFG).


\appendix

\section{Relations for EMT densities}
\label{App:SR-EMT-densities}

Due to EMT conservation we have generically
\ba\label{Eq:App-Z-1}
	\int_V d^3r\,\underbrace{(\nabla^i T^{ij})}_{=0}X^{jklm\dots} = 0
	\quad	\Leftrightarrow \quad
	\int_V d^3r\,T^{ij} (\nabla^iX^{jklm\dots}) = 
	\oint_{S(V)} d a^i\;T^{ij}\,X^{jklm\dots}
\ea
where $X^{jklm\dots}$ is some differentiable tensor which can be used
in integration by parts. $V$ is some finite or infinite volume.
In our situation
(no spin effects for spin 0 and spin 1/2, spherical symmetry) it 
makes sense to choose one free index: we define $X^j = r^j\,f(r)$
where $f(r)$ is an arbitrary function which can be used in integration
by parts. 

Let us choose now the volume to be infinite and $f(r)$ such
that the surface term in Eq.~(\ref{Eq:App-Z-1}) can be neglected.
After performing the integration by parts and inserting the 
expression (\ref{Eq:stress-tensor-p-s}) for the stress tensor
we obtain
\be
	I[f(r)] \equiv \int d^3r\,T^{ij} [\nabla^i(x^j f(r))] =
	\int d^3r\biggl(\frac23\,r\,s(r)\,f^\prime(r)+r\,p(r)\,f^\prime(r)
	+3\,p(r)\,f(r)\biggr) = 0\,.
\ee
This is a ``master integral'' for deriving 
relations for the densities of the stress tensor. Let us explore different
choices for the function $f(r)$:
\begin{itemize}
\item 	{\bf\boldmath choice $f(r)=1$:}
 	We recover the von Laue condition, $I[1] = 3\int d^3r\, p(r)=0$. 
\item 	{\bf\boldmath choice $f(r)=r^N$:}
	We obtain
	$I[r^N]=4\pi\int_0^\infty  d r\;r^{N+2}[\frac23 \,N\,s(r)+(N+3)\,p(r)]=0$
	which is eqivalent to the ``Mellin-moment-relations'' for $p(r)$
	and $s(r)$ derived in App.~B or Ref.~\cite{Goeke:2007fp}.
	For $N=0$ we obtain the previous relation, for $N=-3$ we 
	reproduce the Kelvin relation (\ref{Eq:generelized-Kelvin-relation}).
\item	{\bf\boldmath special case $N=-1$:} this yields
	$I[1/r]=4\times2\pi\int_0^\infty dr\ r\left[-\frac 13 s(r)+p(r)\right]=0$
	and we recover the sum rule for tangential forces 
	(\ref{Eq:tang-stability}) obtained from physics consideration 
	of forces acting on spherical slice of a hadron.
\item	{\bf\boldmath special case $N=-2$:} this case yields
	$I[1/r^2]=4\pi\int_0^\infty dr\ \left[-\frac 43 s(r)+p(r)\right]=0$,
   	which corresponds to the stability condition (\ref{Eq:1Dstability}) 
	for the 1D subsystems in a hadron. 
	
\item 	{\bf\boldmath choice $f(r)=\Theta(R-r)$:} we obtain the
	relation $\int_{V(R)} d^3r\;p(r) = V(R)\,[\frac23s(R)+p(R)]$
	derived in Eq.~(\ref{Eq:finite-Laue-integral}) of this work.
	Here we defined $V(R)=\frac43\,\pi\,R^3$. Notice that this
	``trial function'' is not differentiable, but it does not
	matter. All we need is that it does not contribute to the
	integral at infinity!
\item	{\bf\boldmath choice $f(r)=\delta(r-R)$:} this choice yields the 
	well-known relation $\frac23\,s^\prime(R)+\frac2R\,s(R)+p^\prime(R)=0$ 
	which is equivalent to EMT conservation $\nabla^i T^{ik}=0$.

\item	{\bf\boldmath choice $f(r)=p(r)$:}
      	we obtain the following fascinating {\it non-linear}
	integral sum rule for EMT densities, which was shown 
	above in Eq.~(\ref{Eq:spectacular-rewritten}) in equivalent form, 
	\be
	\label{Eq:spectacular}
		\int d^3r \biggl(\frac23\,s(r)^2-\frac32\,p(r)^2\biggr)=0\,.
	\ee
\end{itemize}

\end{document}